\documentclass[final,3p,times, onecolumn]{elsarticle}

\usepackage{latexsym}
\usepackage{bm}

\usepackage{amsmath}
\usepackage{graphicx}   
\usepackage{enumerate}

\usepackage{amssymb}

\begin{document}

\begin{frontmatter}



\title{Non-coordinates basis in General Relativity and Cartan's structure equations}


\author{Wytler Cordeiro dos Santos}
\ead{wytler@fis.unb.br}
\address{Universidade de
Bras\'\i lia, CEP 70910-900, DF, Brasil}

\begin{abstract}
The basic and fundamental aspects of General Relativity are in general analysed in mathematical level of coordinate basis or holonomic frame by several authors in the literature. However, for many purposes it is more convenient to use a general basis, often called in four dimensions, a tetrad or vierbein, very useful in a local frame with orthonormal basis or pseudo-orthonormal basis. 
This text presents an introduction to non-coordinate basis and the two  Cartan's structure equations that are mathematical implements in Riemannian geometry that facilitate the calculation of curvature tensors.
The purpose of this text is to approach the language and the notation of tetrad field or vierbein  with conceptual and calculational details.
\end{abstract}


\end{frontmatter}


\section{Introduction}

The foundations for a precise and mathematical formulation of General Relativity
is obtained with basic ideas of
differential geometry, i.e.: basic properties of manifolds and tensor fields.
However in a level of basic courses of General Relativity, such as of references
\cite{DInverno, Landau, Hobson, Schutz, Wald, Weinberg}, we have worked just only with coordinate basis or world indices. In some advanced topics of General Relativity it is necessary to work with objects defined on manifold as torsion and curvature in other perspective, i.e.: in a non-coordinate basis or local indices. In a non-coordinate basis, it arises the structure equations of Cartan. The Cartan formalism works with two equations that approach torsion and curvature in the differential forms language.

The discussion of this text begins with vectors and tensors on manifolds,  perhaps it is util to revisite some discussions and approaches in differential manifolds in references \cite{Hawking, Nakahara, Einsenhart, Boothby, Nash}. Coming after the differential forms, exterior product an exterior derivatives,  definition of torsion and curvature tensors and metric.  
Here we focus on vectors and tensors on Riemannian  manifold with Lorentzian signature of spacetime metric: (-+++), and then we introduce the local frame with orthonormal basis and pseudo-orthornormal basis. The pseudo-orthonormal basis is commonly known by complex null tetrad that is very useful to Newman-Penrose formalism \cite{Stephani, Stewart, Newman}. The structure equations of Cartan are presented as a compact and efficient method for calculating the components of curvature tensor with respect to a non-coordinate basis, and then as examples are presented the Schwarzschild and Brinkmann spacetimes.

\section{Riemannian Geometry}

In a nutshell, a manifold ${\cal M}$ is a topological space which locally resembles to  $\mathbb{R}^n$, where a homeomorphism enable us to obtain in each point in a manifold a set of numbers called the local coordinate. The inner product between two vectors in a tangent space of a manifold is defined with aid of a metric tensor, in this case the manifold is named Riemannian manifold.

\subsection{Tangent vectors and differential forms}

In general a vector can not be considered as an arrow connecting two points of
the manifold. For a consistent generalization 
of the concept of vector in  $\mathbb{R}^n$, we can identifies vector on
manifold $\cal M$ with tangent vector. 
So, a tangent vector $\bm V$ at $p \in {\cal M}$ is a linear operator which
assigns to each differentiable function $f$
on $\cal M$ a real number $\bm V(f)$. This operator satisfies the axioms:
\begin{enumerate}
 \item $\bm V(f+g) = \bm V(f) + \bm V(g)$
 \item $\bm V(fg) = g\bm V(f) + f\bm V(g)$
  \item $\bm V(cf) = c\bm V(f)$, $c$ is a constant.
\end{enumerate}
A tangent vector is just a directional derivative along a curve $\gamma(t)$
through $p \in {\cal M}$
\begin{equation}
\label{directionl_derivative_1}
 \left( \frac{\partial f}{\partial t} \right)_\gamma \Bigg\arrowvert_{t=t_0} =  
 \frac{\partial f}{\partial x^{\mu}} \Bigg\arrowvert_{\gamma(t_0)}\frac{d
x^{\mu}}{dt} \Bigg\arrowvert_{t=t_0} =
 \frac{d x^{\mu}}{dt}  \frac{\partial f}{\partial x^{\mu}}
\Bigg\arrowvert_{\gamma(t_0)}.
\end{equation}
The directional derivative of a function $f$ is obtained by applying the
differential operator $\bm V$ to $f$, where
\begin{equation}
\label{directionl_derivative_2}
 \bm V = V^{\mu}\frac{\partial}{\partial x^{\mu}}, \hspace*{1cm}\mbox{where}
\,\, 
 V^{\mu}=\frac{d x^{\mu}}{dt} \Bigg\arrowvert_{t=t_0} 
\end{equation}
that is 
\begin{equation}
\label{directionl_derivative_3}
 \left( \frac{\partial f}{\partial t} \right)_{\gamma} \Bigg\arrowvert_{t=t_0} =
 V^{\mu}\frac{\partial f}{\partial x^{\mu}} = \bm V[f].
\end{equation}
Thus every tangent vector at a point $p$ can be expressed as a linear
combination of the coordinate derivatives
$\dfrac{\partial}{\partial x^{\mu}}$. The directional derivatives along the
coordinate lines at $p$ form a basis of an
$n$-dimensional vector space whose elements are the tangent vector at $p$. This
space is called the tangent space
$\bm T_p$ and the basis $\left\{\dfrac{\partial}{\partial x^{\mu}}\right\}$ is
called a coordinate basis or 
holonomic frame. Thus every tangent vector at a point $p$ can be expressed as a
linear combination of the coordinate 
derivatives $\left\{\dfrac{\partial}{\partial x^{\mu}}\right\}$. Any tangent
vectors  to $\cal M$ at $p$ of 
the $n$-dimensional  vector space $\bm T_p$ are referred as contravariant vector.

The commutator, $[\bm U, \bm V ]$, of two vector fields $\bm U$ and  $\bm V$ is defined by
\begin{equation}
 [\bm U, \bm V ](f) = \bm U(\bm V(f)) - \bm V(\bm U(f)), \nonumber
\end{equation}
for a given coordinates basis  $\bm U = U^{\kappa}\dfrac{\partial}{\partial x^{\kappa}}$ and  
$\bm V = V^{\lambda}\dfrac{\partial}{\partial x^{\lambda}}$ we have that
\begin{eqnarray}
 [\bm U, \bm V]&=&U^{\kappa}\frac{\partial}{\partial x^{\kappa}}\left(V^{\lambda}\frac{\partial}{\partial x^{\lambda}}\right)-
 V^{\lambda}\frac{\partial}{\partial x^{\lambda}}\left(U^{\kappa}\frac{\partial}{\partial x^{\kappa}}\right)
\cr
&=& U^{\kappa}\left(\frac{\partial V^{\lambda}}{\partial x^{\kappa}}\right)\frac{\partial}{\partial x^{\lambda}}+
U^{\kappa}V^{\lambda}\frac{\partial^2}{\partial x^{\kappa}\partial x^{\lambda}}-
 V^{\lambda}\left(\frac{\partial U^{\kappa}}{\partial x^{\lambda}}\right)\frac{\partial}{\partial x^{\kappa}}
 -V^{\lambda}U^{\kappa}\frac{\partial^2}{\partial x^{\lambda} \partial x^{\kappa}}
\cr
&=& U^{\kappa}\left(\frac{\partial V^{\lambda}}{\partial x^{\kappa}}\right)\frac{\partial}{\partial x^{\lambda}}-
 V^{\lambda}\left(\frac{\partial U^{\kappa}}{\partial x^{\lambda}}\right)\frac{\partial}{\partial x^{\kappa}}
 \cr
&=& \left( U^{\kappa}\frac{\partial V^{\lambda}}{\partial x^{\kappa}}-
 V^{\kappa}\frac{\partial U^{\lambda}}{\partial x^{\kappa}}
 \right)\frac{\partial}{\partial x^{\lambda}} = W^{\lambda}\frac{\partial}{\partial x^{\lambda}} ,\nonumber
\end{eqnarray}
where $W^{\lambda} = U^{\kappa}\dfrac{\partial V^{\lambda}}{\partial x^{\kappa}}- V^{\kappa}\dfrac{\partial U^{\lambda}}{\partial x^{\kappa}}$. Thus, the 
commutator defines a new vector
\begin{equation}
\label{comutador_UV}
 [\bm U, \bm V ] = \bm W.
\end{equation}
The commutators satisfy the Jacobi identity
\begin{equation}
 [\bm U,[\bm V, \bm W]] + [\bm V,[\bm W, \bm U]] + [\bm W,[\bm U, \bm V]] = 0
\end{equation}
for arbitrary $\bm U$, $\bm V$ and $\bm W$.


A one-form (1-form) or covariant vector or dual vector $\bm\omega$ at $p$ maps a tangent
vector $\bm V$ into a real number,
the contraction, denoted by $\langle \bm\omega,\bm V \rangle$, and this mapping
is linear:
\begin{equation}
 \langle \bm\omega,a\bm U + b\bm V \rangle = a \langle\bm\omega,\bm U\rangle + b
 \langle\bm\omega,\bm V\rangle
\end{equation}
holds for all $a,b\in \mathbb{R}$ and $\bm U, \bm V \in \bm T_p$. Linear
combinations of 1-forms 
$\bm\omega,\bm\tau$ are defined by the rule
\begin{equation}
\langle a\bm\omega+ b\bm\tau, \bm V \rangle = a \langle \bm\omega \bm V \rangle
+ b \langle \bm\tau, \bm V \rangle.
\end{equation}
An arbitrary 1-form $\bm \omega$ is written as a linear combination,
\begin{equation}
 \bm \omega = \omega_{\mu} dx^{\mu},
\end{equation}
where $\left\{dx^{\mu}\right\}$ is a dual basis to the basis
$\left\{\dfrac{\partial}{\partial x^{\mu}}\right\}$ of 
tangent vectors, being called cotangent basis.
The 1-forms form a basis of $n$-dimensional dual vector space called the dual
or cotangent space $\bm T_p^{*}$ with
basis $\left\{dx^{\mu}\right\}$ with conditions
\begin{equation}
 \left\langle dx^{\mu}, \dfrac{\partial}{\partial x^{\nu}}\right\rangle = \frac{\partial
x^{\mu}}{\partial x^{\nu}} =
 {\delta^{\mu}}_{\nu}.
\end{equation}
The contraction $\langle\bm\omega, \bm V \rangle$ is the inner product between a
contravariant vector $\bm V$ and a 
covariant or dual vector $\bm\omega$. For $\bm\omega \in \bm T_p^{*}$ and $\bm V
\in \bm T_p$ 
we have that inner product is
\begin{equation}
  \langle \bm\omega,\bm V \rangle =  \left\langle \omega_{\mu} d x^{\mu} ,
  V^{\nu}\frac{\partial}{\partial x^{\nu}} \right\rangle
  =\omega_{\mu}V^{\nu}  \left\langle d x^{\mu} ,\frac{\partial}{\partial
x^{\nu}} \right\rangle 
  = \omega_{\mu}V^{\nu}  {\delta^{\mu}}_{\nu} = \omega_{\mu}V^{\mu}.
\end{equation}
In terms of the dual basis, the differential $df$ on an arbitrary function $f$
is given by
\begin{equation}
 df = \frac{\partial f}{\partial x^{\mu}} d x^{\mu}.
\end{equation}

\subsection{Tensors}
A tensor of type $(q,r)$ at a point $p$ is a multilinear object which maps $q$
elements of $\bm T_p$ and $r$
elements of $\bm T_p^{*}$ to a real number. ${\cal T}_r^q({\cal M})$ at $p$
denotes the set of type $(q,r)$ tensors
at $p\in {\cal M}$, by tensor product
\begin{equation}
 {\cal T}_r^q({\cal M})_{p} = \underbrace{{\bm T}_p \otimes \cdots
\otimes {\bm T}_p}_{q\,\mbox{factors}}
 \otimes \underbrace{{\bm T}_p^{*} \otimes \cdots \otimes {\bm
T}_p^{*}}_{r\,\mbox{factors}}.
\end{equation}
In particular, $ {\cal T}_0^1 = \bm T_p$ and $ {\cal T}_1^0 = \bm T_p^{*}$. An
element of 
${\cal T}_r^q({\cal M})$ is written in terms of the basis
$\left\{\dfrac{\partial}{\partial x^{\mu}}\right\}$
and  $\left\{dx^{\nu}\right\}$ as
\begin{equation}
 \mbox{\bf T} = {T^{\mu 1\cdots \mu q}}_{\nu 1 \cdots \nu r}
\frac{\partial}{\partial x^{\,\mu 1}}\otimes \cdots
\otimes  \frac{\partial}{\partial x^{\,\mu q}}\otimes dx^{\nu 1} \otimes \cdots
\otimes dx^{\nu r},
\end{equation}
where all indices run from 1 to $n$. The coefficients ${T^{\mu 1\cdots \mu
q}}_{\nu 1 \cdots \nu r}$ with
contravariant indices $\mu 1 \cdots \mu q $ and covariant indices $\nu 1 \cdots
\nu r $ are the components of
{\bf T} with respect to the basis  $\left\{\dfrac{\partial}{\partial
x^{\mu}}\right\}$
and  $\left\{dx^{\nu}\right\}$.

\subsection{Differential forms, exterior product and exterior derivatives}
A differential form of order $r$, or $r$-form, is a totally antisymmetric tensor
of type $(0,r)$. 
If $\bm \rho$ and $\bm\sigma$ are $r$-form and $s$-form respectively, one can
define a $(r+s)$-form
$\bm\rho \wedge \bm\sigma$ from them, where $\wedge$ is the skew-symmetrized
tensor product $\otimes$.
The exterior product or wedge product $\bm\rho \wedge \bm\sigma$ is the tensor
of type $(0,r+s)$ \cite{Flanders}. For example, if
\begin{equation}
 (\mbox{\bf i})\hspace*{1cm} \bm \rho = r_{\mu}dx^{\mu} 
 \hspace*{1cm} \mbox{and} \hspace*{1cm} \bm \sigma = s_{\nu}dx^{\nu} \nonumber
\end{equation}
$\bm \rho$ and $\bm\sigma$ are 1-form, then
\begin{equation}
 \bm\rho \wedge \bm\sigma =  r_{\mu} s_{\nu}(dx^{\mu}\otimes dx^{\nu} - dx^{\nu}
\otimes dx^{\mu}) =  r_{\mu} s_{\nu}\, dx^{\mu}\wedge dx^{\nu} = -\bm\sigma \wedge \bm\rho, \nonumber
\end{equation}
$\bm\rho \wedge \bm\sigma$ is a 2-form.

Now if
\begin{equation}
 (\mbox{\bf ii})\hspace*{1cm} \bm \rho = r_{\lambda}dx^{\lambda} ,\hspace*{1cm}
\bm \sigma = s_{\mu}dx^{\mu}
 \hspace*{1cm} \mbox{and} \hspace*{1cm} \bm \tau = t_{\nu}dx^{\nu} \nonumber
\end{equation}
then
\begin{equation}
 \bm\rho \wedge \bm\sigma \wedge \bm\tau =  r_{\lambda} s_{\mu}t_{\nu}
 (dx^{\lambda}\otimes dx^{\mu}\otimes dx^{\nu} + dx^{\nu}\otimes
dx^{\lambda}\otimes dx^{\mu}+ 
 dx^{\mu}\otimes dx^{\nu}\otimes dx^{\lambda} - dx^{\lambda}\otimes
dx^{\nu}\otimes dx^{\mu} -
 dx^{\nu}\otimes dx^{\mu}\otimes dx^{\lambda}- dx^{\mu}\otimes
dx^{\lambda}\otimes dx^{\nu}) 
 \nonumber
\end{equation}
or 
\begin{equation}
 \bm\rho \wedge \bm\sigma \wedge \bm\tau = r_{\lambda} s_{\mu}t_{\nu}\, dx^{\lambda} \wedge dx^{\mu} \wedge dx^{\nu}, \nonumber
\end{equation}
$\bm\rho \wedge \bm\sigma \wedge \bm\tau$ is a 3-form.

For the skew-symmetrized tensor product $\wedge$, it is verified that the
exterior product satisfies the 
fllowing axioms:
\begin{enumerate}
 \item is linear in each $dx^{\mu}$
 \item vanish if any two factors coincide
 \item changes sign if any two factors are interchanged.
\end{enumerate}
If we denote the vector space of $r$-forms at point $p\in{\cal M}$ 
by $\bm \Omega^{r}({\cal M})$ and an element $\bm\rho\in \bm \Omega^{r}({\cal
M})$ is expaned as
\begin{eqnarray}
 \bm\rho = r_{\mu 1,\mu 2,\cdots, \mu r}\,dx^{\mu 1}\wedge dx^{\mu 2}\wedge
\cdots \wedge dx^{\mu r},\cr
 1\leq \mu 1 < \mu 2< \cdots < \mu r \leq n,\hspace*{1cm} \mbox{with}\,\, r\leq
n.\nonumber 
\end{eqnarray} 
The above axiom 2 implies that these exterior products vanish for  $r>n$.

We have already stated that
$r_{\mu 1,\mu 2\cdots \mu r}$ are totally antisymmetric, then there are $r$-combinations
$\begin{pmatrix} n\cr r\end{pmatrix}$  of choices of set 
(${\mu 1,\mu 2,\cdots,\mu r}$), thus the dimensional of the vector space 
$\bm \Omega^{r}({\cal M})$ is $\begin{pmatrix} n\cr r\end{pmatrix}$. For example, a manifold  which has
$\dim({\cal M})= 3$ with $dx$, $dy$ and $dz$. The 2-forms: $dx\wedge dy$,  
$dx\wedge dz$ and $dy\wedge dz$ are oriented area elements. $dx\wedge dy \wedge dz$ is oriented
volume element. We have:
\begin{enumerate}
 \item $\{dx,dy,dz\} \in \bm \Omega^{1}({\cal M}) = \bm T^{*}({\cal M})$ where $\Omega^{0}({\cal
M})$ is vector space of smooth functions. $\bm \Omega^{1}({\cal M}) $ has dimension 
$\begin{pmatrix} 3\cr 1\end{pmatrix}=3$.
  \item $\{dx\wedge dy,\, dx\wedge dz,\, dy\wedge dz\} \in \bm \Omega^{2}({\cal M})$ where 
$\bm \Omega^{2}({\cal M})$ has dimension $\begin{pmatrix} 3\cr 2\end{pmatrix}=3$.
  \item $\{dx\wedge dy \wedge dz\} \in \bm \Omega^{3}({\cal M})$  where 
$\dim (\bm \Omega^{3}({\cal M}))= \begin{pmatrix} 3\cr 3\end{pmatrix}=1$.
\end{enumerate}

Let $\bm \rho \in \bm \Omega^{r}({\cal M})$,  $\bm \sigma \in \bm \Omega^{s}({\cal M})$ and 
 $\bm \tau \in \bm \Omega^{t}({\cal M})$, the exterior product is associative:
\begin{equation}
 (\bm\rho \wedge \bm\sigma)\wedge \bm\tau =  \bm\rho \wedge (\bm\sigma\wedge \bm\tau).
\end{equation} 
However, the commutative law is slighty changed:
\begin{equation}
 \bm\rho \wedge \bm\sigma = (-1)^{rs} \bm\sigma \wedge  \bm\rho.
\end{equation} 
The definition of an $r$-form is extended to include ordinary function  $f$ on manifold $\cal M$. Functions on $\cal M$ are called 0-form; 1-form is a covariant vector, 2-form is an antisymmetric covariant tensor of rank 2 and so on.

If $\bm \rho$ is an $r$-form given by
\begin{equation}
\bm\rho = r_{\mu 1,\mu 2,\cdots, \mu r}\,dx^{\mu 1}\wedge dx^{\mu 2}\wedge
\cdots \wedge dx^{\mu r}, \nonumber
\end{equation}
then the exterior derivative of $\bm \rho$ is written $d\bm \rho$ and is defined by
\begin{equation}
d\bm\rho = \frac{\partial r_{\mu 1,\mu 2,\cdots, \mu r}}{\partial x^{\nu}}\,dx^{\nu}\wedge dx^{\mu 1}\wedge dx^{\mu 2}\wedge
\cdots \wedge dx^{\mu r},
\end{equation}
The exterior derivative $d$ is a map $\bm \Omega^{r}({\cal M}) \rightarrow \bm \Omega^{r+1}({\cal M})$.
For example: if ${\cal M} = \mathbb{R}^3$ and
\begin{enumerate}
\item 0-form: $f = f(x,y,z)$ then the action of exterior derivative is 
\begin{equation}
df = \frac{\partial f}{\partial x} dx + \frac{\partial f}{\partial y} dy + \frac{\partial f}{\partial z} dz,
\end{equation}
it is identified with gradient;
\item 1-form: $\bm \rho = \rho_x(x,y,z)\, dx+  \rho_y(x,y,z)\, dy +  \rho_z(x,y,z)\, dz$ then the action of exterior derivative is 
\begin{equation}
d\bm\rho = \left(\frac{\partial \rho_y}{\partial x} -  \frac{\partial \rho_x}{\partial y}\right)  dx \wedge dy  +  \left(\frac{\partial \rho_z}{\partial y} -  \frac{\partial \rho_y}{\partial z}\right)  dy \wedge dz +  \left(\frac{\partial \rho_x}{\partial z} -  \frac{\partial \rho_z}{\partial x}\right)  dz \wedge dx 
\end{equation}
it is identified with rotational;
\item 2-form: $\bm \sigma = \sigma_{xy}(x,y,z)\, dx\wedge dy +  \sigma_{yz}(x,y,z)\, dy \wedge dz+  \sigma_{zx}(x,y,z)\, dz\wedge dx$ then the action of exterior derivative is 
\begin{equation}
d\bm\sigma = \left(\frac{\partial \sigma_{yz}}{\partial x} + \frac{\partial \sigma_{zx}} {\partial y} + \frac{\partial \sigma_{xy}} {\partial z} \right) dx \wedge dy \wedge dz  
\end{equation}
it is identified with divergence.
\end{enumerate}
The second exterior derivative of an $r$-form, it is verified that $d^2 \bm\rho=0$. 
If $\bm \rho$ is an $r$-form given by
\begin{equation}
\bm\rho = r_{\mu 1,\mu 2,\cdots, \mu r}\,dx^{\mu 1}\wedge dx^{\mu 2}\wedge
\cdots \wedge dx^{\mu r}, \nonumber
\end{equation}
then the $d^2\bm \rho$  is 
\begin{equation}
d^2\bm\rho = \frac{\partial^2 r_{\mu 1,\mu 2,\cdots, \mu r}}{\partial x^{\lambda}\partial x^{\nu}}\,dx^{\nu}\wedge dx^{\lambda}\wedge dx^{\mu 1}\wedge dx^{\mu 2}\wedge
\cdots \wedge dx^{\mu r}, \nonumber
\end{equation}
since $\dfrac{\partial^2 r_{\mu 1,\mu 2,\cdots, \mu r}}{\partial x^{\lambda}\partial x^{\nu}}$ is symmetric with respect $\nu$ and $\lambda$ while $dx^{\nu}\wedge dx^{\lambda}$ is antisymmetric.
Another important property is 
\begin{equation}
\label{exterior_derivative}
 d(\rho \wedge \sigma) = d\rho \wedge \sigma + (-1)^r(\rho\wedge d\sigma) 
\end{equation}
 where $\rho$ is an $r$-form.

For example, the electromagnetic potential
\begin{equation}
(A_{\mu}) = \begin{pmatrix}
\phi \cr A_x \cr A_y \cr A_z
\end{pmatrix}, \nonumber
\end{equation}
can be assigned by 1-form $\bm A = A_{\nu} dx^{\nu}$. It remains for us to describe the electromagnetic tensor $\bm F = d\bm A$ as a 2-form,
\begin{eqnarray}
d\bm A &=& \frac{\partial A_{\nu}}{\partial x^{\mu}} dx^{\mu} \wedge dx^{\nu}  = \frac{1}{2} \left(  \frac{\partial A_{\nu}}{\partial x^{\mu}} +  \frac{\partial A_{\nu}}{\partial x^{\mu}} \right) dx^{\mu} \wedge dx^{\nu}  \cr
&=& \frac{1}{2} \left(  \frac{\partial A_{\nu}}{\partial x^{\mu}} dx^{\mu} \wedge dx^{\nu}+  \frac{\partial A_{\mu}}{\partial x^{\nu}} dx^{\nu} \wedge dx^{\mu} \right) \cr 
&=& \frac{1}{2} \left(  \frac{\partial A_{\nu}}{\partial x^{\mu}} -  \frac{\partial A_{\mu}}{\partial x^{\nu}} \right) dx^{\mu} \wedge dx^{\nu}\cr
&=& \frac{1}{2} F_{\mu\nu}\, dx^{\mu} \wedge dx^{\nu}, \nonumber
\end{eqnarray}
where $\bm F = \dfrac{1}{2}F_{\mu\nu}\, dx^{\mu} \wedge dx^{\nu} $. The components of $F_{\mu\nu}$ is given by:
\begin{equation}
(F_{\mu\nu}) = \begin{pmatrix}
 0 & -E_x & -E_y & -E_z \cr
E_x & 0 & B_z & - B_y \cr
E_y & -B_z & 0 & B_x \cr
E_z & B_y & -B_x & 0 
\end{pmatrix}. \nonumber
\end{equation}
The action of exterior derivative on $\bm F$ results in
\begin{equation}
d\bm F = \frac{1}{2} \frac{\partial F_{\mu\nu}}{\partial x^{\lambda}} dx^{\lambda} \wedge dx^{\mu} \wedge dx^{\nu} , \nonumber
\end{equation}
where $d\bm F = d^2 \bm A =0$, then
\begin{equation}
\partial_{\lambda} F_{\mu\nu}\, dx^{\lambda} \wedge dx^{\mu} \wedge dx^{\nu} = \frac{1}{3}\left(\partial_{\lambda} F_{\mu\nu}\, dx^{\lambda} \wedge dx^{\mu} \wedge dx^{\nu}+\partial_{\mu} F_{\nu\lambda}\, dx^{\mu} \wedge dx^{\nu} \wedge dx^{\lambda} + \partial_{\nu} F_{\lambda\mu}\, dx^{\nu} \wedge dx^{\lambda} \wedge dx^{\mu} \right)=0, \nonumber
\end{equation}
we can interchange $ dx^{\mu} \wedge dx^{\nu} \wedge dx^{\lambda}$ and $dx^{\nu} \wedge dx^{\lambda} \wedge dx^{\mu} $ of above equation,
\begin{equation}
 dx^{\mu} \wedge dx^{\nu} \wedge dx^{\lambda} = dx^{\lambda} \wedge dx^{\mu} \wedge dx^{\nu}\nonumber 
\end{equation}
and
\begin{equation}
 dx^{\nu} \wedge dx^{\lambda} \wedge dx^{\mu}  = dx^{\lambda} \wedge dx^{\mu} \wedge dx^{\nu},\nonumber 
\end{equation}
such as
\begin{equation}
\left(\partial_{\lambda} F_{\mu\nu} + \partial_{\mu} F_{\nu\lambda} +  \partial_{\nu} F_{\lambda\mu} \right) dx^{\lambda} \wedge dx^{\mu} \wedge dx^{\nu} = 0,\nonumber 
\end{equation}
which is known as the Bianchi identity, that follows from this, the two homogeneous Maxwell's equations,
\begin{equation}
\nabla \cdot \bm B =0 \hspace*{1cm} \mbox{and} \hspace*{1cm} \frac{\partial \bm B}{\partial t} + \nabla \times \bm E =0.\nonumber
\end{equation}

\subsection{ The covariant derivative}

The exterior derivative is a limited generalization acting only on forms. So, it is necessary to introduce covariant derivative $\nabla_{\bm X}$ in the direction of the vector $\bm X$ at $p$ on manifold $\cal M$, that maps an arbitrary tensor into a tensor of same type.  The symbol $\nabla$ is called affine connection. Let $\bm X$, $\bm Y$ and $\bm Z$ vector fields and $f$ a scalar function. The covariant derivative satisfies the following conditions:
\begin{equation}
\label{covariant_derivative_1}
\nabla_{\bm X}(\bm Y + \bm Z) = \nabla_{\bm X} \bm Y + \nabla_{\bm X} \bm Z;
\end{equation}
\begin{equation}
\label{covariant_derivative_2}
\nabla_{(\bm X+\bm Y)}  \bm Z = \nabla_{\bm X} \bm Z + \nabla_{\bm Y} \bm Z;
\end{equation}
\begin{equation}
\label{covariant_derivative_3}
\nabla_{f \bm X}\bm Y  = f\,\nabla_{\bm X} \bm Y ;
\end{equation}
\begin{equation}
\label{covariant_derivative_4}
\nabla_{\bm X}(f\bm Y)  = \bm X[f] \bm Y + f\nabla_{\bm X} \bm Y .
\end{equation}
where from (\ref{directionl_derivative_3}) we have that
\begin{equation}
 \nabla_{\bm X}f= \bm X[f] = X^{\mu}\frac{\partial f}{\partial x\,^{\mu}}.\nonumber
\end{equation}

The covariant derivative of the basis vector $\bm E_{\nu}=\dfrac{\partial}{\partial x^{\nu}}$ in the direction of basis vector  $\bm E_{\mu}=\dfrac{\partial}{\partial x\,^{\mu}}$ can be expanded in terms of basis vectors:
\begin{equation}
\label{connection_1}
\nabla_{\bm E_{\mu}}\bm E_{\nu} = {\Gamma^{\rho}}_{\mu\nu} \bm E_{\rho}
\end{equation}
where
\begin{equation}
\label{connection_2}
 {\Gamma^{\rho}}_{\mu\nu} = \left\langle dx\,^{\rho}, \nabla_{\bm E_{\mu}}\bm E_{\nu} \right\rangle
\end{equation}
is called connection coefficients. 

The covariant derivative $\nabla_{\bm X} \bm Y $, where $\bm X = X^{\mu} \bm E_{\mu}$ and $\bm Y = Y^{\nu}\bm E_{\nu}$ is given by vector
\begin{equation}
\nabla_{\bm X} \bm Y = \nabla_{(X^{\mu}\bm E_{\mu})} ( Y^{\nu}\bm E_{\nu}) \nonumber
\end{equation}
with conditions  (\ref{covariant_derivative_3}) and (\ref{covariant_derivative_4}) we have
\begin{equation}
\nabla_{\bm X} \bm Y = X^{\mu}\nabla_{\bm E_{\mu}} ( Y^{\nu}\bm E_{\nu}) =  X^{\mu}\left( \bm E_{\mu} [Y^{\nu}] \bm E_{\nu}+ Y^{\nu}\nabla_{\bm E_{\mu}}\bm E_{\nu}\right)\nonumber
\end{equation}
with (\ref{connection_1}) we have
\begin{equation}
\nabla_{\bm X} \bm Y =  X^{\mu}\left( \frac{\partial Y^{\nu}}{\partial x^{\mu}}\bm E_{\nu} + Y^{\nu}\,{\Gamma^{\rho}}_{\mu\nu}\bm E_{\rho}\right)\nonumber
\end{equation}
thus, this covariant derivative is given by the vector
\begin{equation}
\label{covariant_derivative_5}
\nabla_{\bm X} \bm Y =  X^{\mu}\left(\partial_{\mu} Y^{\rho} + Y^{\nu}\,{\Gamma^{\rho}}_{\mu\nu}\right) \bm E_{\rho}.
\end{equation}
By definition, the affine connection $\nabla$ maps two vectors $\bm X$ and $\bm Y$ to a new vector given by of the right hand side of (\ref{covariant_derivative_5}), whose $\rho$th component is
\begin{equation}
 X^{\mu}\left(\partial_{\mu} Y^{\rho} + Y^{\nu}\,{\Gamma^{\rho}}_{\mu\nu}\right) \equiv X^{\mu}\nabla_{\mu} Y^{\rho}. \nonumber
\end{equation}
If we compute the covariant derivative $\nabla_{\bm E_{\mu}} \bm Y$, we have:
\begin{equation}
\nabla_{\bm E_{\mu}} \bm Y = \bm E_{\mu}[Y^{\nu}] \bm E_{\nu} + Y^{\nu} \nabla_{\bm E_{\mu}} \bm E_{\nu}  = \left( \partial_{\mu}Y^{\rho} +Y^{\nu}{\Gamma^{\rho}}_{\mu\nu}\right)\bm E_{\rho} = \left(\nabla_{\mu} Y^{\rho}\right) \bm E_{\rho}, \nonumber
\end{equation}
where $\nabla_{\mu} Y^{\rho}$ is the $\rho$th component of vector $\nabla_{\bm E_{\mu}} \bm Y$.

The covariant derivative of 1-form $\bm\omega = \omega_{\nu}dx^{\nu}$ is obtained from contraction of $\bm \omega$ with a tangent vector $\bm Y$,
\begin{equation}
\left\langle \bm\omega,\bm Y\right\rangle =f.\nonumber
\end{equation}
If we put $\bm Y=\bm E_{\nu}$ and $\bm\omega = dx^\rho$ in above equation, it becomes
\begin{equation}
\left\langle dx^{\rho},\bm E_{\nu}\right\rangle ={\delta^{\rho}}_{\nu},\nonumber
\end{equation}
then the covariant derivative of above equation with respect to $\bm E_{\mu}$ is
\begin{equation}
\nabla_{\bm E_{\mu}}\left\langle dx^{\rho},\bm E_{\nu}\right\rangle = 0 \nonumber
\end{equation}
\begin{equation}
\left\langle \nabla_{\bm E_{\mu}}dx^{\rho},\bm E_{\nu}\right\rangle  + \left\langle dx^{\rho},\nabla_{\bm E_{\mu}}\bm E_{\nu}\right\rangle = 0 \nonumber
\end{equation}
we can use (\ref{connection_2})
\begin{equation}
\left\langle \nabla_{\bm E_{\mu}}dx^{\rho},\bm E_{\nu}\right\rangle  = - {\Gamma^{\rho}}_{\mu\nu}, \nonumber
\end{equation}
and as consequence of above equation we can identify 
\begin{equation}
\label{connection_3}
\nabla_{\bm E_{\mu}}dx^{\rho}  = - {\Gamma^{\rho}}_{\mu\sigma} dx^{\sigma}. 
\end{equation}
Furthermore, if we calculate derivative of 1-form $\bm\omega = \omega_{\nu}dx^{\nu}$ with respect to $\bm E_{\mu}$, we have that,
\begin{eqnarray}
\nabla_{\bm E_{\mu}} \bm\omega &=& \bm E_{\mu}[\omega_{\nu}] dx^{\nu}+ \omega_{\nu} \nabla_{\bm E_{\mu}} dx^{\nu}\cr
&=& \partial_{\mu}\omega_{\nu}dx^{\nu} - \omega_{\nu} {\Gamma^{\nu}}_{\mu\sigma} dx^{\sigma}\cr
&=& \left(\partial_{\mu}\omega_{\sigma} - \omega_{\nu} {\Gamma^{\nu}}_{\mu\sigma}\right) dx^{\sigma}\cr
&=& \left(\nabla_{\mu}\omega_{\sigma}\right) dx^{\sigma},\nonumber
\end{eqnarray}
where we can identify the $\sigma$th component of 1-form $\nabla_{\bm E_{\mu}} \bm\omega$ as
\begin{equation}
\nabla_{\mu}\omega_{\sigma} = \partial_{\mu}\omega_{\sigma} - \omega_{\nu} {\Gamma^{\nu}}_{\mu\sigma}.
\end{equation}

We require that the Leibnitz rule be true for any tensor products,
\begin{equation}
\nabla_{\bm E_{\mu}}  \left( \bm T_1 \otimes \bm T_2 \right) =  \left( \nabla_{\bm E_{\mu}}\bm T_1 \right) \otimes \bm T_2 
+\bm T_1 \otimes \nabla_{\bm E_{\mu}}   \bm T_2 .
\end{equation}
If we have a tensor of type (1,1) $\bm T = {T_{\mu}}^{\nu} \bm E_{\nu} \otimes dx^{\mu}$, under the action of covariant derivative and Leibnitz rule we have that
\begin{eqnarray}
\nabla_{\bm E_{\rho}} \bm T &=& \nabla_{\bm E_{\rho}} \left({T_{\mu}}^{\nu} \bm E_{\nu} \otimes dx^{\mu}\right) \cr
&=&  E_{\rho}[{T_{\mu}}^{\nu}] \bm E_{\nu} \otimes dx^{\mu}+ {T_{\mu}}^{\nu} \left( \nabla_{\bm E_{\rho}} \bm E_{\nu} \right) \otimes dx^{\mu} + 
 {T_{\mu}}^{\nu}  \bm E_{\nu}  \otimes \nabla_{\bm E_{\rho}} dx^{\mu} \cr
&=& \partial_{\rho} {T_{\mu}}^{\nu} \bm E_{\nu} \otimes dx^{\mu} +  {T_{\mu}}^{\nu} {\Gamma^{\sigma}}_{\rho\nu} \bm E_{\sigma} \otimes dx^{\mu} +
{T_{\mu}}^{\nu}  \bm E_{\nu}  \otimes \left(- {\Gamma^{\mu}}_{\rho\sigma} dx^{\sigma}\right)\cr
&=& \left( \partial_{\rho} {T_{\mu}}^{\nu} + {\Gamma^{\nu}}_{\rho\sigma} {T_{\mu}}^{\sigma} - {\Gamma^{\sigma}}_{\rho\mu} {T_{\sigma}}^{\nu}\right) \bm E_{\nu}\otimes dx^{\mu}, \nonumber
\end{eqnarray}
where we have that:
\begin{equation}
\nabla_{\rho}  {T_{\mu}}^{\nu}  =   \partial_{\rho} {T_{\mu}}^{\nu} + {\Gamma^{\nu}}_{\rho\sigma} {T_{\mu}}^{\sigma} - {\Gamma^{\sigma}}_{\rho\mu} {T_{\sigma}}^{\nu}
\end{equation}

\subsection{ The torsion and curvature}

Intrinsic objects defined on manifolds that measure the bends on manifolds are torsion tensor and curvature (Riemann) tensor. Let $\bm X$, $\bm Y$ and $\bm Z$ be vectors of tangent space ${\bm T}_{p}$, then the torsion tensor and curvature tensor are defined by
\begin{equation}
 \label{torsion_1}
 \bm T(\bm X,\bm Y) \equiv \nabla_{\bm X} \bm Y - \nabla_{\bm Y} \bm X - [\bm X, \bm Y] \hspace*{2cm}
 \mbox{\bf torsion tensor}
\end{equation}
\begin{equation}
 \label{curvature_1}
 \bm R(\bm X,\bm Y)\bm Z \equiv \nabla_{\bm X} \nabla_{\bm Y} \bm Z -  \nabla_{\bm Y} \nabla_{\bm X} \bm Z  
 - \nabla_{[\bm X, \bm Y]}\bm Z \hspace*{2cm}
 \mbox{\bf curvature tensor}.
\end{equation}
The torsion tensor gives an intrinsic characterization of how tangent spaces twist about a curve when they are parallel transported. The curvature or Riemann tensor describes how the tangent spaces roll along the curve. 

With respect to coordinate basis $\left\{\bm E_{\mu}=\dfrac{\partial}{\partial x\,^{\mu}}\right\}$ and the dual basis $\left\{dx\,^{\mu}\right\}$, the components of torsion tensor is given by
\begin{equation}
 {T^{\lambda}}_{\mu\nu} = \langle dx^{\lambda}, \bm T( \bm E_{\mu},\bm E_{\nu})\rangle =  \langle dx^{\lambda},\left(\nabla_{\bm E_{\mu}}\bm E_{\nu}- \nabla_{\bm E_{\nu}}\bm E_{\mu} - [\bm E_{\mu},\bm E_{\nu}]\right)\rangle =  \langle dx^{\lambda}, \left({\Gamma^{\kappa}}_{\mu\nu}\bm E_{\kappa}-  {\Gamma^{\kappa}}_{\nu\mu}\bm E_{\kappa}- 0\right)\rangle \nonumber
\end{equation}
where it follows that
\begin{equation}
\label{torsion_2}
 {T^{\lambda}}_{\mu\nu} = {\Gamma^{\lambda}}_{\mu\nu} -  {\Gamma^{\lambda}}_{\nu\mu}.
\end{equation}
And the components of curvature tensor with respect to coordinate basis is given by
\begin{equation}
\label{curvature_1.1}
 {R^{\kappa}}_{\lambda\mu\nu} = \langle dx\,^{\kappa},\bm R(\bm E_{\mu}, \bm E_{\nu})\bm E_{\lambda} \rangle, 
\end{equation}
where we have
\begin{eqnarray}
\bm R(\bm E_{\mu}, \bm E_{\nu})\bm E_{\lambda} &=& \nabla_{\bm E_{\mu}}\left(\nabla_{\bm E_{\nu}}\bm E_{\lambda}\right) -
 \nabla_{\bm E_{\nu}}\left(\nabla_{\bm E_{\mu}}\bm E_{\lambda}\right)- \nabla_{[\bm E_{\mu},\bm E_{\nu}]}\bm E_{\lambda} \cr
  &=& \nabla_{\bm E_{\mu}}\left({\Gamma\,^{\rho}}_{\nu\lambda}\bm E_{\rho}\right) -
\nabla_{\bm E_{\nu}}\left({\Gamma\,^{\rho}}_{\mu\lambda}\bm E_{\rho}\right) - 0, \nonumber
\end{eqnarray}
we recall (\ref{covariant_derivative_4}) where 
$\nabla_{\bm X}(f\bm Y)  = \bm X[f] \bm Y + f\nabla_{\bm X} \bm Y$, such as,
\begin{eqnarray}
\bm R(\bm E_{\mu}, \bm E_{\nu},\bm E_{\lambda}) &=& \bm E_{\mu}\left[{\Gamma\,^{\rho}}_{\nu\lambda}\right]\bm E_{\rho}
+{\Gamma\,^{\rho}}_{\nu\lambda}\nabla_{\bm E_{\mu}}\bm E_{\rho} -
\bm E_{\nu}\left[{\Gamma\,^{\rho}}_{\mu\lambda}\right]\bm E_{\rho} -{\Gamma\,^{\rho}}_{\mu\lambda}\nabla_{\bm E_{\nu}}\bm E_{\rho} \cr
&=& \partial_{\mu}{\Gamma\,^{\rho}}_{\nu\lambda}\bm E_{\rho} +{\Gamma\,^{\rho}}_{\nu\lambda}{\Gamma^{\sigma}}_{\mu\rho}\bm E_{\sigma} -
\partial_{\nu}{\Gamma\,^{\rho}}_{\mu\lambda}\bm E_{\rho} -{\Gamma\,^{\rho}}_{\mu\lambda}{\Gamma^{\sigma}}_{\nu\rho}\bm E_{\sigma}\cr
&=&\left(\partial_{\mu}{\Gamma\,^{\rho}}_{\nu\lambda} - \partial_{\nu}{\Gamma\,^{\rho}}_{\mu\lambda} 
+{\Gamma\,^{\sigma}}_{\nu\lambda}{\Gamma^{\rho}}_{\mu\sigma} -{\Gamma\,^{\sigma}}_{\mu\lambda}{\Gamma^{\rho}}_{\nu\sigma}\right)\bm E_{\rho} , \nonumber
\end{eqnarray}
it follows that
\begin{equation}
\label{curvature_2}
 {R^{\kappa}}_{\lambda\mu\nu} =\partial_{\mu}{\Gamma\,^{\kappa}}_{\nu\lambda} - \partial_{\nu}{\Gamma\,^{\kappa}}_{\mu\lambda} 
+{\Gamma\,^{\sigma}}_{\nu\lambda}{\Gamma^{\kappa}}_{\mu\sigma} -{\Gamma\,^{\sigma}}_{\mu\lambda}{\Gamma^{\kappa}}_{\nu\sigma}.
\end{equation}
We readily find
\begin{equation}
 {T^{\lambda}}_{\mu\nu} = -{T^{\lambda}}_{\nu\mu} \hspace*{1cm}\mbox{and} \hspace*{1cm}   {R^{\kappa}}_{\lambda\mu\nu} = - {R^{\kappa}}_{\lambda\nu\mu} .
\end{equation}

\subsection{Metric tensors}

We have defined the contraction $\langle\bm\omega, \bm V \rangle$ as the inner product between a
contravariant vector $\bm V$ and a 
covariant or dual vector $\bm\omega$. For $\bm\omega \in \bm T_p^{*}$ and $\bm V
\in \bm T_p$ we have that inner product is
\begin{equation}
  \langle \bm\omega,\bm V \rangle = \omega_{\mu}V^{\mu}. \nonumber
\end{equation}

Now, we can define a linear map $\bm g_p(\bm U, \bm V)$ as the inner product between two contravariant vectors $\bm U, \,\bm V \in \bm T_p$.
The Riemannian metric $\bm g_p$ is a type (0,2) tensor field on manifold $\cal M$ which satisfies
\begin{enumerate}[(i)]
 \item $ \bm g_p(\bm U,\bm V) = \bm g_p(\bm V,\bm U) \in  \mathbb{R} $;
 \item $ \bm g_p(\bm U,\bm U) \geq 0$ assigns a magnitude of vector $\bm U\in \bm T_p$ at point $p$.  $ \bm g_p(\bm U,\bm U) = 0$ is true only when $\bm U =0$.
\end{enumerate}
 $\bm g_p$ is called pseudo-Riemannian metric when the metric is no longer positive definite.
Since $\bm g_p$ is a type (0,2) tensor field on $\cal M$, it is expaned in terms of $dx^{\mu} \otimes dx^{\nu}$ as
\begin{equation}
\label{metric_tensor}
 \bm g_p = g_{\mu\nu}(p)\, dx^{\mu} \otimes dx^{\nu}.
\end{equation}
The components of $\bm g_p$ with respect a coordinate basis $\left\{\bm E_{\mu}=\dfrac{\partial}{\partial x^{\mu}}\right\}$ are
\begin{equation}
\label{metric_tensor_2}
 g_{\mu\nu}(p) = \bm g_p(\bm E_{\mu},\bm E_{\nu}) = \bm g_p(\bm E_{\nu},\bm E_{\mu}),
\end{equation} 
where we can see that $(g_{\mu\nu})$ is  a symmetric matrix (we can omit $p$ in metric tensor). The matrix $(g_{\mu\nu})$ has an inverse denoted by $(g^{\mu\nu})$ where we have
\begin{equation}
 g_{\mu\nu}g^{\nu\lambda} = g^{\lambda\nu} g_{\nu\mu} ={\delta_{\mu}}^{\lambda}.
\end{equation}

The map $\bm g(\bm U,\,\,\,\,\,)$ : $\bm T_p{\cal M} \rightarrow \mathbb{R}$ by $\bm V \mapsto \bm g(\bm U,\bm V)$ is identified with  1-form.
Thus, if $U^{\mu}$ are components of a contravariant vector $\bm U\in \bm T_p$, then $U_{\mu}$ are the components of a uniquely associated covariant vector, where
\begin{equation}
 U_{\mu} = g_{\mu\nu} U^{\nu},
\end{equation}
such as
\begin{equation}
\label{inner_product}
 \bm g(\bm U,\bm V) = \bm g( U^{\mu}\bm E_{\mu},V^{\nu}\bm E_{\nu}) =U^{\mu}V^{\nu}\bm g( \bm E_{\mu},\bm E_{\nu})= g_{\mu\nu} U^{\mu}V^{\nu}
 = U_{\nu}V^{\nu} = U^{\mu}V_{\mu}.
\end{equation}
Due to isomorphism between tangent space $\bm T_p$ and cotangent sapce $\bm T^{*}_p$ we have
\begin{equation}
 U^{\mu} = g^{\mu\nu}U_{\nu}.
\end{equation}

In Special Relativity, two separated events with coordinates $(t,x,y,z)$ and $(t+dt,x+dx,y+dy,z+dz)$ in a particular inertial frame have the square of the infinitesimal interval between them given by
\begin{equation}
\label{line_Minkowski_1}
 ds^2= -dt^2+dx^2+dy^2+dz^2,
\end{equation}
which is known as the line element of Minkowski spacetime and is invariant under any Lorentz transformation \cite{DInverno, Landau, Hobson, Schutz, Wald, Weinberg}.
We can rewrite this line element in matrix form
\begin{equation}
 ds^2= \begin{pmatrix} dt & dx & dy & dz \end{pmatrix}
 \begin{pmatrix} -1 & 0 & 0 & 0 \cr
		  0 & 1 & 0 & 0 \cr
		  0 & 0 & 1 & 0 \cr
		  0 & 0 & 0 & 1 \end{pmatrix}
\begin{pmatrix}
 dt\cr
 dx\cr
 dy\cr
 dz
\end{pmatrix}
,\nonumber
\end{equation}
or
\begin{equation}
\label{line_Minkowski_2}
 ds^2 = \eta_{\mu\nu} dx^{\mu} dx^{\nu}.
\end{equation}
It is common to regard $(\eta_{\mu\nu})$ as $\bm \eta = \mbox{diag}(-1,1,1,1) $ whose signature of Minkowski spacetime metric is $(-+++)$.
The Minkowski spacetime is denoted by $(\mathbb{R}^4,\bm\eta)$.

Any metric $\bm g$ in four-dimensional differentiable manifold $\cal M$ with signatures $(-+++)$ is called {\it Lorentzian metric}
(a pseudo-Riemannian metric tensor).
General Relativity is based on the concept of spacetime, which is a four-dimensional differentiable manifold $\cal M$, free of torsion with Lorentzian metric,
denoted by $({\cal M},\bm g)$ or just $V_4$. There are others Theories of Gravitation such as Einstein-Cartan(-Sciama-Kibble) Theory of Gravity concepted in a Riemann-Cartan spacetime with non-vanishing torsion and curvature. Riemann-Cartan spacetimes are denoted by $U_4$ \cite{Blagojevic}.

In accordance with the concept of line element of Minkowski spacetime (\ref{line_Minkowski_2}), if we take an infinitesimal displacement $dx\,^{\mu} \dfrac{\partial}{\partial x\,^{\mu}} \in \bm T_p$ to $\cal M$ and plug into $\bm g$, we have
\begin{equation}
 ds^2 = \bm g\left(dx\,^{\mu} \dfrac{\partial}{\partial x\,^{\mu}}, dx\,^{\nu} \dfrac{\partial}{\partial x\,^{\nu}}\right) = 
 dx\,^{\mu} dx\,^{\nu} \bm g\left( \dfrac{\partial}{\partial x\,^{\mu}}, \dfrac{\partial}{\partial x\,^{\nu}}\right) =
 g_{\mu\nu}\, dx\,^{\mu} dx\,^{\nu}
\end{equation}
that represents the length of the infinitesimal arc determined by the coordinate displacement $x\,^{\mu} \rightarrow x\,^{\mu}+dx\,^{\mu}$.

In spacetime  $({\cal M},\bm g)$ where the metric is Lorentzian, the vectors are divided into three classes:
\begin{enumerate}[(i)]
 \item $\bm g(\bm U,\bm U) > 0$,\hspace*{0.5cm} $\bm U$ is spacelike,
 \item $\bm g(\bm U,\bm U) < 0$,\hspace*{0.5cm} $\bm U$ is timelike,
 \item $\bm g(\bm U,\bm U) = 0$,\hspace*{0.5cm} $\bm U$ is lightlike or null.
\end{enumerate}

Now that a manifold $\cal M$ can be endowed with a metric $\bm g$, we can put some restrictions on the possible form of connections ${\Gamma^{\lambda}}_{\mu\nu}$. We demand that the metric $g_{\mu\nu}$ be covariantly constant,
\begin{equation}
\label{covariance_1} 
 \nabla_{\lambda}g_{\mu\nu} = \partial_{\lambda} g_{\mu\nu} - {\Gamma^{\kappa}}_{\lambda\mu}g_{\kappa\nu} - {\Gamma^{\kappa}}_{\lambda\nu}g_{\kappa\mu} = 0.
\end{equation}
Then, the affine connection $\nabla$ is said to be a metric connection. In a coordinate basis, with (\ref{connection_1}) and (\ref{metric_tensor_2}), we obtain that
\begin{equation}
 \Gamma_{\lambda\mu\nu} = \bm g\left( \bm E_{\lambda}, \nabla_{\bm E_{\mu}}\bm E_{\nu}\right) = \bm g\left( \bm E_{\lambda}, {\Gamma^{\kappa}}_{\mu\nu}\bm E_{\kappa}\right) = {\Gamma^{\kappa}}_{\mu\nu}\, g_{\lambda\kappa}.
\end{equation}
In General Relativity the spacetime is $V_4$ where the manifold is free of torsion, then with equation (\ref{covariance_1}) we can obtain the Christoffel relations
\begin{equation}
 \label{Christoffel}
  \Gamma_{\lambda\mu\nu} = \frac{1}{2}\left(\partial_{\nu}g_{\lambda\mu}+\partial_{\mu}g_{\lambda\nu} - \partial_{\lambda}g_{\mu\nu}\right)
\end{equation}
for the coordinate components of the connection.

From equation (\ref{covariance_1}) we have that
\begin{equation}
\label{covariance_2} 
 \partial_{\lambda} g_{\mu\nu} - \Gamma_{\nu\lambda\mu} -\Gamma_{\mu\lambda\nu} = 0,
\end{equation}
where we can rewrite above equation to 1-form language
\begin{equation}
 \frac{\partial g_{\mu\nu}}{\partial x^{\lambda}} dx^{\lambda} - \Gamma_{\nu\lambda\mu}\,dx^{\lambda} -\Gamma_{\mu\lambda\nu}\,dx^{\lambda} = 0 \nonumber
\end{equation}
or
\begin{equation}
d(g_{\mu\nu})  -\Gamma_{\mu\lambda\nu}\,dx^{\lambda} - \Gamma_{\nu\lambda\mu}\,dx^{\lambda} = 0.\nonumber
\end{equation}
Introducing the {\it connection 1-forms}
\begin{equation}
 \label{connection_1_form}
 \bm\Gamma_{\mu\nu} \equiv \Gamma_{\mu\lambda\nu}\,dx^{\lambda},
\end{equation}
we have that
\begin{equation}
\label{metric_condition_1}
 d(g_{\mu\nu}) - \bm\Gamma_{\mu\nu} -\bm\Gamma_{\nu\mu}=0.
\end{equation}
If we have constant metric or a {\it rigid frame}, $\partial_{\lambda}g_{\mu\nu}=0$, then 
\begin{equation}
\label{metric_condition_2}
 \bm\Gamma_{\mu\nu}= -\bm\Gamma_{\nu\mu}.
\end{equation}
We will see in the next sections that the orthonormal and pseudo-orthonormal basis are rigid frames with the connection 1-forms obeying the
above propertie.


\section{Non-coordinate basis}

In General Relativity there is a possibility that to set up a system of locally
inertial coordinates, valid in a sufficiently small region of spacetime.
Strictly speaking, this must be an infinitesimal region surrounding at a point $p \in {\cal M}$ with coordinates $x^{\mu}$. We can denote these local coordinates by $y^{\alpha}$. We can denote the transformation matrix, which relates the two sets of coordinates - coordinate basis and 
locally inertial coordinates - by 
\begin{equation}
 {e_{\alpha}}^{\mu} (p) = \frac{\partial x^{\mu}}{\partial y^{\alpha}} \hspace*{2cm} \mbox{and the inverse} \hspace*{2cm}
 {\omega^{\alpha}}_{\mu}(p) = \frac{\partial y^{\alpha}}{\partial x^{\mu}}
\end{equation}
If we set up a locally inertial frame of reference at each point of spacetime, in such a way that the directions of their axes vary smoothly from one point to another, then we obtain a set of four vector fields $ {e_{0}}^{\mu},\cdots,{e_{3}}^{\mu} $
which specify, at each point, the directions of these axes. 

For instance, a tangent vector at a point $p\in {\cal M}$ can be expressed as a linear combination of  a coordinate basis or holonomic frame in accordance with equation (\ref{directionl_derivative_2}).
In coordinate basis, $\bm T_p$ is spanned by $\{\bm E_{\mu}\}=\left\{\dfrac{\partial}{\partial x^{\mu}}\right\}$ and $\bm T^{*}_p$ by
$\{dx\,^{\mu}\}$. Now, we can express a new basis of vectors in a system of locally inertial coordinates (local Lorentz spacetime) by linear combinations
\begin{equation}
\label{base_NC1}
{\hat{\bm e}}_{\alpha}={e_{\alpha}}^{\mu}\bm E_{\mu}.
\end{equation} 
for a  new basis of $\bm T_p$ and a new dual basis of $\bm T^{*}_p$,
\begin{equation}
\label{base_NC2}
 \tilde{\bm\theta}^{\beta}={\omega^{\beta}}_{\nu}dx^{\nu}.
\end{equation}
From (\ref{base_NC1}) we can see that new basis $\{{\hat{\bm e}}_{\alpha} \}$ is the frame of basis vectors which is obtained by a rotation of the basis $\{\bm E_{\mu}\}$ preserving the orietation, where $\{{e_{\alpha}}^{\mu}\}\in GL(4,\mathbb{R})$. The new basis $\{{\hat{\bm e}}_{\alpha} \}$  is called non-coordinate basis of the tangent space $\bm T_p$. 

The directional derivative in the general spacetime coordinate $\bm E_{\mu}[f] =\dfrac{\partial f}{\partial x^{\mu}}$ implies that ${\hat{\bm e}}_{\alpha} [f]$ also is a directional derivative, from (\ref{base_NC1}) it follows that
\begin{equation}
\label{directionl_derivative_3.1}
{\hat{\bm e}}_{\alpha}[f]={e_{\alpha}}^{\mu}\frac{\partial f}{\partial x^{\mu}}, 
\end{equation} 
or
\begin{equation}
\label{directionl_derivative_4}
{\hat{\bm e}}_{\alpha}[f]=\partial_{\alpha} f,
\end{equation} 
the directional derivative in the local Lorentz spacetime.

Since $\{\tilde{\bm\theta}^{\beta}\}$ is a new basis of the dual space $\bm T^{*}_p$ of the tangent space $\bm T_p$ in local Lorentz coordinates, we require that
\begin{equation}
 \langle {\hat{\bm e}}_{\alpha},\tilde{\bm\theta}^{\beta}\rangle= {\delta_{\alpha}}^{\beta},\nonumber
\end{equation}
where it follows
\begin{equation}
\label{id_n_coordenada2}
   \langle {e_{\alpha}}^{\mu}\bm E_{\mu},{\omega^{\beta}}_{\nu}dx^{\nu} \rangle
= {e_{\alpha}}^{\mu} {\omega^{\beta}}_{\nu}\langle \bm E_{\mu},dx^{\nu} \rangle =  {e_{\alpha}}^{\mu} {\omega^{\beta}}_{\mu} = {\delta_{\alpha}}^{\beta},
\end{equation}
where ${\omega^{\alpha}}_{\mu}$ is the inverse of  ${e_{\alpha}}^{\mu}$. Also, it is valid that ${\omega^{\alpha}}_{\nu}{e_{\alpha}}^{\mu}= {\delta^{\mu}}_{\nu} $.
The matrix ${e_{\alpha}}^{\mu}$ is called the tetrad or vierbein field.
The vierbein field, ${e_{\alpha}}^{\mu}$, has two kinds of indices: $\mu=1,2,3,4$, labels the general spacetime coordinate and $\alpha = 1,2,3,4$  labels the local spacetime. 
We use $\kappa,\lambda,\mu,\nu,\rho,\sigma,\tau,\phi,$ etc to denote the coordinate basis  $\{\bm{E_{\mu}}\}$ and $\{dx^{\nu}\}$, while  
$\alpha,\beta,\gamma,\delta,\epsilon,$ etc, to denote the non-coordinate basis $\{{\hat{\bm e}}_{\alpha}\}$ and $\{\tilde{\bm\theta}^{\beta}\}$.

From expression (\ref{base_NC1}) where ${\hat{\bm e}}_{\alpha}={e_{\alpha}}^{\mu}\bm E_{\mu}$, we can obtain a reverse expression by using ${\omega^{\alpha}}_{\nu}$,
\begin{equation}
{\omega^{\alpha}}_{\nu}{\hat{\bm e}}_{\alpha}={\omega^{\alpha}}_{\nu}{e_{\alpha}}^{\mu}\bm E_{\mu} = {\delta^{\mu}}_{\nu} \bm E_{\mu} \nonumber
\end{equation}
that it gives
\begin{equation}
\label{base_NC3}
\bm E_{\mu} =  {\omega^{\alpha}}_{\mu}{\hat{\bm e}}_{\alpha}.
\end{equation}
In the same way, the reverse expression for (\ref{base_NC2}), where $\tilde{\bm\theta}^{\alpha}={\omega^{\alpha}}_{\mu}dx^{\mu}$,
results that
\begin{equation}
\label{base_NC4}
 dx^{\mu} = {e_{\alpha}}^{\mu}\tilde{\bm\theta}^{\alpha}.
\end{equation}

We can verify that the differential of an arbitrary scalar function $f$ given by $df =\dfrac{\partial f}{\partial x^{\mu}}dx^{\mu}$ in coordinate basis is given in non-coordinate basis by substituting $dx^{\mu}$ by use of the equation (\ref{base_NC4}),
\begin{equation}
 df =\dfrac{\partial f}{\partial x^{\mu}}{e_{\alpha}}^{\mu}\tilde{\bm\theta}^{\alpha}, \nonumber
\end{equation}
we can use (\ref{directionl_derivative_3.1}) and (\ref{directionl_derivative_4}) where it results in
\begin{equation}
\label{differential_f}
 df = \tilde{\bm\theta}^{\alpha}\partial_{\alpha} f. 
\end{equation}

Let a tensor field of type (1,1) given by ${\bm T}={T^{\mu}}_{\nu}{\bm E}_{\mu}\otimes dx^{\nu}$  expressed in non-coordinate basis as
\begin{equation}
 {\bm T} = {T^{\mu}}_{\nu}({\omega^{\alpha}}_{\mu}{\hat{\bm e}}_{\alpha})\otimes
 ({e_{\beta}}^{\nu}\,\,{\tilde{\bm\theta}}^{\beta}) 
 = ({T^{\mu}}_{\nu}{\omega^{\alpha}}_{\mu}\,\,{e_{\beta}}^{\nu})\,\,
 {\hat{\bm e}}_{\alpha}\otimes \tilde{\bm\theta}^{\beta} \nonumber
\end{equation}
where we have 
\begin{equation}
 {\bm T} = {T^{\alpha}}_{\beta} {\hat{\bm e}}_{\alpha}\otimes \tilde{\bm\theta}^{\beta}, \nonumber
\end{equation}
with a way  to transit tensors from coordinate basis to non-coordinate basis by
\begin{equation}
\label{tensor_non_coordinate_basis}
 {T^{\alpha}}_{\beta}  = {T^{\mu}}_{\nu}\,\,{\omega^{\alpha}}_{\mu}\,\,{e_{\beta}}^{\nu}.
\end{equation}

The metric tensor ${\bm g}$ defined by (\ref{metric_tensor}) in terms of non-coordinate basis is
\begin{eqnarray}
\label{tensor_metrico_n_coordenadas}
 \bm g&=&g_{\mu\nu}dx^{\mu} \otimes dx^{\nu} = g_{\alpha\beta} {\omega^{\alpha}}_{\mu}{\omega^{\beta}}_{\nu}
 dx^{\mu} \otimes dx^{\nu}=g_{\alpha\beta}({\omega^{\alpha}}_{\mu}dx^{\mu})\otimes
 ({\omega^{\beta}}_{\nu} dx^{\nu})\cr
 \bm g &=&  g_{\alpha\beta} \tilde{\bm\theta}^{\alpha}\otimes \tilde{\bm\theta}^{\beta}.
\end{eqnarray}
The components of $\bm g$ with respect a coordinate basis $\{{\hat{\bm e}}_{\alpha} \}$ are
\begin{equation}
  \bm g({\hat{\bm e}}_{\alpha},{\hat{\bm e}}_{\beta})=g_{\alpha\beta}
\end{equation}
where we have
\begin{eqnarray}
\label{tensor_metrico_n_coordenadas_2}
 \bm g({\hat{\bm e}}_{\alpha},{\hat{\bm e}}_{\beta})=\bm g({e_{\alpha}}^{\mu}{\bm E_{\mu}},
 {e_{\beta}}^{\nu}{\bm E_{\nu}})={e_{\alpha}}^{\mu}{e_{\beta}}^{\nu}
 \bm g(E_{\mu},E_{\nu})={e_{\alpha}}^{\mu}{e_{\beta}}^{\nu}g_{\mu\nu},
\end{eqnarray}
that follows that
\begin{equation}
\label{vielbeins0}
 {e_{\alpha}}^{\mu}{e_{\beta}}^{\nu}g_{\mu\nu}=g_{\alpha\beta}.
\end{equation}
In the inverse way it results that
\begin{equation}
\label{vielbeins}
 g_{\mu\nu} = {\omega^{\alpha}}_{\mu}{\omega^{\beta}}_{\nu}g_{\alpha\beta}.
\end{equation}

The inner product is preserved by vierbein ${e_{\alpha}}^{\mu}$ that maps the tangent space to local frame. Let $\bm U$ and $\bm V$ two vectors in tangent space, the inner product in coordinate basis is given by (\ref{inner_product}),
\begin{equation}
 \bm g(\bm U,\bm V) =  g_{\mu\nu} U^{\mu}V^{\nu}. \nonumber
\end{equation}
The above inner product in local frame is obtained by substituting (\ref{vielbeins}) into above equation,
\begin{equation}
 \bm g(\bm U,\bm V) = {\omega^{\alpha}}_{\mu}{\omega^{\beta}}_{\nu}g_{\alpha\beta} U^{\mu}V^{\nu} =
 g_{\alpha\beta}({\omega^{\alpha}}_{\mu}U^{\mu})({\omega^{\beta}}_{\nu}V^{\nu})  =
g_{\alpha\beta}\, U^{\alpha} U^{\beta}. \nonumber
\end{equation}
From (\ref{base_NC1}) we can see that the non-coordinate basis $\{{\hat{\bm e}}_{\alpha} \}$ is the frame of basis vectors which is obtained by a rotation of the basis $\{\bm E_{\mu}\}$ preserving the orietation and inner product.

The action of covariant derivative in (\ref{id_n_coordenada2})  results in
\begin{equation}
\label{id_n_coordenada3}
     \left(\nabla_{\nu} {e_{\alpha}}^{\mu}\right) {\omega^{\beta}}_{\mu}+{e_{\alpha}}^{\mu} \nabla_{\nu}{\omega^{\beta}}_{\mu} = 0
\end{equation}
that it follows
\begin{equation}
\label{id_n_coordenada4}
     {\omega^{\beta}}_{\mu}\nabla_{\nu} {e_{\alpha}}^{\mu} =-{e_{\alpha}}^{\mu} \nabla_{\nu}{\omega^{\beta}}_{\mu}. 
\end{equation}
To lower the coordinate indice of the vierbein field, ${e_{\alpha}}^{\mu}$, we ca use
\begin{equation}
  g_{\mu\nu}{e_{\alpha}}^{\mu}=e_{\alpha\nu} \nonumber
\end{equation}
and when we use (\ref{vielbeins}),
$ g_{\mu\nu} = {\omega^{\alpha}}_{\mu}{\omega^{\beta}}_{\nu}g_{\alpha\beta}$, on above equation, it becomes
\begin{eqnarray}
 e_{\alpha\nu} = {\omega^{\beta}}_{\mu}{\omega^{\gamma}}_{\nu}g_{\beta\gamma}{e_{\alpha}}^{\mu}
 = ({\omega^{\beta}}_{\mu}{e_{\alpha}}^{\mu})(g_{\beta\gamma}\,{\omega^{\gamma}}_{\nu})
 = {\delta^{\beta}}_{\alpha} \omega_{\beta\nu}\nonumber
\end{eqnarray}
from which we find 
\begin{equation}
 e_{\alpha\mu} = \omega_{\alpha\mu}. 
\end{equation}
Now, we can rewrite (\ref{vielbeins}) such as
\begin{equation}
\label{vielbeins_2}
 g_{\mu\nu} = {\omega^{\alpha}}_{\mu}{\omega}_{\alpha\nu},
\end{equation}
we can see that the vierbein fields can be interpreted as `square root' of metric.

From the identification of vectors with directional derivatives 
we can conclude that the result of the successive application of two vectors to a function depends on 
the order in which operators are applied. Let  $ \hat{\bm e}_{\alpha}$ and $\hat{\bm e}_{\beta}$ two vectors of a non-coordinate basis 
$\{{\hat{\bm e}}_{\alpha} \}$,
\begin{equation}
 \hat{\bm e}_{\alpha} = {e_{\alpha}}^{\mu}\frac{\partial}{\partial x^{\mu}} \hspace{1cm}\mbox{and}
  \hspace{1cm}\hat{\bm e}_{\beta} = {e_{\beta}}^{\nu}\frac{\partial}{\partial x^{\nu}},\nonumber
\end{equation}
then the commutation  operation applied to a function $f$ follows
\begin{eqnarray}
[\hat{\bm e}_{\alpha},\hat{\bm e}_{\beta}]f&=&\left({e_{\alpha}}^{\mu}\frac{\partial}{\partial x^{\mu}} 
{e_{\beta}}^{\nu}\frac{\partial}{\partial x^{\nu}} - {e_{\beta}}^{\nu}\frac{\partial}{\partial x^{\nu}}
{e_{\alpha}}^{\mu}\frac{\partial}{\partial x^{\mu}}\right)f\cr
&=&\left({e_{\alpha}}^{\mu}\frac{\partial{e_{\beta}}^{\nu}}{\partial x^{\mu}}\frac{\partial f}{\partial x^{\nu}}+
{e_{\alpha}}^{\mu}{e_{\beta}}^{\nu}\frac{\partial^2 f}{\partial x^{\mu}\partial x^{\nu}} - 
{e_{\beta}}^{\nu}\frac{\partial {e_{\alpha}}^{\mu}}{\partial x^{\nu}}\frac{\partial f}{\partial x^{\mu}} -
{e_{\beta}}^{\nu}{e_{\alpha}}^{\mu}\frac{\partial^2 f}{\partial x^{\nu}\partial x^{\mu}}
\right)\cr
&=&\left({e_{\alpha}}^{\mu}\frac{\partial{e_{\beta}}^{\nu}}{\partial x^{\mu}}\frac{\partial }{\partial x^{\nu}} - 
{e_{\beta}}^{\nu}\frac{\partial {e_{\alpha}}^{\mu}}{\partial x^{\nu}}\frac{\partial }{\partial x^{\mu}}
\right)f\cr
&=&\left({e_{\alpha}}^{\mu}\frac{\partial{e_{\beta}}^{\nu}}{\partial x^{\mu}}\frac{\partial }{\partial x^{\nu}} - 
{e_{\beta}}^{\mu}\frac{\partial {e_{\alpha}}^{\nu}}{\partial x^{\mu}}\frac{\partial }{\partial x^{\nu}}
\right)f\cr
&=&\left({e_{\alpha}}^{\mu}\frac{\partial{e_{\beta}}^{\nu}}{\partial x^{\mu}} - 
{e_{\beta}}^{\mu}\frac{\partial {e_{\alpha}}^{\nu}}{\partial x^{\mu}}
\right)\frac{\partial }{\partial x^{\nu}}f. \nonumber
\end{eqnarray}
It is util to introduce the commutator coefficients
\begin{equation}
\label{commutator_coefficients_1}
 \left({e_{\alpha}}^{\mu}\frac{\partial{e_{\beta}}^{\nu}}{\partial x^{\mu}} - 
{e_{\beta}}^{\mu}\frac{\partial {e_{\alpha}}^{\nu}}{\partial x^{\mu}}
\right)={D^{\gamma}}_{\alpha\beta}\,\,{e_{\gamma}}^{\nu},
\end{equation}
where the commutator of  $ \hat{\bm e}_{\alpha}$ and $ \hat{\bm e}_{\beta}$ becomes
\begin{equation}
 \left[\hat{\bm e}_{\alpha},\hat{\bm e}_{\beta}\right] ={D^{\gamma}}_{\alpha\beta}\,\,{e_{\gamma}}^{\nu}\frac{\partial }{\partial x^{\nu}},
 \nonumber
\end{equation}
or simply
\begin{equation}
\label{comutador_base}
 \left[\hat{\bm e}_{\alpha},\hat{\bm e}_{\beta}\right] ={D^{\gamma}}_{\alpha\beta}\,\,\hat{\bm e}_{\gamma},
\end{equation}
similar to Lie algebra. 
Let us now isolate the commutator coefficients from (\ref{commutator_coefficients_1}) by using
${\omega^{\delta}}_{\nu}$
\begin{eqnarray}
\label{coeficiente_comutacao}
 {D^{\gamma}}_{\alpha\beta}{e_{\gamma}}^{\nu}{\omega^{\delta}}_{\nu} &=& {\omega^{\delta}}_{\nu}\left(
  {e_{\alpha}}^{\mu}\partial_{\mu}{e_{\beta}}^{\nu} - {e_{\beta}}^{\mu}\partial_{\mu}{e_{\alpha}}^{\nu}\right)\cr
  {D^{\gamma}}_{\alpha\beta}{\delta_{\gamma}}^{\delta} &=& {\omega^{\delta}}_{\nu}\left(
   {e_{\alpha}}^{\mu}\partial_{\mu}{e_{\beta}}^{\nu} - {e_{\beta}}^{\mu}\partial_{\mu}{e_{\alpha}}^{\nu}\right)\cr
 {D^{\delta}}_{\alpha\beta} &=& {\omega^{\delta}}_{\nu}\left(
   {e_{\alpha}}^{\mu}\partial_{\mu}{e_{\beta}}^{\nu} - {e_{\beta}}^{\mu}\partial_{\mu}{e_{\alpha}}^{\nu}\right).
\end{eqnarray}
Due to the commutator from this equation, it is important to notice that
\begin{equation}
\label{coeficiente_comutacao_2}
 {D^{\delta}}_{\alpha\beta} = -{D^{\delta}}_{\beta\alpha},
\end{equation}
where the commutator coefficients have antisymmetry in the index pair $(\alpha\beta)$.


\subsection{Conection coefficients}

In the coordinate basis the connection coefficients are given by (\ref{connection_2}). Now, let  $\{{\hat{\bm e}}_{\alpha} \}$ be the non-coordinate basis and $\{\tilde{\bm\theta}^{\beta}\}$ the dual basis, thus the connection coefficients with respect to the non-coordinates is
obtained in the same way that it is obtained in coordinate basis (\ref{connection_1}). Thus $\nabla_{\hat{\bm e}_{\beta}}\hat{\bm e}_{\gamma} = {\Gamma^{\alpha}}_{\beta\gamma}\hat{\bm e}_{\alpha} $ where it get
\begin{equation}
\label{connection_4}
 {\Gamma^{\alpha}}_{\beta\gamma}=\langle \tilde{\bm\theta}^{\alpha},
 \nabla_{\hat{\bm e}_{\beta}}\hat{\bm e}_{\gamma} \rangle,
\end{equation}
where we can use equations (\ref{base_NC1}) e (\ref{base_NC2}), such that the connection coefficients equation follows
\begin{equation}
\label{conexão_NC1}
{\Gamma^{\alpha}}_{\beta\gamma} =  \langle {\omega^{\alpha}}_{\mu}dx^{\mu}, \nabla_{{(e_{\beta}}^{\nu}\bm E_{\nu})}
 {e_{\gamma}}^{\rho}\bm E_{\rho}\rangle.
\end{equation}
Let us use the properties (\ref{covariant_derivative_3}) of covariant derivative, $\nabla_{f\bm X}\bm Y=f\nabla_{\bm X}\bm Y$,
where we can obtain from equation (\ref{conexão_NC1}) the following result
\begin{equation}
\label{conexão_NC2}
{\Gamma^{\alpha}}_{\beta\gamma} =  \langle {\omega^{\alpha}}_{\mu}dx^{\mu}, {e_{\beta}}^{\nu}\nabla_{(\bm E_{\nu})}
 {e_{\gamma}}^{\rho}\bm E_{\rho}\rangle,
\end{equation}
and now we can use the other propertie of covariant derivative (\ref{covariant_derivative_4}), $\nabla_{\bm X}(f\bm Y)= \bm X[f]\bm Y + f\nabla_{\bm X} \bm Y$,
to rewrite the equation (\ref{conexão_NC2}) as
\begin{equation}
\label{conexão_NC3}
{\Gamma^{\alpha}}_{\beta\gamma} =  \langle {\omega^{\alpha}}_{\mu}dx^{\mu}, 
{e_{\beta}}^{\nu}(\bm E_{\nu}[{e_{\gamma}}^{\rho}]\bm E_{\rho}+{e_{\gamma}}^{\rho} \nabla_{(\bm E_{\nu})}
 \bm E_{\rho})\rangle.
\end{equation}
From identity (\ref{connection_1}), where $\nabla_{(\bm E_{\nu})} \bm E_{\rho} = {\Gamma^{\sigma}}_{\nu\rho}\bm E_{\sigma} $,
we can calculate the scalar product 
\begin{equation}
\label{conexão_NC4}
{\Gamma^{\alpha}}_{\beta\gamma} =  {\omega^{\alpha}}_{\mu}
{e_{\beta}}^{\nu}(\partial_{\nu}[{e_{\gamma}}^{\rho}]{\delta^{\mu}}_{\rho}+
{e_{\gamma}}^{\rho} {\Gamma^{\sigma}}_{\nu\rho} {\delta^{\mu}}_{\sigma} ),
\end{equation}
that results in
\begin{equation}
\label{conexão_NC5}
{\Gamma^{\alpha}}_{\beta\gamma} =  {\omega^{\alpha}}_{\mu}
{e_{\beta}}^{\nu}(\partial_{\nu}{e_{\gamma}}^{\mu}+
{e_{\gamma}}^{\rho} \Gamma^{\mu}_{\nu\rho}),
\end{equation}
or
\begin{equation}
\label{conexão_NC6}
{\Gamma^{\alpha}}_{\beta\gamma} =  {\omega^{\alpha}}_{\mu}
{e_{\beta}}^{\nu}\nabla_{\nu}{e_{\gamma}}^{\mu}. 
\end{equation}
In addition, it is important to observe that the $\nu$th component of covariant derivative of vierbein field ${e_{\gamma}}^{\mu}$ acts only
on indice $\mu$ of coordinate basis $\{\bm E_{\mu}\}$.

We have seen in equation (\ref{connection_3}) that $\nabla_{\bm E_{\mu}}dx^{\rho}  = - {\Gamma^{\rho}}_{\mu\sigma} dx^{\sigma}$,
similarly we have
\begin{equation}
 {\Gamma^{\gamma}}_{\beta\alpha}\tilde{\bm\theta}^{\alpha}=-\nabla_{(\tilde{\bm e_{\beta}})}\tilde{\bm\theta}^{\gamma},
\end{equation}
where the connection coefficients in term of (\ref{base_NC1}) e (\ref{base_NC2}) reduce to,
\begin{eqnarray}
 {\Gamma^{\alpha}}_{\beta\gamma}&=&-\langle \bm e_{\gamma},
 \nabla_{({\bm e_{\beta}})}\bm\theta^{\alpha}\rangle\cr
 &=&-\langle {e_{\gamma}}^{\mu}\bm E_{\mu},
  \nabla_{({e_{\beta}}^{\nu}\bm E_{\nu})}({\omega^{\alpha}}_{\rho}dx^{\rho})\rangle\cr
 &=&-\langle {e_{\gamma}}^{\mu}\bm E_{\mu},
 {e_{\beta}}^{\nu}\nabla_{\bm E_{\nu}}({\omega^{\alpha}}_{\rho}dx^{\rho})\rangle\cr
 &=&-\langle {e_{\gamma}}^{\mu}\bm E_{\mu},
 {e_{\beta}}^{\nu}(\bm E_{\nu}[{\omega^{\alpha}}_{\rho}]dx^{\rho}+
 {\omega^{\alpha}}_{\rho}\nabla_{\bm E_{\nu}}dx^{\rho})\rangle \cr
 &=&-\langle {e_{\gamma}}^{\mu}\bm E_{\mu},
 {e_{\beta}}^{\nu}({\partial_{\nu}\omega^{\alpha}}_{\rho} dx^{\rho} -
 {\omega^{\alpha}}_{\rho}\Gamma_{\nu\sigma}^{\rho} dx^{\sigma})\rangle \cr
 &=&-\langle {e_{\gamma}}^{\mu}\bm E_{\mu},
 {e_{\beta}}^{\nu}(\partial_{\nu}{\omega^{\alpha}}_{\rho} - 
 {\omega^{\alpha}}_{\sigma}\Gamma_{\nu\rho}^{\sigma} ) dx^{\rho}\rangle \cr
 &=&-{e_{\gamma}}^{\mu} {e_{\beta}}^{\nu}\,\nabla_{\nu}{\omega^{\alpha}}_{\rho}\,{\delta_{\mu}}^{\rho},
\end{eqnarray}
such that
\begin{equation}
\label{conexão_NC7}
 {\Gamma^{\alpha}}_{\beta\gamma} = - {e_{\gamma}}^{\mu} {e_{\beta}}^{\nu}\nabla_{\nu} {\omega^{\alpha}}_{\mu}.
\end{equation} 
It is important to note that from (\ref{conexão_NC6}), 
\begin{equation}
{\Gamma^{\alpha}}_{\beta\gamma} = {e_{\beta}}^{\nu}\underbrace{ {\omega^{\alpha}}_{\mu}\nabla_{\nu}{e_{\gamma}}^{\mu}},\nonumber
\end{equation}
where the underbraced term can be replaced with aid of the identity (\ref{id_n_coordenada4}), where we have that  
${\omega^{\alpha}}_{\mu} \nabla_{\nu}{e_{\gamma}}^{\mu} =-{e_{\gamma}}^{\mu} \nabla_{\nu}{\omega^{\alpha}}_{\mu}$, 
and consequently the connection coefficients are given by (\ref{conexão_NC7}).


\section{Torsion tensor in non-coordinate basis and the first Cartan equation}
The first structure equation of Cartan can be obtained from definition of torsion tensor (\ref{torsion_1}),
 \begin{equation}
  \bm T(\bm X, \bm Y)=\nabla_{\bm X}\bm Y-\nabla_{\bm Y}\bm X - [\bm X, \bm Y], \nonumber
 \end{equation}
where the components of torsion tensor in non-coordinate basis follow that
\begin{equation}
 {T^{\alpha}}_{\beta\gamma}=\langle \tilde{\bm\theta}^{\alpha},
 T({\hat{\bm e}}_{\beta},{\hat{\bm e}}_{\gamma})\rangle, \nonumber
\end{equation}
such that
\begin{equation}
  {T^{\alpha}}_{\beta\gamma} = \langle \tilde{\bm\theta}^{\alpha},
  \nabla_{{\hat{\bm e}}_{\beta}}{\hat{\bm e}}_{\gamma}-
  \nabla_{{\hat{\bm e}}_{\gamma}}{\hat{\bm e}}_{\beta} - [{\hat{\bm e}}_{\beta}, {\hat{\bm e}}_{\gamma}]
  \rangle.\nonumber
\end{equation}
With aid of 
\begin{equation}
 \nabla_{{\hat{\bm e}}_{\beta}}{\hat{\bm e}}_{\gamma}=
 {\Gamma^{\delta}}_{\beta\gamma}\hat{\bm e}_{\delta} \nonumber
\end{equation}
and from (\ref{comutador_base}) where
\begin{equation}
 [\hat{\bm e}_{\beta},\hat{\bm e}_{\gamma}] ={D^{\delta}}_{\beta\gamma}\,\,\hat{\bm e}_{\delta},\nonumber
\end{equation}
the torsion tensor becomes
\begin{equation}
  {T^{\alpha}}_{\beta\gamma} = \langle \tilde{\bm\theta}^{\alpha},
     {\Gamma^{\delta}}_{\beta\gamma}\hat{\bm e}_{\delta} -
      {\Gamma^{\delta}}_{\gamma\beta}\hat{\bm e}_{\delta}-
   {D^{\delta}}_{\beta\gamma}\,\,\hat{\bm e}_{\delta}  \rangle \nonumber
   \end{equation}
from which we find
\begin{equation}
\label{torcao_2}
  {T^{\alpha}}_{\beta\gamma} = {\Gamma^{\alpha}}_{\beta\gamma}-
   {\Gamma^{\alpha}}_{\gamma\beta} -
   {D^{\alpha}}_{\beta\gamma}.
\end{equation}
It is straightforward to notice that the difference between torsion tensor in coordinate basis of equation (\ref{torsion_2}) and torsion tensor in non-coordinate basis of equation (\ref{torcao_2}) is the presence of commutator coefficient, ${D^{\alpha}}_{\beta\gamma}$.

We can obtain an expression that make a transition of torsion tensor, ${T^{\mu}}_{\nu\rho}$, from coordinate basis to torsion tensor ${T^{\alpha}}_{\beta\gamma}$
in non-coordinate basis. From equation (\ref{torcao_2}) we have
\begin{eqnarray}
 {T^{\alpha}}_{\beta\gamma} &=& {\Gamma^{\alpha}}_{\beta\gamma}  - {\Gamma^{\alpha}}_{\gamma\beta}
 - {D^{\alpha}}_{\beta\gamma} \cr
 &=& -{\omega^{\alpha}}_{\mu} {e_{\gamma}}^{\nu} \nabla_{\nu}{e_{\beta}}^{\mu}+
 {\omega^{\alpha}}_{\mu} {e_{\beta}}^{\nu} \nabla_{\nu}{e_{\gamma}}^{\mu} -{\omega^{\alpha}}_{\nu}\left({e_{\beta}}^{\mu}\partial_{\mu}{e_{\gamma}}^{\nu} - {e_{\gamma}}^{\mu}\partial_{\mu}{e_{\beta}}^{\nu}\right)\cr
 &=& -{\omega^{\alpha}}_{\mu} {e_{\gamma}}^{\nu}\left(\partial_{\nu}{e_{\beta}}^{\mu}+{e_{\beta}}^{\rho}{\Gamma^{\mu}}_{\nu\rho}\right)+
 {\omega^{\alpha}}_{\mu} {e_{\beta}}^{\nu}\left(\partial_{\nu}{e_{\gamma}}^{\mu} + {e_{\gamma}}^{\rho}{\Gamma^{\mu}}_{\nu\rho}\right) -{\omega^{\alpha}}_{\nu}\left({e_{\beta}}^{\mu}\partial_{\mu}{e_{\gamma}}^{\nu} - {e_{\gamma}}^{\mu}\partial_{\mu}{e_{\beta}}^{\nu}\right)\cr
 &=& - {\omega^{\alpha}}_{\mu} {e_{\gamma}}^{\nu}{e_{\beta}}^{\rho}{\Gamma^{\mu}}_{\nu\rho}
 +  {\omega^{\alpha}}_{\mu} {e_{\beta}}^{\nu}{e_{\gamma}}^{\rho}{\Gamma^{\mu}}_{\nu\rho}\cr
 &=&
  {\omega^{\alpha}}_{\mu} {e_{\beta}}^{\nu}{e_{\gamma}}^{\rho}\left({\Gamma^{\mu}}_{\nu\rho}-  {\Gamma^{\mu}}_{\rho\nu}\right),\nonumber
\end{eqnarray}
we have from (\ref{torsion_2}) that
${T^{\mu}}_{\nu\rho}= {\Gamma^{\mu}}_{\nu\rho} -{\Gamma^{\mu}}_{\rho\nu}$ 
which it follows
\begin{equation}
 \label{torcao_2.1}
  {T^{\alpha}}_{\beta\gamma} = {\omega^{\alpha}}_{\mu} {e_{\beta}}^{\nu}{e_{\gamma}}^{\rho}
  {T^{\mu}}_{\nu\rho}.
\end{equation}


If a manifold is free of torsion we have from (\ref{torcao_2}) that
\begin{equation}
\label{torcao3}
 {\Gamma^{\alpha}}_{\beta\gamma}-
   {\Gamma^{\alpha}}_{\gamma\beta} =
   {D^{\alpha}}_{\beta\gamma}
\end{equation}
or $2{\Gamma^{\alpha}}_{[\beta\gamma]}={D^{\alpha}}_{\beta\gamma}$.
As a consequence, in coordinate basis we have that 
${D^{\alpha}}_{\beta\gamma}= 0 $ such that
\begin{equation}
 {\Gamma^{\alpha}}_{\beta\gamma} =  {\Gamma^{\alpha}}_{\gamma\beta}.
\end{equation}


If the torsion tensor vanishes and from requirement that the metric tensor must be covariantly constant, $\nabla_{\alpha} g_{\beta\gamma}=0$, we have that the cycle permutation of $(\alpha,\beta,\gamma)$ yields
\begin{eqnarray}
 \nabla_{\alpha} g_{\beta\gamma} &=& 
 \partial_{\alpha} g_{\beta\gamma} - \Gamma_{\gamma\alpha\beta} -  \Gamma_{\beta\alpha\gamma} = 0\cr
  \nabla_{\beta} g_{\alpha\gamma} &=& 
  \partial_{\beta} g_{\alpha\gamma} - \Gamma_{\gamma\beta\alpha} -  \Gamma_{\alpha\beta\gamma} = 0\cr
   \nabla_{\gamma} g_{\alpha\beta} &=& 
  \partial_{\gamma} g_{\alpha\beta} - \Gamma_{\beta\gamma\alpha} -  \Gamma_{\alpha\gamma\beta} = 0\nonumber
\end{eqnarray}
Let us compute $\nabla_{\alpha} g_{\beta\gamma}+ \nabla_{\beta} g_{\alpha\gamma} -\nabla_{\gamma} g_{\alpha\beta}$, 
to obtain an expression for the connection coefficients
\begin{eqnarray}
  \partial_{\alpha} g_{\beta\gamma} + \partial_{\beta}g_{\alpha\gamma} - \partial_{\gamma} g_{\alpha\beta}
  -\Gamma_{\gamma\alpha\beta} -  \Gamma_{\beta\alpha\gamma} - \Gamma_{\gamma\beta\alpha} -  \Gamma_{\alpha\beta\gamma}
 +\Gamma_{\beta\gamma\alpha} + \Gamma_{\alpha\gamma\beta} = 0.\nonumber
\end{eqnarray}
We require that the torsion tensor is vanish, then from equation (\ref{torcao3}), $\Gamma_{\alpha\gamma\beta}-\Gamma_{\alpha\beta\gamma} = D_{\alpha\gamma\beta}$, we have
\begin{eqnarray}
  \partial_{\alpha} g_{\beta\gamma} + \partial_{\beta}g_{\alpha\gamma} - \partial_{\gamma} g_{\alpha\beta}
  -\Gamma_{\gamma\alpha\beta} - \Gamma_{\gamma\alpha\beta}+\Gamma_{\gamma\alpha\beta} 
  -\Gamma_{\gamma\beta\alpha} +D_{\beta\gamma\alpha} +D_{\alpha\gamma\beta} = 0,\nonumber
\end{eqnarray}
with $D_{\alpha\gamma\beta} = -D_{\alpha\beta\gamma}$ and $D_{\beta\gamma\alpha} = -D_{\beta\alpha\gamma}$, the above equation results in
\begin{equation}
\label{conexão_NC13}
 \Gamma_{\gamma\alpha\beta} =\frac{1}{2}
 \left(   \partial_{\alpha} g_{\beta\gamma} + \partial_{\beta}g_{\alpha\gamma} - \partial_{\gamma} g_{\alpha\beta}
+ D_{\gamma\alpha\beta} -D_{\alpha\beta\gamma}-D_{\beta\alpha\gamma} \right).
\end{equation}


\subsection{The first structure equation of Cartan}

Let us introduce the torsion 2-form
\begin{equation}
\label{torca_2_forma_1}
{\bm T}^{\alpha} = \frac{1}{2}{T^{\alpha}}_{\beta\gamma} \tilde{\bm\theta}^{\beta}
\wedge \tilde{\bm\theta}^{\gamma}.
\end{equation}
From equation (\ref{torcao_2}) where we have
\begin{equation}
  {T^{\alpha}}_{\beta\gamma} = \Gamma^{\alpha}_{\beta\gamma} -  \Gamma^{\alpha}_{\gamma\beta} - {D^{\alpha}}_{\beta\gamma}, \nonumber
\end{equation}
we can calculate the torsion two-form (\ref{torca_2_forma_1}) as it follows
\begin{eqnarray}
\label{torca_2_forma_2}
 {\bm T}^{\alpha} &=& \frac{1}{2}\left[{\Gamma^{\alpha}}_{\beta\gamma}\tilde{\bm\theta}^{\beta}
\wedge \tilde{\bm\theta}^{\gamma} - {\Gamma^{\alpha}}_{\gamma\beta}\tilde{\bm\theta}^{\beta}
\wedge \tilde{\bm\theta}^{\gamma} -
   {D^{\alpha}}_{\beta\gamma}\tilde{\bm\theta}^{\beta}
\wedge \tilde{\bm\theta}^{\gamma}\right]\cr
&=& \frac{1}{2}\left[{\Gamma^{\alpha}}_{\beta\gamma}\tilde{\bm\theta}^{\beta}
\wedge \tilde{\bm\theta}^{\gamma} + {\Gamma^{\alpha}}_{\gamma\beta}\tilde{\bm\theta}^{\gamma}
\wedge \tilde{\bm\theta}^{\beta} -
   {D^{\alpha}}_{\beta\gamma}\tilde{\bm\theta}^{\beta}
\wedge \tilde{\bm\theta}^{\gamma}\right]\cr
&=& \frac{1}{2}\left[ 2  {\Gamma^{\alpha}}_{\beta\gamma}\tilde{\bm\theta}^{\beta}
\wedge \tilde{\bm\theta}^{\gamma} -
   {D^{\alpha}}_{\beta\gamma}\tilde{\bm\theta}^{\beta}
\wedge \tilde{\bm\theta}^{\gamma}\right].
\end{eqnarray}
We can compute the second term in above equation by using (\ref{coeficiente_comutacao}) that follows
\begin{eqnarray}
\label{coeficinte_comutacao_2_forma_1}
 \frac{1}{2} {D^{\alpha}}_{\beta\gamma}\tilde{\bm\theta}^{\beta}
\wedge \tilde{\bm\theta}^{\gamma} &=& \frac{1}{2}{\omega^{\alpha}}_{\nu}\left(
 {e_{\beta}}^{\mu}\partial_{\mu}{e_{\gamma}}^{\nu}-{e_{\gamma}}^{\mu}\partial_{\mu}{e_{\beta}}^{\nu}\right)
 \tilde{\bm\theta}^{\beta}
\wedge \tilde{\bm\theta}^{\gamma}\cr
&=&  {\omega^{\alpha}}_{\nu}
 {e_{\beta}}^{\mu}\partial_{\mu}{e_{\gamma}}^{\nu}\,\,\tilde{\bm\theta}^{\beta}
\wedge \tilde{\bm\theta}^{\gamma}.
\end{eqnarray}
 We can notice that
\begin{eqnarray}
 \partial_{\mu}({e_{\gamma}}^{\nu} {\omega^{\alpha}}_{\nu}) = \partial_{\mu}({\delta_{\gamma}}^{\alpha})
 =0,\nonumber
\end{eqnarray}
where it results in
\begin{equation}
 {\omega^{\alpha}}_{\nu}\partial_{\mu}{e_{\gamma}}^{\nu}= - {e_{\gamma}}^{\nu} \partial_{\mu}{\omega^{\alpha}}_{\nu}, \nonumber
\end{equation}
hence, the equation (\ref{coeficinte_comutacao_2_forma_1}) with above identity becomes
\begin{equation}
 \label{coeficinte_comutacao_2_forma_2}
 \frac{1}{2} {D^{\alpha}}_{\beta\gamma}\tilde{\bm\theta}^{\beta}
\wedge \tilde{\bm\theta}^{\gamma} = -\left(\partial_{\mu}{\omega^{\alpha}}_{\nu}\right) {e_{\gamma}}^{\nu}  {e_{\beta}}^{\mu}
\tilde{\bm\theta}^{\beta}
\wedge \tilde{\bm\theta}^{\gamma} =  -\left(\partial_{\mu}{\omega^{\alpha}}_{\nu}\right)
( {e_{\beta}}^{\mu}\tilde{\bm\theta}^{\beta})\wedge ({e_{\gamma}}^{\nu}\tilde{\bm\theta}^{\gamma}).\nonumber
\end{equation}
With aid of expression (\ref{base_NC4}) it follows that
\begin{equation}
 \label{coeficinte_comutacao_2_forma_3}
 \frac{1}{2} {D^{\alpha}}_{\beta\gamma}\tilde{\bm\theta}^{\beta}
\wedge \tilde{\bm\theta}^{\gamma} =  -\left(\partial_{\mu}{\omega^{\alpha}}_{\nu}\right) 
(dx^{\mu})\wedge (dx^{\nu}).
\end{equation}
Now, if we compute an exterior derivative of expression (\ref{base_NC2}) we have that 
\begin{eqnarray}
\label{eq_Cartan1}
d \tilde{\bm\theta}^{\alpha}=\left(\partial_{\mu}{\omega^{\alpha}}_{\nu}\right) dx^{\mu}\wedge dx^{\nu}.
\end{eqnarray}
Hence we have that the equation (\ref{coeficinte_comutacao_2_forma_3}) becomes 
\begin{equation}
 \label{coeficinte_comutacao_2_forma_4}
 \frac{1}{2} {D^{\alpha}}_{\beta\gamma}\tilde{\bm\theta}^{\beta}
\wedge \tilde{\bm\theta}^{\gamma} = -d \tilde{\bm\theta}^{\alpha},
\end{equation}
where it can be returned in expression (\ref{torca_2_forma_2}), that results in
\begin{equation}
{\bm T}^{\alpha} =  \frac{1}{2}\left[ 2 {\Gamma^{\alpha}}_{\beta\gamma}\tilde{\bm\theta}^{\beta} \wedge \tilde{\bm\theta}^{\gamma} -
   {D^{\alpha}}_{\beta\gamma}\tilde{\bm\theta}^{\beta}
\wedge \tilde{\bm\theta}^{\gamma}\right] =  \left({\Gamma^{\alpha}}_{\beta\gamma}\tilde{\bm\theta}^{\beta}\right)
\wedge \tilde{\bm\theta}^{\gamma} + d \tilde{\bm\theta}^{\alpha}.
\end{equation}
We have introduced the {\it conection 1-form} in equation (\ref{connection_1_form}) in coordinate basis, similarly we have the conection 1-form in non-coordinate basis
\begin{equation}
 \label{conexao_1forma}
{\bm\Gamma^{\alpha}}_{\gamma} \equiv {\Gamma^{\alpha}}_{\beta\gamma}\tilde{\bm\theta}^{\beta}.
\end{equation}
With this conection 1-form in equation of torsion 2-form, we arrive in 
\begin{eqnarray}
\label{eq_Cartan5}
{\bm T}^{\alpha} = d \tilde{\bm\theta}^{\alpha}+{\bm\Gamma^{\alpha}}_{\beta} \wedge 
\tilde{\bm\theta}^{\beta}\hspace{1cm}\mbox{\bf first structure equation of Cartan }.
\end{eqnarray}
If the manifold is free of torsion then it follows that
\begin{eqnarray}
\label{eq_Cartan4}
d \tilde{\bm\theta}^{\alpha}=  -{\bm\Gamma^{\alpha}}_{\beta} \wedge 
\tilde{\bm\theta}^{\beta}.
\end{eqnarray}
For non-coordinate basis, the connection coefficients can be computed from (\ref{eq_Cartan4}).


\section{Curvature tensor  in non-coordinate basis and the second Cartan equation}

Let us examine the curvature tensor given by (\ref{curvature_1})
\begin{equation}
 {\bm R}(\bm X, \bm Y)\bm Z=\nabla_{\bm X}\nabla_{\bm Y}\bm Z- 
 \nabla_{\bm Y}\nabla_{\bm X}\bm Z- \nabla_{[\bm X,\bm Y]}\bm Z.\nonumber
\end{equation}
From this definition we can obtain the components of curvature tensor in non-coordinate basis with
$\bm X=\hat{\bm e}_{\gamma}$,  $\bm Y=\hat{\bm e}_{\delta}$ and  $\bm Z=\hat{\bm e}_{\beta}$, such as
\begin{equation}
 {\bm R}(\hat{\bm e}_{\gamma}, \hat{\bm e}_{\delta})\hat{\bm e}_{\beta}
 =\nabla_{\hat{\bm e}_\gamma}(\nabla_{\delta}\hat{\bm e}_{\beta}) - \nabla_{\hat{\bm e}\delta}(\nabla_{\gamma}\hat{\bm e}_{\beta})
 - \nabla_{[\hat{\bm e}_{\gamma},\hat{\bm e}_{\delta}]}
 \hat{\bm e}_{\beta}, \nonumber
\end{equation}
with $\nabla_{\hat{\bm e}_{\beta}}\hat{\bm e}_{\gamma} = {\Gamma^{\alpha}}_{\beta\gamma}\hat{\bm e}_{\alpha} $ and (\ref{comutador_base}) it results in
\begin{equation}
 {\bm R}(\hat{\bm e}_{\gamma}, \hat{\bm e}_{\delta})\hat{\bm e}_{\beta}
 =\nabla_{\hat{\bm e}_\gamma}({\Gamma^{\alpha}}_{\delta\beta}\hat{\bm e}_{\alpha}) -
  \nabla_{\hat{\bm e}_\delta}({\Gamma^{\alpha}}_{\gamma\beta}\hat{\bm e}_{\alpha})
 - \nabla_{({D^{\epsilon}}_{\gamma\delta}\,\,\hat{\bm e}_{\epsilon})}
 \hat{\bm e}_{\beta}. \nonumber
\end{equation}
We can use covariant derivative properties (\ref{covariant_derivative_3}) and (\ref{covariant_derivative_4}),
\begin{equation}
 \nabla_{f\bm X}\bm Y=f\nabla_{\bm X}\bm Y,\nonumber
\end{equation}
for the third term and
\begin{equation}
 \nabla_{\bm X}(f\bm Y)=\bm X[f]\bm Y+f\nabla_{\bm X}\bm Y \nonumber
\end{equation}
for the first and second terms.
Thus
\begin{equation}
 {\bm R}(\hat{\bm e}_{\gamma}, \hat{\bm e}_{\delta})\hat{\bm e}_{\beta}
 =\hat{\bm e}_{\gamma}[{\Gamma^{\alpha}}_{\delta\beta}]\hat{\bm e}_{\alpha} + 
 {\Gamma^{\alpha}}_{\delta\beta} \nabla_{\hat{\bm e}_\gamma}\hat{\bm e}_{\alpha} -
 \hat{\bm e}_{\delta}[{\Gamma^{\alpha}}_{\gamma\beta}]\hat{\bm e}_{\alpha} - 
 {\Gamma^{\alpha}}_{\gamma\beta} \nabla_{\hat{\bm e}_\delta}\hat{\bm e}_{\alpha}
 - {D^{\epsilon}}_{\gamma\delta}\nabla_{\hat{\bm e}_\epsilon}
 \hat{\bm e}_{\beta}, \nonumber
\end{equation}
that results in 
\begin{equation}
 {\bm R}(\hat{\bm e}_{\gamma}, \hat{\bm e}_{\delta})\hat{\bm e}_{\beta}
 =\hat{\bm e}_{\gamma}[{\Gamma^{\alpha}}_{\delta\beta}]\hat{\bm e}_{\alpha} -
 \hat{\bm e}_{\delta}[{\Gamma^{\alpha}}_{\gamma\beta}]\hat{\bm e}_{\alpha}+ 
 {\Gamma^{\alpha}}_{\delta\beta} {\Gamma^{\epsilon}}_{\gamma\alpha}\hat{\bm e}_{\epsilon} -
  {\Gamma^{\alpha}}_{\gamma\beta}{\Gamma^{\epsilon}}_{\delta\alpha}\hat{\bm e}_{\epsilon}
 - {D^{\epsilon}}_{\gamma\delta}{\Gamma^{\alpha}}_{\epsilon\beta}
 \hat{\bm e}_{\alpha}, \nonumber
\end{equation}
or 
\begin{equation}
 {\bm R}(\hat{\bm e}_{\gamma}, \hat{\bm e}_{\delta})\hat{\bm e}_{\beta}
 =\left\{\hat{\bm e}_{\gamma}[{\Gamma^{\alpha}}_{\delta\beta}] -
 \hat{\bm e}_{\delta}[{\Gamma^{\alpha}}_{\gamma\beta}]+ 
 {\Gamma^{\epsilon}}_{\delta\beta} {\Gamma^{\alpha}}_{\gamma\epsilon} -
  {\Gamma^{\epsilon}}_{\gamma\beta}\Gamma^{\alpha}_{\delta\epsilon}
 - {D^{\epsilon}}_{\gamma\delta}{\Gamma^{\alpha}}_{\epsilon\beta}\right\}\hat{\bm e}_{\alpha}. \nonumber
\end{equation}
From equation (\ref{curvature_1.1}) we have that
\begin{equation}
 {\bm R}(\hat{\bm e}_{\gamma}, \hat{\bm e}_{\delta})\hat{\bm e}_{\beta}
 ={R^{\alpha}}_{\beta\gamma\delta}\hat{\bm e}_{\alpha}, \nonumber
\end{equation}
it follows that the components of curvature tensor with respect to non-coordinate basis is given by
\begin{equation}
 {R^{\alpha}}_{\beta\gamma\delta}
 =\hat{\bm e}_{\gamma}[{\Gamma^{\alpha}}_{\delta\beta}] -
 \hat{\bm e}_{\delta}[{\Gamma^{\alpha}}_{\gamma\beta}]+ 
 {\Gamma^{\epsilon}}_{\delta\beta} {\Gamma^{\alpha}}_{\gamma\epsilon} -
  {\Gamma^{\epsilon}}_{\gamma\beta}\Gamma^{\alpha}_{\delta\epsilon}
 - {D^{\epsilon}}_{\gamma\delta}{\Gamma^{\alpha}}_{\epsilon\beta} \nonumber
\end{equation}
with (\ref{directionl_derivative_4}) where we have $\hat{\bm e}_{\gamma}[{\Gamma^{\alpha}}_{\delta\beta}] = \partial_{\gamma}{\Gamma^{\alpha}}_{\delta\beta}$
\begin{equation}
\label{curvatura_NC}
 {R^{\alpha}}_{\beta\gamma\delta}
 =\partial_{\gamma}{\Gamma^{\alpha}}_{\delta\beta} -
 \partial_{\delta}{\Gamma^{\alpha}}_{\gamma\beta}+ 
 {\Gamma^{\epsilon}}_{\delta\beta} {\Gamma^{\alpha}}_{\gamma\epsilon} -
  {\Gamma^{\epsilon}}_{\gamma\beta}{\Gamma^{\alpha}}_{\delta\epsilon}
 - {D^{\epsilon}}_{\gamma\delta}{\Gamma^{\alpha}}_{\epsilon\beta}.
\end{equation}
Furthermore, similar to torsion tensor, we have to notice that the principal difference between curvature tensor in coordinate basis of equation (\ref{curvature_2}) and curvature tensor in non-coordinate basis of equation (\ref{curvatura_NC}) is the presence of commutator coefficient, ${D^{\epsilon}}_{\gamma\delta}$.

\subsection{The second structure equation of Cartan}

Similarly as we have worked with the torsion 2-form, we can introduce the curvature 2-form
\begin{equation}
\label{curvatura_2forma}
 {\bm \Theta^{\alpha}}_{\beta}=\frac{1}{2}  {R^{\alpha}}_{\beta\gamma\delta}\tilde{\bm\theta}^{\gamma}
\wedge \tilde{\bm\theta}^{\delta}.
\end{equation}
Let us substitute ${R^{\alpha}}_{\beta\gamma\delta}$ from equation (\ref{curvatura_NC}) into curvature two-form
\begin{equation}
 {\bm \Theta^{\alpha}}_{\beta} = \frac{1}{2}\left[ 
\partial_{\gamma}{\Gamma^{\alpha}}_{\delta\beta} \tilde{\bm\theta}^{\gamma}\wedge \tilde{\bm\theta}^{\delta}
- \partial_{\delta}{\Gamma^{\alpha}}_{\gamma\beta}\tilde{\bm\theta}^{\gamma}\wedge \tilde{\bm\theta}^{\delta}  
+{\Gamma^{\epsilon}}_{\delta\beta} {\Gamma^{\alpha}}_{\gamma\epsilon}\tilde{\bm\theta}^{\gamma}
\wedge \tilde{\bm\theta}^{\delta}  -
  {\Gamma^{\epsilon}}_{\gamma\beta}{\Gamma^{\alpha}}_{\delta\epsilon}\tilde{\bm\theta}^{\gamma}\wedge \tilde{\bm\theta}^{\delta}
 - {D^{\epsilon}}_{\gamma\delta}{\Gamma^{\alpha}}_{\epsilon\beta}\tilde{\bm\theta}^{\gamma}
\wedge \tilde{\bm\theta}^{\delta}
 \right], \nonumber
\end{equation}
with $\tilde{\bm\theta}^{\gamma}\wedge \tilde{\bm\theta}^{\delta}=-\tilde{\bm\theta}^{\delta}
\wedge \tilde{\bm\theta}^{\gamma}$ and from equation (\ref{torcao3}) when the manifold is free of torsion, where
${\Gamma^{\alpha}}_{\beta\gamma}-{\Gamma^{\alpha}}_{\gamma\beta} = {D^{\alpha}}_{\beta\gamma}$,
we have
\begin{equation}
 {\bm \Theta^{\alpha}}_{\beta} = \frac{1}{2}\left[ 
 \partial_{\gamma}{\Gamma^{\alpha}}_{\delta\beta} \tilde{\bm\theta}^{\gamma}\wedge \tilde{\bm\theta}^{\delta}
- \partial_{\delta}{\Gamma^{\alpha}}_{\gamma\beta}(-\tilde{\bm\theta}^{\delta}\wedge \tilde{\bm\theta}^{\gamma})  
+{\Gamma^{\epsilon}}_{\delta\beta} {\Gamma^{\alpha}}_{\gamma\epsilon}\tilde{\bm\theta}^{\gamma}
\wedge \tilde{\bm\theta}^{\delta}  -
  {\Gamma^{\epsilon}}_{\gamma\beta}{\Gamma^{\alpha}}_{\delta\epsilon}(-\tilde{\bm\theta}^{\delta}
\wedge \tilde{\bm\theta}^{\gamma})
 -({\Gamma^{\epsilon}}_{\gamma\delta}-{\Gamma^{\epsilon}}_{\delta\gamma}){\Gamma^{\alpha}}_{\epsilon\beta}\tilde{\bm\theta}^{\gamma}
\wedge \tilde{\bm\theta}^{\delta}
 \right],\nonumber
\end{equation}
that it follows
\begin{equation}
 {\bm \Theta^{\alpha}}_{\beta} = \frac{1}{2}\left[ 2\,
 \partial_{\gamma}{\Gamma^{\alpha}}_{\delta\beta} \tilde{\bm\theta}^{\gamma}\wedge \tilde{\bm\theta}^{\delta}
+2\,{\Gamma^{\epsilon}}_{\delta\beta} {\Gamma^{\alpha}}_{\gamma\epsilon}\tilde{\bm\theta}^{\gamma}
\wedge \tilde{\bm\theta}^{\delta}  -{\Gamma^{\alpha}}_{\epsilon\beta}{\Gamma^{\epsilon}}_{\gamma\delta}\tilde{\bm\theta}^{\gamma}
\wedge \tilde{\bm\theta}^{\delta}+{\Gamma^{\alpha}}_{\epsilon\beta}{\Gamma^{\epsilon}}_{\delta\gamma}(-\tilde{\bm\theta}^{\delta}
\wedge \tilde{\bm\theta}^{\gamma})
 \right],\nonumber
\end{equation}
or
\begin{eqnarray}
 {\bm \Theta^{\alpha}}_{\beta} &=& \partial_{\gamma}{\Gamma^{\alpha}}_{\delta\beta} \tilde{\bm\theta}^{\gamma}\wedge \tilde{\bm\theta}^{\delta}
+ {\Gamma^{\epsilon}}_{\delta\beta} {\Gamma^{\alpha}}_{\gamma\epsilon}\tilde{\bm\theta}^{\gamma}
\wedge \tilde{\bm\theta}^{\delta} - {\Gamma^{\alpha}}_{\epsilon\beta}{\Gamma^{\epsilon}}_{\gamma\delta}\tilde{\bm\theta}^{\gamma}
\wedge \tilde{\bm\theta}^{\delta}\cr
 &=& \partial_{\gamma}{\Gamma^{\alpha}}_{\delta\beta} \tilde{\bm\theta}^{\gamma}\wedge \tilde{\bm\theta}^{\delta} 
+({\Gamma^{\alpha}}_{\gamma\epsilon}\tilde{\bm\theta}^{\gamma})
\wedge ( {\Gamma^{\epsilon}}_{\delta\beta} \tilde{\bm\theta}^{\delta})
 -{\Gamma^{\alpha}}_{\epsilon\beta}({\Gamma^{\epsilon}}_{\gamma\delta}\tilde{\bm\theta}^{\gamma})
\wedge \tilde{\bm\theta}^{\delta}.\nonumber
\end{eqnarray}
From definition from conection 1-form from (\ref{conexao_1forma}), we have
\begin{equation}
\label{curvatura_2forma3}
 {\bm \Theta^{\alpha}}_{\beta} =   \underbrace{
\partial_{\gamma}{\Gamma^{\alpha}}_{\delta\beta} \tilde{\bm\theta}^{\gamma}\wedge \tilde{\bm\theta}^{\delta} 
 +{\Gamma^{\alpha}}_{\epsilon\beta}\tilde{\bm\theta}^{\delta}
\wedge{\bm\Gamma^{\epsilon}}_{\delta}}+
{\bm\Gamma^{\alpha}}_{\epsilon} \wedge {\bm \Gamma^{\epsilon}}_{\beta}. 
\end{equation}
The underbraced terms can be simplifyied by use of exterior derivative rule (\ref{exterior_derivative}). If $f$ is a 0-form and $\xi$ is one-form we have that
\begin{equation}
 d(f\wedge\bm \xi)= (df)\wedge\bm\xi + f\wedge d \bm\xi,
\end{equation}
then
\begin{equation}
 d{\bm\Gamma^{\alpha}}_{\beta}=d({\Gamma^{\alpha}}_{\gamma\beta}\tilde{\bm\theta}^{\gamma})=
 (\partial_{\delta}{\Gamma^{\alpha}}_{\gamma\beta})\tilde{\bm\theta}^{\delta}\wedge\tilde{\bm\theta}^{\gamma}
 +{\Gamma^{\alpha}}_{\gamma\beta}d\tilde{\bm\theta}^{\gamma}. \nonumber
\end{equation}
In a manifold free of torsion we have from first Cartan equation (\ref{eq_Cartan4}), 
$d \tilde{\bm\theta}^{\gamma}= -{\bm\Gamma^{\gamma}}_{\beta}\wedge\tilde{\bm\theta}^{\beta}$, 
where it can be inserted into above equation that follows
\begin{equation}
 d{\bm\Gamma^{\alpha}}_{\beta}=
(\partial_{\delta}{\Gamma^{\alpha}}_{\gamma\beta})\tilde{\bm\theta}^{\delta}\wedge\tilde{\bm\theta}^{\gamma}
 +{\Gamma^{\alpha}}_{\gamma\beta}(-{\bm\Gamma^{\gamma}}_{\delta}\wedge\tilde{\bm\theta}^{\delta})
 =(\partial_{\gamma}{\Gamma^{\alpha}}_{\delta\beta})\tilde{\bm\theta}^{\gamma}\wedge\tilde{\bm\theta}^{\delta}
 +{\Gamma^{\alpha}}_{\gamma\beta}\tilde{\bm\theta}^{\delta}\wedge{\bm\Gamma^{\gamma}}_{\delta}.\nonumber
\end{equation}
Now we can change the underbraced terms from equation (\ref{curvatura_2forma3}) by above result, that it results in the following curvature two-form
\begin{equation}
 \label{curvatura_2forma4}
 {\bm \Theta^{\alpha}}_{\beta}= d{\bm\Gamma^{\alpha}}_{\beta}+ 
 {\bm\Gamma^{\alpha}}_{\gamma} \wedge {\bm \Gamma^{\gamma}}_{\beta}
 \hspace{1cm}\mbox{\bf second structure equation of Cartan}
\end{equation}
The second structure equation of Cartan is an efficient method for calculating the components of curvature tensor (\ref{curvatura_NC}) with respect to a non-coordinate basis. The second Cartan's equation gives an algororithm for the calculation of curvature from the conection coefficients ${\bm\Gamma^{\alpha}}_{\beta}$.

We have seen that the second exterior derivative of an $r$-form is zero, $d^2 {\bm\Gamma^{\alpha}}_{\beta} = 0$, thus in this sense we can write the Bianchi identity for the curvature 2-form,
\begin{equation}
 d{\bm\Theta^{\alpha}}_{\beta} - {\bm\Theta^{\alpha}}_{\gamma}\wedge {\bm\Gamma^{\gamma}}_{\beta} + {\bm\Gamma^{\alpha}}_{\gamma} \wedge {\bm \Theta^{\gamma}}_{\beta} = 0.\nonumber
\end{equation}
with aid of exterior derivative property (\ref{exterior_derivative}).

\section{Local Lorentz transformations}

In an $m$-dimensional Riemannian manifold, the metric tensor $g_{\mu\nu}$ has $\frac{m(m+1)}{2}$ degrees of freedom, in 4-dimensional spacetime it has 10 degrees of freedom. While the vierbein field ${e_{\alpha}}^{\mu}$ has $m^2$ degrees of freedom, in 4-dimensional spacetime it has 16 degrees of freedom. There are many non-coordinate bases which yield the same metric $\bm g$, for example, the the two main non-coordinate bases are (i) orthonormal basis and (ii) pseudo-orthonormal basis. Each of them is related to the other by local rotation,
\begin{equation}
\label{transformation_1}
 \tilde{\bm\theta} '^{\alpha}(p) = {\Lambda^{\alpha}}_{\beta} (p)\tilde{\bm\theta}^{\beta}(p)
\end{equation}
at each point $p$. From equation  (\ref{base_NC2}), where we have,
\begin{equation}
 \tilde{\bm\theta}^{\alpha}={\omega^{\alpha}}_{\mu}dx^{\mu},\nonumber
\end{equation}
and where it yields 
\begin{equation}
 \tilde{\bm\theta} '^{\alpha}(p) ={\omega '^{\alpha}}_{\mu}dx^{\mu} =
 {\Lambda^{\alpha}}_{\beta} (p)\tilde{\bm\theta}^{\beta}(p) \nonumber
\end{equation}
or
\begin{equation}
{\omega '^{\alpha}}_{\mu}dx^{\mu} =
 {\Lambda^{\alpha}}_{\beta}\,\, {\omega^{\beta}}_{\mu} dx^{\mu}\nonumber
\end{equation}
where it results in
\begin{equation}
\label{transformação_nao_coordenadas_1}
 {\omega '^{\alpha}}_{\mu} = {\Lambda^{\alpha}}_{\beta}\,\, {\omega^{\beta}}_{\mu}.
\end{equation}
Under the local transformation, $ {\Lambda^{\alpha}}_{\beta} $, the indices 
$\alpha,\beta\cdots$ rotated while  $\mu,\nu\cdots$, from coordinate basis (world indices) are not affected.

The basis vector of non-coordinate basis in a referential frame ${\cal O}$ is given by
\begin{equation}
\label{transformação_nao_coordenadas_3}
{\hat{\bm e}}_{\alpha}={e_{\alpha}}^{\mu}\bm E_{\mu}
\end{equation} 
while
\begin{equation}
\label{transformação_nao_coordenadas_2}
{\hat{\bm e}'}_{\alpha}={e_{\alpha}'}^{\mu}\bm E_{\mu}
\end{equation} 
is given in the referential frame ${\cal O}'$. From expression (\ref{transformação_nao_coordenadas_1}) where
${\omega '^{\alpha}}_{\mu} = {\Lambda^{\alpha}}_{\beta}\,\, {\omega^{\beta}}_{\mu}$ and from identity
(\ref{id_n_coordenada2}) where $ {e_{\alpha}}^{\mu} {\omega^{\beta}}_{\mu} = {\delta_{\alpha}}^{\beta} $, we can obtain 
\begin{eqnarray}
  {e_{\alpha}'}^{\mu} {\omega'^{\beta}}_{\mu} &=& {\delta_{\alpha}}^{\beta}\cr
   {e_{\alpha}'}^{\mu} \,\, {\Lambda^{\beta}}_{\gamma}\,\, {\omega^{\gamma}}_{\mu} &=&  {\delta_{\alpha}}^{\beta}
\end{eqnarray}
it can be multiplied by ${(\Lambda^{-1})^{\delta}}_{\beta}$, such that it follows
\begin{eqnarray}
   {(\Lambda^{-1})^{\delta}}_{\beta} {\Lambda^{\beta}}_{\gamma}  {e_{\alpha}'}^{\mu}
   {\omega^{\gamma}}_{\mu} &=&  {(\Lambda^{-1})^{\delta}}_{\beta}  {\delta_{\alpha}}^{\beta}\cr
   {\delta^{\delta}}_{\gamma} {\omega^{\gamma}}_{\mu} {e_{\alpha}'}^{\mu} &=&  {(\Lambda^{-1})^{\delta}}_{\alpha}\cr
    {\omega^{\delta}}_{\mu} {e_{\alpha}'}^{\mu} &=&  {(\Lambda^{-1})^{\delta}}_{\alpha}
\end{eqnarray}    
and multiplying by ${e_{\delta}}^{\nu}$ we have that
\begin{eqnarray}    
{e_{\delta}}^{\nu} {\omega^{\delta}}_{\mu} {e_{\alpha}'}^{\mu} &=& 
{e_{\delta}}^{\nu} {(\Lambda^{-1})^{\delta}}_{\alpha}\cr
{\delta^{\nu}}_{\mu}  {e_{\alpha}'}^{\mu} &=& {e_{\delta}}^{\nu} {(\Lambda^{-1})^{\delta}}_{\alpha}\nonumber
\end{eqnarray}
that results in
\begin{equation}
  {e_{\alpha}'}^{\mu} = {e_{\beta}}^{\mu}\,\, {(\Lambda^{-1})^{\beta}}_{\alpha}.
\end{equation}

Now, we can return to equation (\ref{transformação_nao_coordenadas_2}), where we have
\begin{equation}
 {\hat{\bm e}'}_{\alpha}={e_{\alpha}'}^{\mu}\bm E_{\mu} = {e_{\beta}}^{\mu}\,\, {(\Lambda^{-1})^{\beta}}_{\alpha}
 \bm E_{\mu}. \nonumber
\end{equation}
In local frame ${\cal O}$, from (\ref{transformação_nao_coordenadas_3}), where
${\hat{\bm e}}_{\alpha}={e_{\alpha}}^{\mu}\bm E_{\mu}$ and from (\ref{base_NC3}) where
$\bm E_{\mu} =  {\omega^{\gamma}}_{\mu} {\hat{\bm e}}_{\gamma} $, the above equation yields
\begin{eqnarray}
  {\hat{\bm e}'}_{\alpha} &=& {e_{\beta}}^{\mu}\,\, {(\Lambda^{-1})^{\beta}}_{\alpha}\,\,
  {\omega^{\gamma}}_{\mu} {\hat{\bm e}}_{\gamma} \cr
  &=&  {e_{\beta}}^{\mu}\,\,  {\omega^{\gamma}}_{\mu} {\hat{\bm e}}_{\gamma}
  {(\Lambda^{-1})^{\beta}}_{\alpha}\,\,{\hat{\bm e}}_{\gamma}\cr
  &=& {\delta_{\beta}}^{\gamma} {(\Lambda^{-1})^{\beta}}_{\alpha}\,\,{\hat{\bm e}}_{\gamma}\nonumber
\end{eqnarray}
so that
\begin{equation}
\label{transformation_2}
  {\hat{\bm e}'}_{\alpha} = {(\Lambda^{-1})^{\beta}}_{\alpha}\,\,{\hat{\bm e}}_{\beta}. 
\end{equation}

Let a tensor field of type (1,1) given by ${\bm T} = {T^{\alpha}}_{\beta} {\hat{\bm e}}_{\alpha}\otimes \tilde{\bm\theta}^{\beta}$
in local frame $\cal O$.
In any rotated local frame  ${\cal O}'$ it is given by
\begin{equation}
 {\bm T}' = {T'^{\alpha}}_{\beta}{\hat{\bm e}'}_{\alpha}\otimes \tilde{\bm\theta'}^{\beta}. \nonumber
\end{equation}
It is possible to obtain a rule how this tensor transforms by use of equation (\ref{transformation_1}) where
$\tilde{\bm\theta} '^{\alpha} = {\Lambda^{\alpha}}_{\beta}\tilde{\bm\theta}^{\beta}$ and the equation (\ref{transformation_2}) where
$ {\hat{\bm e}'}_{\alpha} = {(\Lambda^{-1})^{\beta}}_{\alpha}\,\,{\hat{\bm e}}_{\beta} $, with respective inverses
\begin{equation}
\label{transformation_3}
  \tilde{\bm\theta} ^{\alpha} = {(\Lambda^{-1})^{\alpha}}_{\beta}\tilde{\bm\theta}'^{\beta}
  \hspace*{1cm}\mbox{and} \hspace*{1cm}
{\hat{\bm e}}_{\alpha} = {\Lambda^{\beta}}_{\alpha}\,\,{\hat{\bm e}'}_{\beta}. 
\end{equation}
Thus the tensor ${\bm T} = {T^{\alpha}}_{\beta} {\hat{\bm e}}_{\alpha}\otimes \tilde{\bm\theta}^{\beta} $ in a local frame $\cal O$ is related to the other by local rotation by
\begin{eqnarray}
 {\bm T} &=& {T^{\alpha}}_{\beta} {\hat{\bm e}}_{\alpha}\otimes \tilde{\bm\theta}^{\beta}\cr
         &=& {T^{\alpha}}_{\beta}\left[{\Lambda^{\gamma}}_{\alpha}\,\,{\hat{\bm e}'}_{\gamma}\right]\otimes \left[{(\Lambda^{-1})^{\beta}}_{\delta}\tilde{\bm\theta}'^{\delta} \right]\cr
         &=& \left[{\Lambda^{\gamma}}_{\alpha} {T^{\alpha}}_{\beta} {(\Lambda^{-1})^{\beta}}_{\delta}  \right]{\hat{\bm e}'}_{\gamma} \otimes \tilde{\bm\theta}'^{\delta} \nonumber
\end{eqnarray}
from which we find the transformation rule,
\begin{equation}
\label{transf_tensor_nao_coord}
 {T'^{\alpha}}_{\beta} =
 {\Lambda^{\alpha}}_{\gamma} {T^{\gamma}}_{\delta} {(\Lambda^{-1})^{\delta}}_{\beta}.
\end{equation}
This transformation rule for tensor field of type (1,1) can be written in tensor form,
\begin{equation}
 {\bm T}' = {\bm\Lambda} {\bm T}  {\bm\Lambda}^{-1}.
\end{equation}
The upper non-coordinate indices are rotated by $\bm \Lambda$ while lower non-coordinate indices are rotated by $\bm \Lambda^{-1}$.
However not all objects with indices are the components of a tensor, an important example is provided by the transformation properties of the connection coefficients ${\Gamma^{\alpha}}_{\beta\gamma}$ determined by (\ref{connection_4}). The connection coefficients under the rotation is
\begin{equation}
 {{\Gamma'}^{\alpha}}_{\beta\gamma} = \langle \tilde{\bm\theta}'^{\alpha}, \nabla_{\hat{\bm e}'_{\beta}}\hat{\bm e}'_{\gamma} \rangle 
 = \langle {\Lambda^{\alpha}}_{\delta}  \tilde{\bm\theta}^{\delta},  \nabla_{\left[{(\Lambda^{-1})^{\epsilon}}_{\beta} \hat{\bm e}_{\epsilon}\right]}{(\Lambda^{-1})^{\zeta}}_{\gamma} \hat{\bm e}_{\zeta} \rangle, \nonumber
\end{equation}
withe properties (\ref{covariant_derivative_3}), (\ref{covariant_derivative_4}) and (\ref{directionl_derivative_4}) it follows that
\begin{eqnarray}
 {{\Gamma'}^{\alpha}}_{\beta\gamma} &=&  {\Lambda^{\alpha}}_{\delta} \langle  \tilde{\bm\theta}^{\delta}, {(\Lambda^{-1})^{\epsilon}}_{\beta} \nabla_{ \hat{\bm e}_{\epsilon}}{(\Lambda^{-1})^{\zeta}}_{\gamma} \hat{\bm e}_{\zeta} \rangle \cr
 &=&  {\Lambda^{\alpha}}_{\delta} {(\Lambda^{-1})^{\epsilon}}_{\beta} \langle  \tilde{\bm\theta}^{\delta},  \left[ \partial_{\epsilon}{(\Lambda^{-1})^{\zeta}}_{\gamma} \hat{\bm e}_{\zeta} + {(\Lambda^{-1})^{\zeta}}_{\gamma} {\Gamma^{\eta}}_{\epsilon\zeta} \hat{\bm e}_{\eta}\right] \rangle \cr
 &=& {\Lambda^{\alpha}}_{\delta} {(\Lambda^{-1})^{\epsilon}}_{\beta}     \left[ \partial_{\epsilon}{(\Lambda^{-1})^{\zeta}}_{\gamma} 
 {\delta^{\delta}}_{\zeta} + {(\Lambda^{-1})^{\zeta}}_{\gamma} {\Gamma^{\eta}}_{\epsilon\zeta}{\delta^{\delta}}_{\eta}\right] \cr
  &=& {\Lambda^{\alpha}}_{\delta} {(\Lambda^{-1})^{\epsilon}}_{\beta}  \partial_{\epsilon}{(\Lambda^{-1})^{\delta}}_{\gamma} 
  + {\Lambda^{\alpha}}_{\delta} {(\Lambda^{-1})^{\epsilon}}_{\beta} {(\Lambda^{-1})^{\zeta}}_{\gamma} {\Gamma^{\delta}}_{\epsilon\zeta} , \nonumber
\end{eqnarray}
Because of the first term on the right-hand side of above equation, the connection coefficients do not transform as the components of a tensor.
We can see how to transform the connection 1-form 
${{\bm\Gamma}^{\alpha}}_{\gamma}$ by calculate $ {{\Gamma'}^{\alpha}}_{\beta\gamma}\tilde{\bm\theta}'^{\beta}$,
\begin{equation}
 {{\Gamma'}^{\alpha}}_{\beta\gamma} \tilde{\bm\theta}'^{\beta}
  = {\Lambda^{\alpha}}_{\delta} {(\Lambda^{-1})^{\epsilon}}_{\beta} \tilde{\bm\theta}'^{\beta} \partial_{\epsilon}{(\Lambda^{-1})^{\delta}}_{\gamma} 
  + {\Lambda^{\alpha}}_{\delta} {(\Lambda^{-1})^{\zeta}}_{\gamma} {\Gamma^{\delta}}_{\epsilon\zeta} {(\Lambda^{-1})^{\epsilon}}_{\beta}\tilde{\bm\theta}'^{\beta} , \nonumber
\end{equation}
with transformation ${(\Lambda^{-1})^{\alpha}}_{\beta}\tilde{\bm\theta}'^{\beta} = \tilde{\bm\theta} ^{\alpha}$ from (\ref{transformation_3}) it follows that
\begin{equation}
 {{\bm\Gamma'}^{\alpha}}_{\gamma} 
  = {\Lambda^{\alpha}}_{\delta} \tilde{\bm\theta}^{\epsilon} \partial_{\epsilon}{(\Lambda^{-1})^{\delta}}_{\gamma} 
  + {\Lambda^{\alpha}}_{\delta} {(\Lambda^{-1})^{\zeta}}_{\gamma} {\Gamma^{\delta}}_{\epsilon\zeta} \tilde{\bm\theta}^{\epsilon}, \nonumber
\end{equation}
we can use (\ref{differential_f}), where 
$\tilde{\bm\theta}^{\epsilon} \partial_{\epsilon}{(\Lambda^{-1})^{\delta}}_{\gamma} = d{(\Lambda^{-1})^{\delta}}_{\gamma}$,   to reduce the above equation to
\begin{equation}
\label{conexão_NC11}
 {{\bm\Gamma'}^{\alpha}}_{\gamma} 
  = {\Lambda^{\alpha}}_{\delta} d{(\Lambda^{-1})^{\delta}}_{\gamma} 
  + {\Lambda^{\alpha}}_{\delta} {\bm\Gamma^{\delta}}_{\zeta}  {(\Lambda^{-1})^{\zeta}}_{\gamma}. 
\end{equation}
Also, the above transformation rule of the connection 1-form  ${{\bm\Gamma}^{\alpha}}_{\beta}$, can be obtained by rotate the torsion 2-form from first Cartan equation  (\ref{eq_Cartan5}),
\begin{equation}
 {\bm T}'^{\alpha} = d{\tilde{\bm\theta}}'^{\alpha}+ {{\bm\Gamma}'^{\alpha}}_{\beta} \wedge
  \tilde{\bm\theta}'^{\beta}. \nonumber
\end{equation}
On the left-hand side of the above equation we have ${\bm T}'^{\alpha} ={\Lambda^{\alpha}}_{\gamma}\,{\bm T}^{\gamma}$,
while on the right-hand side we can apply (\ref{transformation_1}),
\begin{eqnarray}
 {\Lambda^{\alpha}}_{\gamma}\,{\bm T}^{\gamma} &=& d({\Lambda^{\alpha}}_{\gamma}{\tilde{\bm\theta}}^{\gamma}) +
 {{\bm\Gamma}'^{\alpha}}_{\beta} \wedge({\Lambda^{\beta}}_{\gamma}\,\tilde{\bm\theta}^{\gamma})\cr
  {\Lambda^{\alpha}}_{\gamma}\left(d\tilde{\bm\theta}^{\gamma} + {{\bm\Gamma}^{\gamma}}_{\beta} \wedge
 \tilde{\bm\theta}^{\beta}\right) &=& d{\Lambda^{\alpha}}_{\gamma} \wedge  {\tilde{\bm\theta}}^{\gamma}
 +{\Lambda^{\alpha}}_{\gamma} d {\tilde{\bm\theta}}^{\gamma}+  
 {{\bm\Gamma}'^{\alpha}}_{\beta} \wedge\, {\Lambda^{\beta}}_{\gamma}\,\tilde{\bm\theta}^{\gamma}\cr
 {\Lambda^{\alpha}}_{\gamma}  {{\bm\Gamma}^{\gamma}}_{\beta} \wedge
 \tilde{\bm\theta}^{\beta} &=& d{\Lambda^{\alpha}}_{\gamma} \wedge  {\tilde{\bm\theta}}^{\gamma}+  
 {{\bm\Gamma}'^{\alpha}}_{\beta} \wedge\, {\Lambda^{\beta}}_{\gamma}\,\tilde{\bm\theta}^{\gamma}.
 \nonumber
\end{eqnarray}
We can isolate $ {{\bm\Gamma}'^{\alpha}}_{\beta} $ to obtain
\begin{equation}
 {{\bm\Gamma}'^{\alpha}}_{\beta} {\Lambda^{\beta}}_{\gamma}\, \wedge\,\tilde{\bm\theta}^{\gamma}
= {\Lambda^{\alpha}}_{\beta}  {{\bm\Gamma}^{\beta}}_{\gamma} \wedge
 \tilde{\bm\theta}^{\gamma} - d{\Lambda^{\alpha}}_{\gamma} \wedge  {\tilde{\bm\theta}}^{\gamma}\nonumber
\end{equation}
where it reduces to
\begin{equation}
 {{\bm\Gamma}'^{\alpha}}_{\beta} {\Lambda^{\beta}}_{\gamma}
= {\Lambda^{\alpha}}_{\beta}  {{\bm\Gamma}^{\beta}}_{\gamma} 
- d{\Lambda^{\alpha}}_{\gamma},\nonumber
\end{equation}
and multiplying both sides by ${\bm\Lambda}^{-1}$ from the right, we have
\begin{equation}
 {{\bm\Gamma}'^{\alpha}}_{\beta} {\Lambda^{\beta}}_{\gamma}{(\Lambda^{-1})^{\gamma}}_{\delta}
= {\Lambda^{\alpha}}_{\beta}  {{\bm\Gamma}^{\beta}}_{\gamma} {(\Lambda^{-1})^{\gamma}}_{\delta}
- (d{\Lambda^{\alpha}}_{\gamma}){(\Lambda^{-1})^{\gamma}}_{\delta}.\nonumber
\end{equation}
We can use 
\begin{equation}
  {{\bm\Gamma}'^{\alpha}}_{\beta} {\Lambda^{\beta}}_{\gamma}{(\Lambda^{-1})^{\gamma}}_{\delta} =  {{\bm\Gamma}'^{\alpha}}_{\beta}\, {\delta^{\beta}}_{\delta} ={{\bm\Gamma}'^{\alpha}}_{\delta} \nonumber
\end{equation}
and
\begin{eqnarray}
d\left[{\Lambda^{\alpha}}_{\gamma})({\Lambda^{-1})^{\gamma}}_{\delta}\right] &=& 
(d{\Lambda^{\alpha}}_{\gamma})({\Lambda^{-1})^{\gamma}}_{\delta}+
{\Lambda^{\alpha}}_{\gamma}d({\Lambda^{-1})^{\gamma}}_{\delta},\cr
d{\delta^{\alpha}}_{\delta} &=& (d{\Lambda^{\alpha}}_{\gamma})({\Lambda^{-1})^{\gamma}}_{\delta}+
{\Lambda^{\alpha}}_{\gamma}d({\Lambda^{-1})^{\gamma}}_{\delta}\nonumber
\end{eqnarray}
where it follows that
\begin{equation}
(d{\Lambda^{\alpha}}_{\gamma})({\Lambda^{-1})^{\gamma}}_{\delta} =
- {\Lambda^{\alpha}}_{\gamma}d({\Lambda^{-1})^{\gamma}}_{\delta}.
\end{equation}
Thus we can express $ {{\bm\Gamma}'^{\alpha}}_{\beta} $ as
\begin{equation}
\label{conexão_NC12}
 {{\bm\Gamma}'^{\alpha}}_{\beta} = 
 {\Lambda^{\alpha}}_{\gamma}  {{\bm\Gamma}^{\gamma}}_{\delta} {(\Lambda^{-1})^{\delta}}_{\beta}
+ {\Lambda^{\alpha}}_{\gamma}d({\Lambda^{-1})^{\gamma}}_{\beta}.
\end{equation}
Thus, the above result is the same one obtained in equation (\ref{conexão_NC11}).

It is easy to notice from expression (\ref{transf_tensor_nao_coord}), that the curvature two-form transforms as 
\begin{equation}
\label{transf_curv_2_forma}
 {\bm\Theta'^{\alpha}}_{\beta} =
 {\Lambda^{\alpha}}_{\gamma} {\bm\Theta^{\gamma}}_{\delta} {(\Lambda^{-1})^{\delta}}_{\beta},
\end{equation}
under a local frame transformation $\bm \Lambda$.


\section{The non-coordinate orthonormal basis}

General Relativity is based on the concept of spacetime, which is a 4-dimensional differentiable manifold $\cal M$ with
Lorentzian metric, denoted by $({\cal M}, \bm g)$ where 
the components of the metric tensor $\bm g$ of spacetime with respect a coordinate basis $\left\{\bm E_{\mu}=\dfrac{\partial}{\partial x^{\mu}}\right\}$ are given by (\ref{metric_tensor_2}),
\begin{equation}
 \bm g(\bm E_{\mu},\bm E_{\nu}) = g_{\mu\nu} . \nonumber
\end{equation} 
However, for many purposes it is more convenient to use a non-coordinate orthonormal basis, or orthonormal tetrad or Lorentz frame
$\left\{\hat{\bm e}_{\alpha}\right\}$, where we have:
\begin{equation}
  g({\hat{\bm e}}_{\alpha},{\hat{\bm e}}_{\beta})=\eta_{\alpha\beta},
\end{equation}
where $(\eta_{\alpha\beta}) = \mbox{diag}(-1, 1, 1, 1)$ are the components of metric tensor of Minkowski reference system. 
The relationship between coordinate basis with non-coordinate orthonormal basis is given by (\ref{tensor_metrico_n_coordenadas_2}), where
\begin{equation}
 g({\hat{\bm e}}_{\alpha},{\hat{\bm e}}_{\beta})={e_{\alpha}}^{\mu}{e_{\beta}}^{\nu}g_{\mu\nu},\nonumber
\end{equation}
that follows that
\begin{equation}
\label{base_ortonormal_4}
\eta_{\alpha\beta} = {e_{\alpha}}^{\mu}{e_{\beta}}^{\nu}g_{\mu\nu}.
\end{equation}
In the inverse way it results that
\begin{equation}
\label{base_ortonormal_5}
 g_{\mu\nu} = {\omega^{\alpha}}_{\mu}{\omega^{\beta}}_{\nu}\eta_{\alpha\beta}.\nonumber
\end{equation}

The {\bf metric tensor} given by (\ref{tensor_metrico_n_coordenadas}) is
\begin{eqnarray}
\label{tensor_metrico_ortonormal}
 \mbox{\bf g} &=&  \eta_{\alpha\beta} \tilde{\bm\theta}^{\alpha}\otimes \tilde{\bm\theta}^{\beta}\cr
  &=& -(\tilde{\bm\theta}^{0}\otimes \tilde{\bm\theta}^{0})+(\tilde{\bm\theta}^{1}\otimes \tilde{\bm\theta}^{1})
+(\tilde{\bm\theta}^{2}\otimes \tilde{\bm\theta}^{2})+(\tilde{\bm\theta}^{3}\otimes \tilde{\bm\theta}^{3}).
\end{eqnarray}
The metric tensor in non-coordinate orthonormal basis is a system of locally inertial coordinates or local Lorentz spacetime.

\subsection{The Ricci rotation coefficients}

In the non-coordinate orthonormal basis the Minkowskian metric tensor is constant and consequently this local reference system is a rigid frame with $\partial_{\gamma}\eta_{\alpha\beta}=0$, then we can review equation (\ref{conexão_NC13}) where we have,
\begin{equation}
\label{conexão_NC14}
 \Gamma_{\gamma\alpha\beta} =\frac{1}{2}
 \left(D_{\gamma\alpha\beta} -D_{\alpha\beta\gamma}-D_{\beta\alpha\gamma}\right). 
\end{equation}
we can interchange $\gamma \leftrightarrow \beta$ into above equation such as
\begin{eqnarray}
 \Gamma_{\beta\alpha\gamma} &=& \frac{1}{2}
 \left(D_{\beta\alpha\gamma} -D_{\alpha\gamma\beta}-D_{\gamma\alpha\beta}\right) \cr
  &=& - \frac{1}{2}
 \left(D_{\gamma\alpha\beta} + D_{\alpha\gamma\beta} -D_{\beta\alpha\gamma} \right), \nonumber
\end{eqnarray}
where we can use equation (\ref{coeficiente_comutacao_2}) for the second term in right hand side of above equation, 
$D_{\alpha\gamma\beta} = -D_{\alpha\beta\gamma}$, that it follows,
\begin{equation}
 \Gamma_{\beta\alpha\gamma} = - \frac{1}{2}  \left(D_{\gamma\alpha\beta} - D_{\alpha\beta\gamma} -D_{\beta\alpha\gamma} \right). \nonumber
\end{equation}
Now we can compare above equation with (\ref{conexão_NC14}) and obtain that
\begin{equation}
 \label{rotacao_Ricci_1}
 \Gamma_{\gamma\alpha\beta} = -  \Gamma_{\beta\alpha\gamma}.
\end{equation}
The connection coefficients in non-coordinate orthonormal basis also referred as the {\it Ricci rotation coefficients} \cite{Wald, Nakahara, Stephani}.
The antisymmetry of the  Ricci rotation coefficients in the index pair $(\beta\gamma)$ yields   $n^2\frac{(n-1)}{2}$, in 4-dimensional spacetime
$V^4$, we have 24 components $\Gamma_{\gamma\alpha\beta}$.

The matrix of Ricci rotation coefficients $(\Gamma_{\alpha\gamma\beta})$ is $4\times 4\times 4$ matrix and we can see this in 4 layers
\begin{equation}
\hspace{-230pt}
\Gamma_{\alpha 0 \beta} =
\begin{pmatrix} 
                       0 & \Gamma_{001}   & \Gamma_{002}   & \Gamma_{003} \cr
			 - \Gamma_{001} & 0  & \Gamma_{102}   & \Gamma_{103} \cr
			- \Gamma_{002} & - \Gamma_{102}   & 0 &  \Gamma_{203} \cr
			- \Gamma_{003} & - \Gamma_{103}   & -\Gamma_{203} & 0 
\end{pmatrix}\nonumber
\end{equation}
\begin{equation}
\hspace{-80pt}
\Gamma_{\alpha 1 \beta} =
\begin{pmatrix} 
                       0 & \Gamma_{011}   & \Gamma_{012}   & \Gamma_{013} \cr
			- \Gamma_{011} & 0  & \Gamma_{112}   & \Gamma_{113} \cr
			- \Gamma_{012} & - \Gamma_{112}   & 0 &  \Gamma_{213} \cr
			- \Gamma_{013} & - \Gamma_{113}   & -\Gamma_{213} & 0 
\end{pmatrix}\nonumber
\end{equation}
\begin{equation}
\hspace{70pt}
\Gamma_{\alpha 2 \beta} =
\begin{pmatrix} 
                       0 & \Gamma_{021}   & \Gamma_{022}   & \Gamma_{023} \cr
			- \Gamma_{021} & 0  & \Gamma_{122}   & \Gamma_{123} \cr
			- \Gamma_{022} & - \Gamma_{122}   & 0 &  \Gamma_{223} \cr
			- \Gamma_{023} & - \Gamma_{123}   & -\Gamma_{223} & 0 
\end{pmatrix}\nonumber
\end{equation}
\begin{equation}
\hspace{220pt}
\Gamma_{\alpha 3 \beta} =
\begin{pmatrix} 
                       0 & \Gamma_{031}   & \Gamma_{032}   & \Gamma_{033} \cr
			- \Gamma_{031} & 0  & \Gamma_{132}   & \Gamma_{133} \cr
			- \Gamma_{032} & - \Gamma_{132}   & 0 &  \Gamma_{233} \cr
			- \Gamma_{033} & - \Gamma_{133}   & -\Gamma_{233} & 0 
\end{pmatrix}\nonumber
\end{equation}
We can see the 24 terms, 6 in each matrix layer.

In a coordinate basis the  Ricci rotation coefficients are substituted by the Christoffel symbols ${\Gamma^{\mu}}_{\nu\rho}$ that are symmetric in the index pair $(\nu\rho)$, ${\Gamma^{\mu}}_{\nu\rho} = {\Gamma^{\mu}}_{\rho\nu} $. It yields  $n^2\frac{(n+1)}{2}$ terms, where in 
in 4-dimensional manifold $V^4$, there are 40 components $ {\Gamma^{\alpha}}_{\beta\gamma} $. Thus, we have advantage in use of non-coordinate orthonormal basis for calculating curvature.
The Cartan's method for the calculation of curvature is more compact and efficient in many applications. The calculation of the connection 1-forms (\ref{connection_1_form}) are obtained from (\ref{rotacao_Ricci_1}),
\begin{eqnarray}
\label{conexão_NC10}
 \Gamma_{\alpha\gamma\beta} \tilde{\bm\theta}^{\gamma}
 &=& -\Gamma_{\beta\gamma\alpha}\tilde{\bm\theta}^{\gamma}\cr
  {\bm\Gamma}_{\alpha\beta} &=& -{\bm\Gamma}_{\beta\alpha},
\end{eqnarray}
with the first Cartan equation determine  ${\bm\Gamma}_{\alpha\beta}$ uniquely. In a non-coordinate orthonormal basis in 4-dimensional manifold, at most six independent connection 1-forms survive.

\subsection{The first equation of Cartan } 

In a non-coordinate orthonormal basis the first Cartan equation yields a system of four equations
\begin{equation}
\label{eq_Cartan6}
 \begin{cases}
   -d\tilde{\bm\theta}^{0}= {\bm\Gamma^{0}}_{1} \wedge \tilde{\bm\theta}^{1}+
  {\bm\Gamma^{0}}_{2} \wedge \tilde{\bm\theta}^{2}+ {\bm\Gamma^{0}}_{3} \wedge \tilde{\bm\theta}^{3} \cr
   -d\tilde{\bm\theta}^{1}= {\bm\Gamma^{1}}_{0} \wedge \tilde{\bm\theta}^{0}+
  {\bm\Gamma^{1}}_{2} \wedge \tilde{\bm\theta}^{2}+ {\bm\Gamma^{1}}_{3} \wedge \tilde{\bm\theta}^{3} \cr 
-d\tilde{\bm\theta}^{2}= {\bm\Gamma^{2}}_{0} \wedge \tilde{\bm\theta}^{0}+
  {\bm\Gamma^{2}}_{1} \wedge \tilde{\bm\theta}^{1}+ {\bm\Gamma^{2}}_{3} \wedge \tilde{\bm\theta}^{3} \cr 
  -d\tilde{\bm\theta}^{3}= {\bm\Gamma^{3}}_{0} \wedge \tilde{\bm\theta}^{0}+
  {\bm\Gamma^{3}}_{1} \wedge \tilde{\bm\theta}^{1}+ {\bm\Gamma^{3}}_{2} \wedge \tilde{\bm\theta}^{2} .\cr 
  \end{cases}
\end{equation}
There are 12  Ricci rotation coefficients in above system, but with the antisymmetry of the connection 1-forms (\ref{conexão_NC10}), ${\bm\Gamma}_{\beta\alpha} = -{\bm\Gamma}_{\alpha\beta}$ the Ricci rotation coefficients are reduced to 6. Let  $\alpha=0$ (timelike) and $\beta=i=1,2,3$ (spacelike), where we have
$ \bm \Gamma_{0i} = -\bm \Gamma_{i0}$, thus,
\begin{equation}
\label{coeficinte_ricci_1_forma_5}
 {\bm \Gamma^{0}}_{i}=\eta^{00} \bm \Gamma_{0i}=(-1)\bm \Gamma_{0i} \hspace{1cm} \mbox{or} 
 \hspace{1cm}  {\bm \Gamma^{0}}_{i} = \bm \Gamma_{i0}.
\end{equation}
Moreover, we have that
\begin{equation}
 {\bm \Gamma^{i}}_{0}=\eta^{ij} \bm \Gamma_{j0}=\delta^{ij} \bm \Gamma_{j0},
\end{equation}
for  $i=1$, it yields
\begin{equation}
 {\bm \Gamma^{1}}_{0}= \bm \Gamma_{10},
\end{equation}
with respect to (\ref{coeficinte_ricci_1_forma_5}) where ${\bm \Gamma^{0}}_{1} = \bm \Gamma_{10}$
we have that
\begin{equation}
 \label{coeficinte_ricci_1_forma_6}
 {\bm \Gamma^{1}}_{0}={\bm \Gamma^{0}}_{1} .
\end{equation}
The next step, let $\alpha=1$ e $\beta=2$ (both spacelike), and from (\ref{conexão_NC10}), we have $\bm \Gamma_{12} = -\bm \Gamma_{21}$,
\begin{equation}
 {\bm \Gamma^{1}}_{2}=\eta^{1k}\bm \Gamma_{k2}=\delta^{11} \bm \Gamma_{12}=\bm \Gamma_{12}.
\end{equation}
For $\bm \Gamma_{21}$ we have
\begin{equation}
 {\bm \Gamma^{2}}_{1}=\eta^{2k}\bm \Gamma_{k1}=\delta^{22} \bm \Gamma_{21}=\bm \Gamma_{21}=-\bm \Gamma_{12},
\end{equation}
comparing the two above expression we have that 
\begin{equation}
 \label{curvatura_2forma9}
 {\bm \Gamma^{1}}_{2}= -{\bm \Gamma^{2}}_{1}.
\end{equation}
With these results we can write the system of equations (\ref{eq_Cartan6}) as
\begin{equation}
\label{equacao_de_Cartan_ortonormal}
 \begin{cases}
   -d\tilde{\bm\theta}^{0}= {\bm\Gamma^{0}}_{1} \wedge \tilde{\bm\theta}^{1}+
  {\bm\Gamma^{0}}_{2} \wedge \tilde{\bm\theta}^{2}+ {\bm\Gamma^{0}}_{3} \wedge \tilde{\bm\theta}^{3} \cr
   -d\tilde{\bm\theta}^{1}= {\bm\Gamma^{0}}_{1} \wedge \tilde{\bm\theta}^{0}+
  {\bm\Gamma^{1}}_{2} \wedge \tilde{\bm\theta}^{2}+ {\bm\Gamma^{1}}_{3} \wedge \tilde{\bm\theta}^{3} \cr 
-d\tilde{\bm\theta}^{2}= {\bm\Gamma^{0}}_{2} \wedge \tilde{\bm\theta}^{0} -
  {\bm\Gamma^{1}}_{2} \wedge \tilde{\bm\theta}^{1}+ {\bm\Gamma^{2}}_{3} \wedge \tilde{\bm\theta}^{3} \cr 
  -d\tilde{\bm\theta}^{3}= {\bm\Gamma^{0}}_{3} \wedge \tilde{\bm\theta}^{0} -
  {\bm\Gamma^{1}}_{3} \wedge \tilde{\bm\theta}^{1}- {\bm\Gamma^{2}}_{3} \wedge \tilde{\bm\theta}^{2} \cr 
  \end{cases}
\end{equation}
with six independent connection 1-forms: ${\bm\Gamma^{0}}_{1},{\bm\Gamma^{0}}_{2},
{\bm\Gamma^{0}}_{3},{\bm\Gamma^{1}}_{2},{\bm\Gamma^{1}}_{3}$ e ${\bm\Gamma^{2}}_{3}$ to be determined.

\subsection{The second equation of Cartan } 
For the calculation of curvature we use the second Cartan equation from  (\ref{curvatura_2forma4}), 
\begin{equation}
  \bm \Theta_{\alpha\beta}= d \bm\Gamma_{\alpha\beta}+
 \bm\Gamma_{\alpha\gamma} \wedge {\bm \Gamma^{\gamma}}_{\beta}
 \nonumber
\end{equation}
where the curvature 2-forms $\bm \Theta_{\alpha\beta}$ are antisymmetric in the index pair $\alpha\beta$. 
From  (\ref{conexão_NC10}) we have that
\begin{eqnarray}
 \bm \Theta_{\alpha\beta} =  -d{\bm\Gamma}_{\beta\alpha}-
 \bm\Gamma_{\gamma\alpha} \wedge {\bm \Gamma^{\gamma}}_{\beta}
 = -d{\bm\Gamma}_{\beta\alpha}-{\bm\Gamma^{\gamma}}_{\alpha} \wedge \bm \Gamma_{\gamma\beta} 
 = -d{\bm\Gamma}_{\beta\alpha}+\bm \Gamma_{\gamma\beta} \wedge {\bm\Gamma^{\gamma}}_{\alpha} 
 = -d{\bm\Gamma}_{\beta\alpha}-\bm \Gamma_{\beta\gamma} \wedge {\bm\Gamma^{\gamma}}_{\alpha}, \nonumber
\end{eqnarray}
it follows that
\begin{equation}
 \label{curvatura_2forma6}
 \bm \Theta_{\alpha\beta} = -\bm \Theta_{\beta\alpha}.
\end{equation}
The antisymmetry of the curvature 2-forms, $\bm \Theta_{\alpha\beta}$, yields a system of six equations
\begin{equation}
 \begin{cases}
    \bm \Theta_{01}= d \bm\Gamma_{01}+ \bm\Gamma_{02} \wedge {\bm \Gamma^{2}}_{1} + 
    \bm\Gamma_{03} \wedge {\bm \Gamma^{3}}_{1} \cr
    \bm \Theta_{02}= d \bm\Gamma_{02}+ \bm\Gamma_{01} \wedge {\bm \Gamma^{1}}_{2} + 
    \bm\Gamma_{03} \wedge {\bm \Gamma^{3}}_{2} \cr
     \bm \Theta_{03}= d \bm\Gamma_{03}+ \bm\Gamma_{01} \wedge {\bm \Gamma^{1}}_{3} + 
    \bm\Gamma_{02} \wedge {\bm \Gamma^{2}}_{3} \cr
     \bm \Theta_{12}= d \bm\Gamma_{12}+ \bm\Gamma_{10} \wedge {\bm \Gamma^{0}}_{2} + 
    \bm\Gamma_{13} \wedge {\bm \Gamma^{3}}_{2} \cr
    \bm \Theta_{13}= d \bm\Gamma_{13}+ \bm\Gamma_{10} \wedge {\bm \Gamma^{0}}_{3} + 
    \bm\Gamma_{12} \wedge {\bm \Gamma^{2}}_{3} \cr
    \bm \Theta_{23}= d \bm\Gamma_{23}+ \bm\Gamma_{20} \wedge {\bm \Gamma^{0}}_{3} + 
    \bm\Gamma_{21} \wedge {\bm \Gamma^{1}}_{3} \nonumber
 \end{cases}
\end{equation}
We have seen that there are six independent connection 1-forms: ${\bm\Gamma^{0}}_{1},{\bm\Gamma^{0}}_{2},
{\bm\Gamma^{0}}_{3},{\bm\Gamma^{1}}_{2},{\bm\Gamma^{1}}_{3}$ e ${\bm\Gamma^{2}}_{3}$ to be determined.  With conditions  (\ref{coeficinte_ricci_1_forma_6}), $ {\bm \Gamma^{1}}_{0}={\bm \Gamma^{0}}_{1} = -\bm \Gamma_{01}$, and (\ref{curvatura_2forma9}), 
$ {\bm \Gamma^{1}}_{2}= -{\bm \Gamma^{2}}_{1} = \bm \Gamma_{12}$, the above system of equations from second Cartan equation becomes 
\begin{equation}
\label{sistema_2a_equacao_de_Cartan}
 \begin{cases}
    \bm \Theta_{01}= d \bm\Gamma_{01} - \bm\Gamma_{02} \wedge \bm \Gamma_{12} - 
    \bm\Gamma_{03} \wedge \bm \Gamma_{13} \cr
    \bm \Theta_{02}= d \bm\Gamma_{02}+ \bm\Gamma_{01} \wedge \bm \Gamma_{12} - 
    \bm\Gamma_{03} \wedge \bm \Gamma_{23} \cr
     \bm \Theta_{03}= d \bm\Gamma_{03}+ \bm\Gamma_{01} \wedge \bm \Gamma_{13} + 
    \bm\Gamma_{02} \wedge \bm \Gamma_{23} \cr
     \bm \Theta_{12}= d \bm\Gamma_{12}+ \bm\Gamma_{01} \wedge \bm \Gamma_{02} - 
    \bm\Gamma_{13} \wedge \bm \Gamma_{23} \cr
    \bm \Theta_{13}= d \bm\Gamma_{13}+ \bm\Gamma_{01} \wedge \bm \Gamma_{03} + 
    \bm\Gamma_{12} \wedge \bm \Gamma_{23} \cr
    \bm \Theta_{23}= d \bm\Gamma_{23}+ \bm\Gamma_{02} \wedge \bm \Gamma_{03} - 
    \bm\Gamma_{12} \wedge \bm \Gamma_{13}. 
 \end{cases}
\end{equation}
Now let us see an example of Cartan's formalism applied to General Relativity.

\subsection{Example of Schwarzschild spacetime}
When the non-null components of metric tensor $\bm g$ are only diagonal components, it is practical to work with the non-coordinate orthonormal basis, as the case may be Schwarzschild spacetime.
In coordinate basis, the metric of the Schwarzschild spacetime is given by
\begin{equation}
 \label{Schwarzschild}
 ds^2 = -\left(1-\frac{2M}{r}\right)dt^2+ \left(\frac{1}{1-\frac{2M}{r}}\right)dr^2+
 r^2(d\theta^2+\sin^2\theta d \phi),
\end{equation}
where the parameters run over the range $r>2M$, 
$0\leq\theta<\pi$ and  $0\leq\phi<2\pi$. The metric of the Schwarzschild spacetime is already in diagonal form, thus it is easy to rewrite it in 
non-coordinate orthonormal basis in accordance with (\ref{tensor_metrico_ortonormal}) 
\begin{equation}
ds^2 = \eta_{\alpha\beta} \tilde{\bm\theta}^{\alpha}\otimes \tilde{\bm\theta}^{\beta}
=-\tilde{\bm\theta}^{0}\otimes \tilde{\bm\theta}^{0}+\tilde{\bm\theta}^{1}\otimes \tilde{\bm\theta}^{1}
+\tilde{\bm\theta}^{2}\otimes \tilde{\bm\theta}^{2}+\tilde{\bm\theta}^{3}\otimes \tilde{\bm\theta}^{3}. \nonumber
\end{equation}
From (\ref{base_NC2}) we have that
\begin{equation}
\label{1-forms-ortonormal}
 \tilde{\bm\theta}^{0} = {\omega^{0}}_{\mu} dx^{\mu}, \hspace*{1cm}
 \tilde{\bm\theta}^{1} = {\omega^{1}}_{\mu} dx^{\mu},  \hspace*{1cm}
  \tilde{\bm\theta}^{2} = {\omega^{2}}_{\mu} dx^{\mu}  \hspace*{1cm} \mbox{and}  \hspace*{1cm}
   \tilde{\bm\theta}^{3} = {\omega^{3}}_{\mu} dx^{\mu}.
\end{equation}
and from (\ref{vielbeins}) where $
 g_{\mu\nu} = {\omega^{\alpha}}_{\mu}{\omega^{\beta}}_{\nu}\eta_{\alpha\beta} $,
it follows that
\begin{eqnarray}
 g_{\mu\nu} &=& {\omega^{0}}_{\mu}{\omega^{0}}_{\nu}\eta_{00}+ {\omega^{1}}_{\mu}{\omega^{1}}_{\nu}\eta_{11}
 + {\omega^{2}}_{\mu}{\omega^{2}}_{\nu}\eta_{22}+ {\omega^{3}}_{\mu}{\omega^{3}}_{\nu}\eta_{33}\cr
 &=& - {\omega^{0}}_{\mu}{\omega^{0}}_{\nu}+ {\omega^{1}}_{\mu}{\omega^{1}}_{\nu}
 + {\omega^{2}}_{\mu}{\omega^{2}}_{\nu}+ {\omega^{3}}_{\mu}{\omega^{3}}_{\nu}, \nonumber
\end{eqnarray}
the diagonal terms of $g_{\mu\nu}$
\begin{equation}
 g_{tt} =  ({\omega^{0}}_{t})^2, \hspace*{1cm}  g_{rr} =  ({\omega^{1}}_{r})^2,\hspace*{1cm} 
 g_{\theta\theta} =  ({\omega^{2}}_{\theta})^2 \hspace*{1cm} \mbox{and}  \hspace*{1cm} g_{\phi\phi} =  ({\omega^{3}}_{\phi})^2\nonumber
\end{equation}
where the vierbein fields are interpreted as ‘square root’ of metric and it results in
\begin{equation}
\label{vierbein_fields_1}
 {\omega^{0}}_{t} = \sqrt{1-\frac{2M}{r}}, \hspace*{1cm} {\omega^{1}}_{r} = \frac{1}{\sqrt{1-\frac{2M}{r}}}, 
 \hspace*{1cm} {\omega^{2}}_{\theta} = r \hspace*{1cm} \mbox{and}  \hspace*{1cm} {\omega^{3}}_{\phi} = r\sin\theta.
\end{equation}
Thus the 1-form terms from (\ref{1-forms-ortonormal}) become
\begin{equation}
\label{1-forms-ortonormal_2}
 \tilde{\bm\theta}^{0} = dt \sqrt{1-\frac{2M}{r}}, \hspace*{1cm}
 \tilde{\bm\theta}^{1} =\frac{dr}{\sqrt{1-\frac{2M}{r}}} ,  \hspace*{1cm}
  \tilde{\bm\theta}^{2} =r d\theta  \hspace*{1cm} \mbox{and}  \hspace*{1cm}
   \tilde{\bm\theta}^{3} = r\sin\theta d\phi.
\end{equation}
The inverse terms are
\begin{equation}
\label{1-forms-ortonormal_3}
 dt = \frac{\tilde{\bm\theta}^{0}}{\sqrt{1-\frac{2M}{r}}}, \hspace*{1cm}
 dr = \tilde{\bm\theta}^{1}{\sqrt{1-\frac{2M}{r}}},  \hspace*{1cm}
  d\theta =\frac{\tilde{\bm\theta}^{2}}{r}   \hspace*{1cm} \mbox{and}  \hspace*{1cm}
  d\phi =\frac{\tilde{\bm\theta}^{3}}{r\sin\theta} .
\end{equation}
With (\ref{1-forms-ortonormal_2}) and (\ref{1-forms-ortonormal_3}) we can calculate 2-form $d\tilde{\bm\theta}^{\alpha}$,
\begin{itemize}
 \item   $d\tilde{\bm\theta}^{0}$ results in
\begin{equation}
 d\tilde{\bm\theta}^{0}= \frac{M/r^2}{\sqrt{1-\frac{2M}{r}}} \left(dr\wedge dt\right) =  \frac{M/r^2}{\sqrt{1-\frac{2M}{r}}} \,
\tilde{\bm\theta}^{1} \wedge \tilde{\bm\theta}^{0}, \nonumber
\end{equation}
\item  $d\tilde{\bm\theta}^{1}$ results in
\begin{equation}
 d\tilde{\bm\theta}^{1}= \frac{-M/r^2}{\left({1-\frac{2M}{r}}\right)^{3/2}} \left(dr\wedge dr\right) =  0, \nonumber
\end{equation}
\item  $d\tilde{\bm\theta}^{2}$ results in
\begin{equation}
 d\tilde{\bm\theta}^{2} = dr\wedge d\theta = \frac{\sqrt{1-\frac{2M}{r}}}{r}\, \tilde{\bm\theta}^{1} \wedge \tilde{\bm\theta}^{2}, \nonumber
\end{equation}
\item  $d\tilde{\bm\theta}^{3}$ results in
\begin{equation}
  d\tilde{\bm\theta}^{3} = \sin\theta dr\wedge d\phi+ r\cos\theta d\theta \wedge d\phi = \frac{\sqrt{1-\frac{2M}{r}}}{r}\, \tilde{\bm\theta}^{1}
  \wedge \tilde{\bm\theta}^{3} + \frac{\cot\theta}{r} \, \tilde{\bm\theta}^{2}   \wedge \tilde{\bm\theta}^{3}.\nonumber
\end{equation}
\end{itemize}
With these results we can solve the first structure equation of Cartan as the system of equations (\ref{equacao_de_Cartan_ortonormal}), 
\begin{eqnarray}
   -\frac{M/r^2}{\sqrt{1-\frac{2M}{r}}} \,
\tilde{\bm\theta}^{1} \wedge \tilde{\bm\theta}^{0} &=& {\bm\Gamma^{0}}_{1} \wedge \tilde{\bm\theta}^{1}, \cr
   0 &=& {\bm\Gamma^{0}}_{1} \wedge \tilde{\bm\theta}^{0}+
  {\bm\Gamma^{1}}_{2} \wedge \tilde{\bm\theta}^{2}+ {\bm\Gamma^{1}}_{3} \wedge \tilde{\bm\theta}^{3}, \cr 
-\frac{\sqrt{1-\frac{2M}{r}}}{r}\, \tilde{\bm\theta}^{1} \wedge \tilde{\bm\theta}^{2} &=&  -{\bm\Gamma^{1}}_{2} \wedge \tilde{\bm\theta}^{1},\cr 
  -\frac{\sqrt{1-\frac{2M}{r}}}{r}\, \tilde{\bm\theta}^{1}
  \wedge \tilde{\bm\theta}^{3} - \frac{\cot\theta}{r} \, \tilde{\bm\theta}^{2}   \wedge \tilde{\bm\theta}^{3} &=&  - {\bm\Gamma^{1}}_{3} \wedge \tilde{\bm\theta}^{1}- {\bm\Gamma^{2}}_{3} \wedge \tilde{\bm\theta}^{2}. \nonumber
\end{eqnarray}
Thus, we have four independent connection 1-forms
\begin{eqnarray}
 {\bm\Gamma^{0}}_{1} &=& \frac{M}{r^2\,\sqrt{1-\frac{2M}{r}}} \,\tilde{\bm\theta}^{0} \,\,=\,\,  -{\bm\Gamma}_{01},\cr
  {\bm\Gamma^{1}}_{2} &=& -\frac{\sqrt{1-\frac{2M}{r}}}{r} \,\tilde{\bm\theta}^{2} \,\,=\,\,  {\bm\Gamma}_{12},\cr
  {\bm\Gamma^{1}}_{3} &=& -\frac{\sqrt{1-\frac{2M}{r}}}{r} \,\tilde{\bm\theta}^{3} \,\,=\,\,  {\bm\Gamma}_{13},\cr
  {\bm\Gamma^{2}}_{3} &=& -\frac{\cot\theta}{r} \,\tilde{\bm\theta}^{3} \,\,=\,\,  {\bm\Gamma}_{23}, \nonumber
\end{eqnarray}
with ${\bm\Gamma^{0}}_{2} = {\bm\Gamma^{0}}_{3} =0$. Now we can calculate $d{\bm\Gamma^{\alpha}}_{\beta}$ with objective to solve the second structure equation of Cartan. The calculation of 2-form $d{\bm\Gamma^{0}}_{1}$ follows that
\begin{eqnarray}
 d{\bm\Gamma^{0}}_{1} &=& \left[-\frac{2M}{r^3 \sqrt{1-\frac{2M}{r}}} -\frac{M\cdot 2M/r^2}{2r^2\left({1-\frac{2M}{r}}\right)^{3/2}}\right]dr\wedge \tilde{\bm\theta}^{0} + \frac{M}{r^2\,\sqrt{1-\frac{2M}{r}}}  \,d\tilde{\bm\theta}^{0} \cr
    &=& -\frac{M}{r^3 \sqrt{1-\frac{2M}{r}}}\left[ \frac{M}{r\left({1-\frac{2M}{r}}\right)} + 2 \right]  \sqrt{1-\frac{2M}{r}}\,\, \tilde{\bm\theta}^{1} \wedge \tilde{\bm\theta}^{0} +  \frac{M}{r^2\,\sqrt{1-\frac{2M}{r}}} \cdot  \frac{M}{r^2\,\sqrt{1-\frac{2M}{r}}} 
    \tilde{\bm\theta}^{1} \wedge \tilde{\bm\theta}^{0}\cr
    &=& \left[-\frac{M}{r^3} \left(\frac{M}{r-2M} + 2\right) + \frac{M^2}{r^4\left(1-\frac{2M}{r}\right)}\right]  \tilde{\bm\theta}^{1} \wedge \tilde{\bm\theta}^{0}\cr
    &=& \frac{2M}{r^3} \tilde{\bm\theta}^{0} \wedge \tilde{\bm\theta}^{1} \nonumber
\end{eqnarray}
or 
\begin{equation}
 d{\bm\Gamma}_{01} = - \frac{2M}{r^3} \tilde{\bm\theta}^{0} \wedge \tilde{\bm\theta}^{1}
\end{equation} 
The calculation of 2-form $d{\bm\Gamma^{1}}_{2}$ follows that
\begin{eqnarray}
 d{\bm\Gamma^{1}}_{2} &=& -d\left(\frac{ \sqrt{1-\frac{2M}{r}}}{r} \right) \wedge \tilde{\bm\theta}^{2} - \frac{ \sqrt{1-\frac{2M}{r}}}{r} d\tilde{\bm\theta}^{2} \cr
  &=&\left({ \sqrt{1-\frac{2M}{r}}}{r^2} -\frac{2M/r^2}{2r \sqrt{1-\frac{2M}{r}}} \right)dr\wedge  \tilde{\bm\theta}^{2} - \frac{ \sqrt{1-\frac{2M}{r}}}{r} \cdot \frac{ \sqrt{1-\frac{2M}{r}}}{r} \tilde{\bm\theta}^{1} \wedge \tilde{\bm\theta}^{2} \cr
  &=&\left[\left(\frac{ \sqrt{1-\frac{2M}{r}}}{r^2} -\frac{M}{r^3 \sqrt{1-\frac{2M}{r}}} \right) \sqrt{1-\frac{2M}{r}} - \frac{1-\frac{2M}{r}}{r^2}\right] \tilde{\bm\theta}^{1} \wedge \tilde{\bm\theta}^{2} \cr
  &=& -\frac{M}{r^3}\tilde{\bm\theta}^{1} \wedge \tilde{\bm\theta}^{2} \nonumber
\end{eqnarray}
or
\begin{equation}
 d{\bm\Gamma}_{12} = - \frac{M}{r^3} \tilde{\bm\theta}^{1} \wedge \tilde{\bm\theta}^{2}.
\end{equation}
The calculation of 2-form $d{\bm\Gamma^{1}}_{3}$ follows that
\begin{eqnarray}
 d{\bm\Gamma^{1}}_{3} &=& \left(\frac{{1-\frac{2M}{r}}}{r^2} -\frac{M}{r^3} \right) \tilde{\bm\theta}^{1} \wedge \tilde{\bm\theta}^{3} 
 - \frac{\sqrt{1-\frac{2M}{r}}}{r}d\tilde{\bm\theta}^{3}  \cr
  &=& \left(\frac{{1-\frac{2M}{r}}}{r^2} -\frac{M}{r^3} \right) \tilde{\bm\theta}^{1} \wedge \tilde{\bm\theta}^{3} - \frac{\sqrt{1-\frac{2M}{r}}}{r} \left( \frac{\sqrt{1-\frac{2M}{r}}}{r}\, \tilde{\bm\theta}^{1}
  \wedge \tilde{\bm\theta}^{3} + \frac{\cot\theta}{r} \, \tilde{\bm\theta}^{2}   \wedge \tilde{\bm\theta}^{3} \right), \nonumber
\end{eqnarray}
where it reduces to 
\begin{equation}
 d{\bm\Gamma}_{13} = - \frac{M}{r^3} \tilde{\bm\theta}^{1} \wedge \tilde{\bm\theta}^{3} - \frac{\sqrt{1-\frac{2M}{r}}}{r^2}\cot\theta\,\, \tilde{\bm\theta}^{2} \wedge \tilde{\bm\theta}^{3} .
\end{equation}
The calculation of 2-form $d{\bm\Gamma^{2}}_{3}$ follows that
\begin{eqnarray}
 d{\bm\Gamma^{2}}_{3} &=& -d\left(\frac{ \cot\theta}{r} \right) \wedge \tilde{\bm\theta}^{3} - \frac{ \cot\theta}{r} d\tilde{\bm\theta}^{3},\nonumber
\end{eqnarray}
where it results is 
\begin{equation}
 d{\bm\Gamma}_{23} = - \frac{1}{r^2} \tilde{\bm\theta}^{2} \wedge \tilde{\bm\theta}^{3}.
\end{equation}
The curvature 2-forms are founf from the second structure Cartan equation (\ref{sistema_2a_equacao_de_Cartan}). 
\begin{eqnarray}
    \bm \Theta_{01}&=& d \bm\Gamma_{01} - \bm\Gamma_{02} \wedge \bm \Gamma_{12} - 
    \bm\Gamma_{03} \wedge \bm \Gamma_{13} = - \frac{2M}{r^3} \,\,\tilde{\bm\theta}^{0} \wedge \tilde{\bm\theta}^{1}, \cr
    \bm \Theta_{02}&=& d \bm\Gamma_{02}+ \bm\Gamma_{01} \wedge \bm \Gamma_{12} - 
    \bm\Gamma_{03} \wedge \bm \Gamma_{23} = \frac{M}{r^3}\,\, \tilde{\bm\theta}^{0} \wedge \tilde{\bm\theta}^{2},\cr
     \bm \Theta_{03}&=& d \bm\Gamma_{03}+ \bm\Gamma_{01} \wedge \bm \Gamma_{13} + 
    \bm\Gamma_{02} \wedge \bm \Gamma_{23} =\frac{M}{r^3}\,\, \tilde{\bm\theta}^{0} \wedge \tilde{\bm\theta}^{3}, \cr
     \bm \Theta_{12}&=& d \bm\Gamma_{12}+ \bm\Gamma_{01} \wedge \bm \Gamma_{02} - 
    \bm\Gamma_{13} \wedge \bm \Gamma_{23} = - \frac{M}{r^3}\,\, \tilde{\bm\theta}^{1} \wedge \tilde{\bm\theta}^{2},  \cr
    \bm \Theta_{13}&=& d \bm\Gamma_{13}+ \bm\Gamma_{01} \wedge \bm \Gamma_{03} + 
    \bm\Gamma_{12} \wedge \bm \Gamma_{23} = - \frac{M}{r^3} \,\,\tilde{\bm\theta}^{1} \wedge \tilde{\bm\theta}^{3},\cr
    \bm \Theta_{23}&=& d \bm\Gamma_{23}+ \bm\Gamma_{02} \wedge \bm \Gamma_{03} - 
    \bm\Gamma_{12} \wedge \bm \Gamma_{13} = \frac{2M}{r^3} \,\,\tilde{\bm\theta}^{2} \wedge \tilde{\bm\theta}^{3} . \nonumber
\end{eqnarray}
We have introduced the curvature two-form given by equation (\ref{curvatura_2forma}),
\begin{equation}
 \bm \Theta_{\alpha\beta}=\frac{1}{2}  R_{\alpha\beta\gamma\delta}\tilde{\bm\theta}^{\gamma}
\wedge \tilde{\bm\theta}^{\delta}. \nonumber
\end{equation}
Thus we can obtain the Riemann curvature tensor in non-coordinate orthonormal basis. For 
\begin{equation}
 \bm \Theta_{01} = - \frac{2M}{r^3} \,\,\tilde{\bm\theta}^{0} \wedge \tilde{\bm\theta}^{1}, \nonumber
\end{equation}
we have
\begin{eqnarray}
\frac{1}{2}  R_{0101}\,\,\tilde{\bm\theta}^{0}
\wedge \tilde{\bm\theta}^{1} + \frac{1}{2}  R_{0110}\,\,\tilde{\bm\theta}^{1}
\wedge \tilde{\bm\theta}^{0}  =   - \frac{2M}{r^3}\,\,\tilde{\bm\theta}^{0} \wedge \tilde{\bm\theta}^{1}, \nonumber
\end{eqnarray}
that results in
\begin{equation}
 R_{0101} = - \frac{2M}{r^3}. \nonumber
\end{equation}
In the same way we have for others components of Riemann tensor,
\begin{eqnarray}
 R_{0202} &=&  \frac{M}{r^3}, \cr
 R_{0303} &=&  \frac{M}{r^3}, \cr
 R_{1212} &=& - \frac{M}{r^3}, \cr
 R_{1313} &=& - \frac{M}{r^3}, \cr
 R_{2323} &=& \frac{2M}{r^3}. \nonumber
\end{eqnarray}
Now we can compute the components of Ricci tensor
\begin{equation}
 R_{\beta\delta} = {R^{\gamma}}_{\beta\gamma\delta} = \eta^{\alpha\gamma}R_{\alpha\beta\gamma\delta}.\nonumber
\end{equation}
The components of above Riemann tensor show that we have readily $R_{\beta\delta} =0$ with $\beta \neq \delta$, however   it is possible to compute $R_{00}$, $R_{11}$, $R_{22}$ and $R_{33}$, but they are null as it follows
\begin{equation}
 R_{00}= {R^{1}}_{010}+{R^{2}}_{020}+{R^{3}}_{030}= -\frac{2M}{r^3}+\frac{M}{r^3}+\frac{M}{r^3}=0, \nonumber
\end{equation}
\begin{equation}
 R_{11}= {R^{0}}_{101}+{R^{2}}_{121}+{R^{3}}_{131}= +\frac{2M}{r^3}-\frac{M}{r^3}-\frac{M}{r^3}=0, \nonumber
\end{equation}
\begin{equation}
 R_{22}= {R^{0}}_{202}+{R^{1}}_{212}+{R^{3}}_{232}= -\frac{M}{r^3}-\frac{M}{r^3}+\frac{2M}{r^3}=0, \nonumber
\end{equation}
\begin{equation}
 R_{33}= {R^{0}}_{303}+{R^{1}}_{313}+{R^{2}}_{323}= -\frac{M}{r^3}-\frac{M}{r^3}+\frac{2M}{r^3}=0. \nonumber
\end{equation}
Thus $R_{\alpha\beta}=0$ and consequently we have that the Einstein tensor $G_{\alpha\beta}=R_{\alpha\beta}-\frac{1}{2}g_{\alpha\beta}R =0$, that is a vacuum solution.

In coordinate basis or world indices we can obtain that
\begin{equation}
 R_{\mu\nu} = {\omega^{\alpha}}_{\mu}  {\omega^{\beta}}_{\nu}\,R_{\alpha\beta} 
\end{equation}
where clearly $R_{\mu\nu}=0$. However we can se that the non null components   of Riemann tensor in coordinate basis are obtained from
\begin{equation}
 R_{\mu\nu\rho\sigma} = {\omega^{\alpha}}_{\mu}  {\omega^{\beta}}_{\nu} {\omega^{\gamma}}_{\rho}{\omega^{\delta}}_{\sigma}\,R_{\alpha\beta\rho\sigma}, 
\end{equation}
where with aid of (\ref{vierbein_fields_1}). It follows that
\begin{equation}
 R_{trtr} =  {\omega^{0}}_{t}  {\omega^{1}}_{r} {\omega^{0}}_{t}{\omega^{1}}_{r}\,R_{0101} = \left( {\omega^{0}}_{t}\right)^2 \left({\omega^{1}}_{r}\right)^2 \left(- \frac{2M}{r^3} \right) = - \frac{2M}{r^3}, \nonumber
\end{equation}
\begin{equation}
 R_{t\theta t \theta} =  {\omega^{0}}_{t}  {\omega^{2}}_{\theta} {\omega^{0}}_{t}{\omega^{2}}_{\theta}\,R_{0202} = \left( {\omega^{0}}_{t}\right)^2 \left({\omega^{2}}_{\theta}\right)^2 \left(\frac{M}{r^3} \right) = \left(1-\frac{2M}{r}\right)r^2 \frac{M}{r^3} =  \frac{M}{r}\left(1-\frac{2M}{r}\right), \nonumber
\end{equation}
\begin{equation}
 R_{t\phi t \phi} =  {\omega^{0}}_{t}  {\omega^{3}}_{\phi} {\omega^{0}}_{t}{\omega^{3}}_{\phi}\,R_{0303} = \left( {\omega^{0}}_{t}\right)^2 \left({\omega^{2}}_{\phi}\right)^2 \left(\frac{M}{r^3} \right) = \left(1-\frac{2M}{r}\right)\left(r^2\sin^2\theta\right) \frac{M}{r^3} =  \frac{M}{r}\left(1-\frac{2M}{r}\right)\sin^2\theta, \nonumber
\end{equation}
\begin{equation}
 R_{r\theta r \theta} =  {\omega^{1}}_{r}  {\omega^{2}}_{\theta} {\omega^{1}}_{r}{\omega^{2}}_{\theta}\,R_{1212} = \left( {\omega^{1}}_{r}\right)^2 \left({\omega^{2}}_{\theta}\right)^2 \left(\frac{-M}{r^3} \right) = -\left(\frac{r^2}{1-\frac{2M}{r}}\right) \frac{M}{r^3} =  -\frac{M}{r}\left(1-\frac{2M}{r}\right)^{-1}, \nonumber
\end{equation}
\begin{equation}
 R_{r\phi r \phi} =  {\omega^{1}}_{r}  {\omega^{3}}_{\phi} {\omega^{1}}_{r}{\omega^{3}}_{\phi}\,R_{1313} = \left( {\omega^{1}}_{r}\right)^2 \left({\omega^{3}}_{\phi}\right)^2 \left(\frac{-M}{r^3} \right) = -\left(\frac{r^2\sin^2\theta}{1-\frac{2M}{r}}\right) \frac{M}{r^3} =  -\frac{M}{r}\left(1-\frac{2M}{r}\right)^{-1}\sin^2\theta, \nonumber
\end{equation}
\begin{equation}
 R_{\theta\phi \theta \phi} =  {\omega^{2}}_{\theta}  {\omega^{3}}_{\phi} {\omega^{2}}_{\theta}{\omega^{3}}_{\phi}\,R_{2323} = \left( {\omega^{2}}_{\theta}\right)^2 \left({\omega^{3}}_{\phi}\right)^2 \left(\frac{2M}{r^3} \right) = \left({r^4\sin^2\theta}\right) \frac{2M}{r^3} =  2Mr \sin^2\theta.\nonumber
\end{equation}

After some practices, we can see that the non-coordinate orthonormal basis to advance in an easy method to calculate the Riemann tensor of spacetimes that have the diagonal shape in the metric tensor $\bm g$ in coordinate basis.


\section{Newman-Penrose null tetrad - pseudo-orthonormal basis}

A pseudo-orthonormal basis or null tetrad formalism is due to Newman and Penrose \cite{Newman}. It has proved very useful in the construction of exact solutions on General Relativity. This formalism is adapted to treatment of the propagation of radiation in spacetime as will be seen at the end of this section with the example of Brinkmann metric. 

We can start with metric of Minkowski spacetime in orthonormal basis where the line element is
\begin{equation}
 ds^2= -dt^2+dx^2+ dy^2 + dz^2.\nonumber
\end{equation}
An alternative coordinate system for Minkowski spacetime can be through the introduction of advanced and retarded null coordinates, $u$ and $v$ with the following coordinate transformations
\begin{equation}
\label{relacao_pseudo_ortogonal}
 u=\frac{1}{\sqrt{2}}\left(t-x\right) \hspace{1cm}\mbox{and} \hspace{1cm} v=\frac{1}{\sqrt{2}}\left(t+x\right), 
 \end{equation}
so that the line element of Minkowski spacetime can be
\begin{equation}
 ds^2=  -2dudv + dy^2 + dz^2.\nonumber
\end{equation}
With these null coordinates, we can see that if $u$ is constant, we have $du=0$, then $ dx = dt $ or $\frac{dx}{dt}=1$, the speed of  light. 
The null coordinate $u$ may be thought of as light ray  at the speed of light in direction $+x$, while the null coordinate $v$ may be thought of as light ray  at the speed of light in direction $-x$. These null coordinates, $v$ and $u$ are  orthogonal to the $yz$-plane. In the similar way, it is convenient to parametrise the $yz$-plane in terms of the complex coordinate
\begin{equation}
\label{relacao_pseudo_ortogonal_2}
 \zeta=\frac{1}{\sqrt{2}}\left(y+iz\right),\hspace{1cm} 
  \bar\zeta=\frac{1}{\sqrt{2}}\left(y-iz\right),
\end{equation}
thus we can see that in terms of these coordinate transformations the line element $ds^2$ of Minkowski spacetime is
\begin{equation}
\label{metrica_pseudo-ortonormal_1}
 ds^2 = -2dudv +2 d\zeta d\bar\zeta.
\end{equation}
Now we can express the 1-form terms of $ds^2$ as
\begin{equation}
 \tilde{\bm\theta}^0 = dv ,\hspace{1cm} 
  \tilde{\bm\theta}^1 = du ,\hspace{1cm} 
  \tilde{\bm\theta}^2 = d\zeta ,\hspace{1cm} 
   \tilde{\bm\theta}^3 = d\bar\zeta, \nonumber
\end{equation}
where we have the metric tensor $\bm g$ in pseudo-orthonormal coordinate basis given by
\begin{equation}
\label{metrica_pseudo-ortonormal_2}
  \mbox{\bf g} =  -2\tilde{\bm\theta}^0 \otimes \tilde{\bm\theta}^1 + 
  2\tilde{\bm\theta}^2\otimes \tilde{\bm\theta}^3 
\end{equation}
that in matrix form is
\begin{equation}
  \mbox{\bf g} = \begin{pmatrix}
                   \tilde{\bm\theta}^0 & \tilde{\bm\theta}^1 & \tilde{\bm\theta}^2 & \tilde{\bm\theta}^3
                 \end{pmatrix}
                 \begin{pmatrix}
                  0 & -1 & 0 & 0 \cr
                  -1 & 0 & 0 & 0\cr
                  0 & 0 & 0 & 1\cr
                  0 & 0 & 1 & 0
                 \end{pmatrix}
 \begin{pmatrix}
    \tilde{\bm\theta}^0 \cr \tilde{\bm\theta}^1 \cr \tilde{\bm\theta}^2 \cr \tilde{\bm\theta}^3
                 \end{pmatrix}\nonumber
\end{equation}
or
\begin{equation}
  \mbox{\bf g} = \gamma_{\alpha\beta} \tilde{\bm\theta}^{\alpha}\otimes \tilde{\bm\theta}^{\beta}
\end{equation}
since
\begin{equation}
\label{pseudo_ortonormal_metrica}
 (\gamma_{\alpha\beta}) =   \begin{pmatrix}
                  0 & -1 & 0 & 0 \cr
                  -1 & 0 & 0 & 0\cr
                  0 & 0 & 0 & 1\cr
                  0 & 0 & 1 & 0
                 \end{pmatrix}
\end{equation}
$\gamma_{\alpha\beta}$ is defined the components of metric tensor with respect to the non-coordinate pseudo-orthonormal basis.

Again, from a tangent space $\bm T_p$ spanned by $\{\bm E_{\mu}\}$ in coordinate basis, we can obtain the non-coordinate pseudo-orthonormal basis by a rotation. From equation (\ref{base_NC1}) we have
\begin{equation}
{\hat{\bm e}}_{\alpha}={e_{\alpha}}^{\mu}\bm E_{\mu}, \nonumber
\end{equation} 
Thus we have a tetrad,
\begin{eqnarray}
\begin{cases}
 {\hat{\bm e}}_{0}={e_{0}}^{\mu}\bm E_{\mu} = k^{\mu} \bm E_{\mu} = \bm{k},\cr
  {\hat{\bm e}}_{1}={e_{1}}^{\mu}\bm E_{\mu} = l^{\mu} \bm E_{\mu} = \bm{l},\cr
   {\hat{\bm e}}_{2}={e_{2}}^{\mu}\bm E_{\mu} = m^{\mu} \bm E_{\mu} = \bm{m},\cr
    {\hat{\bm e}}_{3}={e_{3}}^{\mu}\bm E_{\mu} = \bar{m}^{\mu} \bm E_{\mu} = \bar{\bm{m}}.
    \end{cases}
\end{eqnarray}
The above set of complex null vectors $\{{\hat{\bm e}}_{\alpha}\} = \{\bm{k},\bm{l},\bm{m},\bar{\bm{m}}\}$ is called a Newman-Penrose null tetrad. This is usually abbreviated to NP null tetrad \cite{Stewart}.
The reason for it designation is because the condition (\ref{vielbeins0}), where we have,
\begin{equation}
 {e_{\alpha}}^{\mu}{e_{\beta}}^{\nu}g_{\mu\nu}=\gamma_{\alpha\beta}. \nonumber
\end{equation}
It is straightforward to notice that the diagonal terms of metric tensor $\gamma_{\alpha\beta}$ from (\ref{pseudo_ortonormal_metrica}) are zero. For Example
\begin{equation}
 {e_{0}}^{\mu}{e_{0}}^{\nu}g_{\mu\nu}=\gamma_{00} = 0, \nonumber
\end{equation}
where
\begin{equation}
 k^{\mu}k^{\nu}g_{\mu\nu}= k^{\mu}k_{\mu} = 0. \nonumber
\end{equation}
And the same way the other three contractions are null,
\begin{equation}
 l^{\mu}l_{\mu} = 0, \hspace{1cm} m^{\mu}m_{\mu} = 0 \hspace{1cm} \mbox{and} \hspace{1cm} \bar{m}^{\mu}\bar{m}_{\mu} = 0 \nonumber
\end{equation}
The non-null contractions are given by
\begin{equation}
 {e_{0}}^{\mu}{e_{1}}^{\nu}g_{\mu\nu}=\gamma_{01}, \nonumber
\end{equation}
that results in $k^{\mu} l_{\mu}=-1$ and
\begin{equation}
 {e_{2}}^{\mu}{e_{3}}^{\nu}g_{\mu\nu}=\gamma_{23}, \nonumber
\end{equation}
that results $ m^{\mu}\bar{m}_{\mu} = 1$.

The relationships (\ref{relacao_pseudo_ortogonal}) and (\ref{relacao_pseudo_ortogonal_2}), in flat spacetime implies that,
\begin{equation}
 \bm k = \frac{\partial }{\partial v}, \hspace{1cm}\bm l = \frac{\partial }{\partial u}, \hspace{1cm}
 \bm m = \frac{\partial }{\partial \zeta} \hspace{1cm} \mbox{and} \hspace{1cm} \bar{\bm m}  = \frac{\partial }{\partial \bar\zeta},
\end{equation}
where we see that $\bm m$ and $\bar{\bm m} $ are complex conjugates. In coordinate basis we can see from (\ref{relacao_pseudo_ortogonal}) and (\ref{relacao_pseudo_ortogonal_2}) that,
\begin{equation}
\label{tetrad_NP}
 k^{\mu} =\frac{1}{\sqrt{2}}\begin{pmatrix}
                             1\cr
                             1\cr
                             0\cr
                             0
                            \end{pmatrix}, \hspace{1cm}
 l^{\mu} =\frac{1}{\sqrt{2}}\begin{pmatrix}
                             1\cr
                             -1\cr
                             0\cr
                             0
                            \end{pmatrix}, \hspace{1cm}
 m^{\mu} =\frac{1}{\sqrt{2}}\begin{pmatrix}
                             0\cr
                             0\cr
                             1\cr
                             i
                            \end{pmatrix}, \hspace{1cm}    \mbox{and} \hspace{1cm}
 \bar{m}^{\mu} =\frac{1}{\sqrt{2}}\begin{pmatrix}
                             0\cr
                             0\cr
                             1\cr
                             -i
                            \end{pmatrix}, \hspace{1cm}                                                   
\end{equation}

From equation (\ref{id_n_coordenada2})
\begin{equation}
   \langle {\hat{\bm e}}_{\alpha},\tilde{\bm\theta}^{\beta}\rangle=
   {e_{\alpha}}^{\mu} {\omega^{\beta}}_{\mu} = {\delta_{\alpha}}^{\beta},\nonumber
\end{equation}
we can identify that
\begin{equation}
   \langle {\hat{\bm e}}_{0},\tilde{\bm\theta}^{0}\rangle=
   {e_{0}}^{\mu} {\omega^{0}}_{\mu} = k^{\mu}{\omega^{0}}_{\mu} =1, \nonumber
\end{equation}
where from condition  $k^{\mu} l_{\mu}=-1$, the above equation results for 
${\omega^{0}}_{\mu} = - l_{\mu}$.
In the same way we have
\begin{equation}
   {e_{1}}^{\mu} {\omega^{1}}_{\mu} = l^{\mu}{\omega^{1}}_{\mu} =1, \nonumber
\end{equation}
that from $k_{\mu} l^{\mu}=-1$, it results in
${\omega^{1}}_{\mu} = - k_{\mu}$. For the others terms we have
\begin{equation}
   {e_{2}}^{\mu} {\omega^{2}}_{\mu} = m^{\mu}{\omega^{2}}_{\mu} =1 
  \hspace{1cm} \Rightarrow  \hspace{1cm}  {\omega^{2}}_{\mu} = \bar{m}^{\mu}, \nonumber
\end{equation}
and
\begin{equation}
   {e_{3}}^{\mu} {\omega^{3}}_{\mu} = \bar{m}^{\mu}{\omega^{3}}_{\mu} =1 
  \hspace{1cm} \Rightarrow  \hspace{1cm}  {\omega^{3}}_{\mu} = m^{\mu}. \nonumber
\end{equation}
So, as $\tilde{\bm\theta}^{\alpha} = {\omega^{\alpha}}_{\mu} dx^{\mu}$, we have that the differential 1-forms, the basis for cotangent space in pseudo-orthonormal coordinate, are
\begin{equation}
\label{vielbeins_pseudo-ortonormal}
 \begin{cases}
 \tilde{\bm\theta}^{0} = {\omega^{0}}_{\mu} dx^{\mu} = -l_{\mu}dx^{\mu}\cr
 \tilde{\bm\theta}^{1} = {\omega^{1}}_{\mu} dx^{\mu} = -k_{\mu}dx^{\mu}\cr
  \tilde{\bm\theta}^{2} = {\omega^{2}}_{\mu} dx^{\mu} = \bar{m}_{\mu}dx^{\mu}\cr
   \tilde{\bm\theta}^{3} = {\omega^{3}}_{\mu} dx^{\mu} = m_{\mu}dx^{\mu}
 \end{cases}
\end{equation}
In coordinate basis, from (\ref{tetrad_NP}) we have
\begin{equation}
\label{tetrad_NP_2}
 k_{\mu} =\frac{1}{\sqrt{2}}\begin{pmatrix}
                             -1\cr
                             1\cr
                             0\cr
                             0
                            \end{pmatrix}, \hspace{1cm}
 l_{\mu} =\frac{1}{\sqrt{2}}\begin{pmatrix}
                             -1\cr
                             -1\cr
                             0\cr
                             0
                            \end{pmatrix}, \hspace{1cm}
 m_{\mu} =\frac{1}{\sqrt{2}}\begin{pmatrix}
                             0\cr
                             0\cr
                             1\cr
                             i
                            \end{pmatrix}, \hspace{1cm}    \mbox{and} \hspace{1cm}
 \bar{m}_{\mu} =\frac{1}{\sqrt{2}}\begin{pmatrix}
                             0\cr
                             0\cr
                             1\cr
                             -i
                            \end{pmatrix}, \hspace{1cm}                                                   
\end{equation}
where we have the only nonzero contractions $k^{\mu} l_{\mu}=-1$ and $ m^{\mu}\bar{m}_{\mu} = 1$.

Now we can see the way to write the components of metric tensor in coordinate basis from the terms $l_{\mu}$, $k_{\mu}$, $m_{\mu}$ and  $\bar{m}_{\mu}$. Recall equation (\ref{vielbeins}),
\begin{equation}
 g_{\mu\nu} = {\omega^{\alpha}}_{\mu}{\omega^{\beta}}_{\nu}\gamma_{\alpha\beta}, \nonumber
\end{equation}
it results in
\begin{eqnarray}
 g_{\mu\nu} &=& {\omega^{0}}_{\mu}{\omega^{1}}_{\nu}\gamma_{01}+ {\omega^{1}}_{\mu}{\omega^{0}}_{\nu}\gamma_{10}
 + {\omega^{2}}_{\mu}{\omega^{3}}_{\nu}\gamma_{23}+ {\omega^{3}}_{\mu}{\omega^{2}}_{\nu}\gamma_{32}\cr
 &=&(-l_{\mu})(-k_{\nu})(-1) + (-k_{\mu})(-l_{\nu})(-1)+ (\bar{m}_{\mu})(m_{\nu})(1) + (m_{\mu})(\bar{m}_{\nu})(1)\cr
 &=& -( l_{\mu}k_{\nu}+  l_{\nu}k_{\mu})+ (\bar{m}_{\mu}m_{\nu}+\bar{m}_{\nu}m_{\mu}),\nonumber
\end{eqnarray}
where we can simplify the symmetrization of index pairs by round brackets
\begin{equation}
 g_{\mu\nu} = -2k_{(\mu}l_{\nu)} + 2 m_{(\mu}\bar{m}_{\nu)}. 
\end{equation}

\subsection{Ricci rotation coefficients}

We have seen that in orthonormal basis there are 24 Ricci rotation coefficients. But in pseudo-orthonormal basis, 
because $\bm m$ and $\bar{\bm m} $ are complex conjugates, the number of Ricci rotation coefficients reduces to 16 coefficients.
The Ricci rotation coefficients are given by
\begin{equation}
\Gamma_{\alpha\beta\gamma} = g(\hat{\bm e}_{\alpha}, \nabla_{\hat{\bm e}_{\beta}}  \hat{\bm e}_{\gamma}).\nonumber
\end{equation}
In terms of NP null tetrad we have for $\Gamma_{100}$ the following calculation 
\begin{equation}
 \Gamma_{100} = g(\hat{\bm e}_{1}, \nabla_{\hat{\bm e}_{0}}  \hat{\bm e}_{0})
 = g(l^{\mu}{\bm E}_{\mu}, \nabla_{(k^{\nu}{\bm E}_{\nu})}  k^{\rho}{\bm E}_{\rho}) =
 l^{\mu} k^{\nu} g({\bm E}_{\mu},  \nabla_{\nu} k^{\rho} {\bm E}_{\rho}) = l^{\mu} k^{\nu}  \nabla_{\nu} k^{\rho} g({\bm E}_{\mu}, {\bm E}_{\rho})
 = l^{\mu} k^{\nu}\nabla_{\nu} k^{\rho} g_{\mu\rho},\nonumber
\end{equation}
it follows that
\begin{equation}
 \Gamma_{100} =
  l^{\mu} k^{\nu}\nabla_{\nu} k_{\mu}.\nonumber
\end{equation}
Notice that $ \overline{\Gamma}_{100} =  \Gamma_{100}$ because $l^{\mu}$ and $k^{\mu}$ are real numbers.
Similarly to orthonormal basis, the pseudo-orthonormal basis is a rigid frame with $\partial_{\delta}\gamma_{\alpha\beta}=0$, so the Ricci rotation coefficients are given by (\ref{conexão_NC14}) and results in (\ref{rotacao_Ricci_1}), $\Gamma_{\gamma\alpha\beta} = -  \Gamma_{\beta\alpha\gamma}$, such as $\Gamma_{100} = -\Gamma_{001} = - k^{\mu} k^{\nu}\nabla_{\nu} l_{\mu}$.

The calculation for $ \Gamma_{200}$ is
\begin{equation}
 \Gamma_{200} = g(\hat{\bm e}_{2}, \nabla_{\hat{\bm e}_{0}}  \hat{\bm e}_{0})
 = g(m^{\mu}{\bm E}_{\mu}, \nabla_{(k^{\nu}{\bm E}_{\nu})}  k^{\rho}{\bm E}_{\rho})=
  m^{\mu} k^{\nu}\nabla_{\nu} k_{\mu}.\nonumber
\end{equation}
and for $\Gamma_{300} =  \bar{m}^{\mu} k^{\nu}\nabla_{\nu} k_{\mu}$, where we ca see that
\begin{equation}
 \overline{\Gamma}_{300} = m^{\mu} k^{\nu}\nabla_{\nu} k_{\mu} =  \Gamma_{200}. \nonumber
\end{equation}

The calculation for $ \Gamma_{101}$ is
\begin{equation}
 \Gamma_{101} = g(\hat{\bm e}_{1}, \nabla_{\hat{\bm e}_{0}}  \hat{\bm e}_{1})
 = g(l^{\mu}{\bm E}_{\mu}, \nabla_{(k^{\nu}{\bm E}_{\nu})}  l^{\rho}{\bm E}_{\rho})=
  l^{\mu} k^{\nu}\nabla_{\nu} l_{\mu}.\nonumber
\end{equation}
where we observe that $\overline{\Gamma}_{101} =  \Gamma_{101}$ . 

We have Ricci rotation coefficient $\Gamma_{102} =  l^{\mu} k^{\nu}\nabla_{\nu} m_{\mu}$ and 
${\Gamma}_{103} = l^{\mu} k^{\nu}\nabla_{\nu} \bar{m}_{\mu}$, where we can see ${\Gamma}_{103} =  \overline{\Gamma}_{102}$.
With these calculations we can recognize the 24 Ricci rotation coefficient $\Gamma_{\alpha\beta\gamma}$ as
\begin{equation}
\begin{matrix}
 \Gamma_{100} =  l^{\mu} k^{\nu}\nabla_{\nu} k_{\mu} \hspace*{1cm} &
  \Gamma_{200} =  m^{\mu} k^{\nu}\nabla_{\nu} k_{\mu}  \hspace*{1cm} &
  \Gamma_{300} =  \bar{m}^{\mu} k^{\nu}\nabla_{\nu} k_{\mu} \hspace*{1cm} &
  \Gamma_{201} =  m^{\mu} k^{\nu}\nabla_{\nu} l_{\mu} \cr 
 \Gamma_{301} = \bar{m}^{\mu} k^{\nu}\nabla_{\nu} l_{\mu}\hspace*{1cm} &
  \Gamma_{302} = \bar{m}^{\mu} k^{\nu}\nabla_{\nu} m_{\mu}\hspace*{1cm} &
  \Gamma_{110} =  l^{\mu} l^{\nu}\nabla_{\nu} k_{\mu} \hspace*{1cm} &
  \Gamma_{210} =  m^{\mu} l^{\nu}\nabla_{\nu} k_{\mu}  \cr
  \Gamma_{310} =  \bar{m}^{\mu} l^{\nu}\nabla_{\nu} k_{\mu} \hspace*{1cm} &
  \Gamma_{211} =  m^{\mu} l^{\nu}\nabla_{\nu} l_{\mu}  \hspace*{1cm} &
  \Gamma_{311} =  \bar{m}^{\mu} l^{\nu}\nabla_{\nu} l_{\mu}  \hspace*{1cm} &
  \Gamma_{312} =  \bar{m}^{\mu} l^{\nu}\nabla_{\nu} m_{\mu}   \cr
   \Gamma_{120} =  l^{\mu} m^{\nu}\nabla_{\nu} k_{\mu} \hspace*{1cm} &
   \Gamma_{220} =  m^{\mu} m^{\nu}\nabla_{\nu} k_{\mu} \hspace*{1cm} &
   \Gamma_{320} =  \bar{m}^{\mu} m^{\nu}\nabla_{\nu} k_{\mu} \hspace*{1cm} &
   \Gamma_{221} =  m^{\mu} m^{\nu}\nabla_{\nu} l_{\mu} \cr 
    \Gamma_{321} = \bar{m}^{\mu} m^{\nu}\nabla_{\nu} l_{\mu}\hspace*{1cm} &
    \Gamma_{322} = \bar{m}^{\mu} m^{\nu}\nabla_{\nu} m_{\mu}\hspace*{1cm} &
    \Gamma_{130} =  l^{\mu} \bar{m}^{\nu}\nabla_{\nu} k_{\mu} \hspace*{1cm} &
    \Gamma_{230} =  m^{\mu} \bar{m}^{\nu}\nabla_{\nu} k_{\mu} \cr
     \Gamma_{330} =  \bar{m}^{\mu} \bar{m}^{\nu}\nabla_{\nu} k_{\mu}\hspace*{1cm} &
      \Gamma_{231} = m^{\mu} \bar{m}^{\nu}\nabla_{\nu} l_{\mu}\hspace*{1cm} &
      \Gamma_{331} =  \bar{m}^{\mu} \bar{m}^{\nu}\nabla_{\nu} l_{\mu}\hspace*{1cm} &
      \Gamma_{332} =  \bar{m}^{\mu} \bar{m}^{\nu}\nabla_{\nu} m_{\mu}
  \end{matrix}
\end{equation}

Newman and Penrose classify 12 independent complex linear combination of the above Ricci rotation coefficients and called {\it spin coefficients}
as follows
\begin{eqnarray}
 &-\kappa& = \Gamma_{200} = \bar{\Gamma}_{300} = m^{\mu} k^{\nu}\nabla_{\nu} k_{\mu},\cr\cr
 &-\rho& =  \Gamma_{320} =  \bar{\Gamma}_{230} = \bar{m}^{\mu} m^{\nu}\nabla_{\nu} k_{\mu},\cr\cr
 &-\sigma& = \Gamma_{220} =  \bar{\Gamma}_{330} =  m^{\mu} m^{\nu}\nabla_{\nu} k_{\mu}, \cr\cr
 &-\tau& = \Gamma_{210} = \bar{\Gamma}_{310} =  m^{\mu} l^{\nu}\nabla_{\nu} k_{\mu}, \cr\cr
 &\nu& = \Gamma_{311} = \bar{\Gamma}_{211}= \bar{m}^{\mu} l^{\nu}\nabla_{\nu} l_{\mu},\cr\cr
 &\mu& = \Gamma_{321} = \bar{\Gamma}_{231} =\bar{m}^{\mu} m^{\nu}\nabla_{\nu} l_{\mu},\cr\cr
 &\lambda& =  \Gamma_{331} = \bar{\Gamma}_{211} = \bar{m}^{\mu} \bar{m}^{\nu}\nabla_{\nu} l_{\mu},\cr\cr
 &\pi& = \Gamma_{301} = \bar{\Gamma}_{201} = \bar{m}^{\mu} k^{\nu}\nabla_{\nu} l_{\mu},\cr\cr
 &-\epsilon& =\frac{1}{2}\left(\Gamma_{100} - \Gamma_{302} \right) = 
	    \frac{1}{2}\left ( l^{\mu} k^{\nu}\nabla_{\nu} k_{\mu} - \bar{m}^{\mu} k^{\nu}\nabla_{\nu} m_{\mu} \right),\cr\cr
  &-\beta& =\frac{1}{2}\left(\Gamma_{120} - \Gamma_{322} \right) = 
	    \frac{1}{2}\left ( l^{\mu} m^{\nu}\nabla_{\nu} k_{\mu} - \bar{m}^{\mu} m^{\nu}\nabla_{\nu} m_{\mu} \right),\cr\cr
  &\gamma& =\frac{1}{2}\left(\Gamma_{011} - \Gamma_{213} \right) = 
	    \frac{1}{2}\left ( k^{\mu} l^{\nu}\nabla_{\nu} l_{\mu} - m^{\mu} l^{\nu}\nabla_{\nu} \bar{m}_{\mu} \right),\cr\cr
  &\alpha& = \frac{1}{2}\left(\Gamma_{031} - \Gamma_{233} \right) = 
	    \frac{1}{2}\left ( k^{\mu} \bar{m}^{\nu}\nabla_{\nu} l_{\mu} - m^{\mu} \bar{m}^{\nu}\nabla_{\nu} \bar{m}_{\mu} \right).	    
\end{eqnarray}

From expression (\ref{metric_condition_1}), where we have to connection 1-forms in  pseudo-orthonormal coordinate that
\begin{equation}
 d(\gamma_{\alpha\beta}) - {\bm \Gamma}_{\alpha\beta} - {\bm \Gamma}_{\beta\alpha}=0, \nonumber
\end{equation}
and with a constant metric, a rigid frame,  $d(\gamma_{\alpha\beta})=0 $, it results to connection 1-forms that
\begin{equation}
{\bm \Gamma}_{\alpha\beta} = - {\bm \Gamma}_{\beta\alpha}. \nonumber
\end{equation}
In this frame we have
\begin{equation}
\label{pseudo_ortonormal_connection}
 {\bm\Gamma^{\alpha}}_{\beta} = \gamma^{\alpha\delta}{\bm \Gamma}_{\delta\beta},
\end{equation}
and we have that only non-null components of metric tensor from (\ref{pseudo_ortonormal_metrica}) are $\gamma^{01}= \gamma^{10}=-1$ and $\gamma^{23}= \gamma^{32}=1$ 
where we can use above equation (\ref{pseudo_ortonormal_connection}) to find the connection 1-forms ${\bm\Gamma^{\alpha}}_{\beta}$ from ${\bm \Gamma}_{\delta\beta}$.
The calculation for ${\bm\Gamma^{0}}_{0}$ follows
\begin{equation}
  {\bm\Gamma^{0}}_{0} = \gamma^{00}{\bm \Gamma}_{00}+\gamma^{01}{\bm \Gamma}_{10}
  +\gamma^{02}{\bm \Gamma}_{20} +\gamma^{03}{\bm \Gamma}_{30}, \nonumber
\end{equation}
which is simplified in
\begin{equation}
  {\bm\Gamma^{0}}_{0} = -{\bm \Gamma}_{10}. \nonumber
\end{equation}
The calculations for the others connection 1-forms are displayed below
\begin{eqnarray}
  {\bm\Gamma^{0}}_{1} &=& \gamma^{00}{\bm \Gamma}_{01}+\gamma^{01}{\bm \Gamma}_{11}
  +\gamma^{02}{\bm \Gamma}_{21} +\gamma^{03}{\bm \Gamma}_{31} = 0;\cr
  {\bm\Gamma^{0}}_{2} &=& \gamma^{00}{\bm \Gamma}_{02}+\gamma^{01}{\bm \Gamma}_{12}
  +\gamma^{02}{\bm \Gamma}_{22} +\gamma^{03}{\bm \Gamma}_{32} = - {\bm \Gamma}_{12};\cr
   {\bm\Gamma^{0}}_{3} &=& \gamma^{00}{\bm \Gamma}_{03}+\gamma^{01}{\bm \Gamma}_{13}
  +\gamma^{02}{\bm \Gamma}_{23} +\gamma^{03}{\bm \Gamma}_{33} = - {\bm \Gamma}_{13};\cr
  {\bm\Gamma^{1}}_{0} &=& \gamma^{10}{\bm \Gamma}_{00}+\gamma^{11}{\bm \Gamma}_{10}
  +\gamma^{12}{\bm \Gamma}_{20} +\gamma^{13}{\bm \Gamma}_{30}= 0; \cr
   {\bm\Gamma^{1}}_{1} &=& \gamma^{10}{\bm \Gamma}_{01}+\gamma^{11}{\bm \Gamma}_{11}
  +\gamma^{12}{\bm \Gamma}_{21} +\gamma^{13}{\bm \Gamma}_{31}= -{\bm \Gamma}_{01};\cr
   {\bm\Gamma^{1}}_{2} &=& \gamma^{10}{\bm \Gamma}_{02}+\gamma^{11}{\bm \Gamma}_{12}
  +\gamma^{12}{\bm \Gamma}_{22} +\gamma^{13}{\bm \Gamma}_{32}= -{\bm \Gamma}_{02};\cr
   {\bm\Gamma^{1}}_{3} &=& \gamma^{10}{\bm \Gamma}_{03}+\gamma^{11}{\bm \Gamma}_{13}
  +\gamma^{12}{\bm \Gamma}_{23} +\gamma^{13}{\bm \Gamma}_{33}=  -{\bm \Gamma}_{03};\cr
  {\bm\Gamma^{2}}_{0} &=& \gamma^{20}{\bm \Gamma}_{00}+\gamma^{21}{\bm \Gamma}_{10}
  +\gamma^{22}{\bm \Gamma}_{20} +\gamma^{23}{\bm \Gamma}_{30}=  -{\bm \Gamma}_{03}; \cr
   {\bm\Gamma^{2}}_{1} &=& \gamma^{20}{\bm \Gamma}_{01}+\gamma^{21}{\bm \Gamma}_{11}
  +\gamma^{22}{\bm \Gamma}_{21} +\gamma^{23}{\bm \Gamma}_{31}= -{\bm \Gamma}_{13};\cr
   {\bm\Gamma^{2}}_{2} &=& \gamma^{20}{\bm \Gamma}_{02}+\gamma^{21}{\bm \Gamma}_{12}
  +\gamma^{22}{\bm \Gamma}_{22} +\gamma^{23}{\bm \Gamma}_{32}= -{\bm \Gamma}_{23};\cr
   {\bm\Gamma^{2}}_{3} &=& \gamma^{20}{\bm \Gamma}_{03}+\gamma^{21}{\bm \Gamma}_{13}
  +\gamma^{22}{\bm \Gamma}_{23} +\gamma^{23}{\bm \Gamma}_{33}=  0;\cr
  {\bm\Gamma^{3}}_{0} &=& \gamma^{30}{\bm \Gamma}_{00}+\gamma^{31}{\bm \Gamma}_{10}
  +\gamma^{32}{\bm \Gamma}_{20} +\gamma^{33}{\bm \Gamma}_{30}=  -{\bm \Gamma}_{02}; \cr
   {\bm\Gamma^{3}}_{1} &=& \gamma^{30}{\bm \Gamma}_{01}+\gamma^{31}{\bm \Gamma}_{11}
  +\gamma^{32}{\bm \Gamma}_{21} +\gamma^{33}{\bm \Gamma}_{31}= -{\bm \Gamma}_{12};\cr
   {\bm\Gamma^{3}}_{2} &=& \gamma^{30}{\bm \Gamma}_{02}+\gamma^{31}{\bm \Gamma}_{12}
  +\gamma^{32}{\bm \Gamma}_{22} +\gamma^{33}{\bm \Gamma}_{32}= 0;\cr
   {\bm\Gamma^{3}}_{3} &=& \gamma^{30}{\bm \Gamma}_{03}+\gamma^{31}{\bm \Gamma}_{13}
  +\gamma^{32}{\bm \Gamma}_{23} +\gamma^{33}{\bm \Gamma}_{33}=  {\bm \Gamma}_{23}.
\end{eqnarray}
The above terms can be rearranged in a matrix form
\begin{equation}
\label{conexoes_pseudo-ortonomais_5}
 ({\bm\Gamma^{\alpha}}_{\beta}) = \begin{pmatrix}
               -{\bm \Gamma}_{01} & 0 & -{\bm \Gamma}_{12} & -{\bm \Gamma}_{13} \cr
                0 & -{\bm \Gamma}_{01} &  -{\bm \Gamma}_{02} & -{\bm \Gamma}_{03} \cr
                 -{\bm \Gamma}_{03} &  -{\bm \Gamma}_{13} & -{\bm \Gamma}_{23} & 0 \cr
                  -{\bm \Gamma}_{02} &  -{\bm \Gamma}_{12} & 0 & {\bm \Gamma}_{23}  \cr
                                  \end{pmatrix}.
\end{equation}
Thus with these connection 1-forms we can see the first structure equation of Cartan (\ref{eq_Cartan4}),
\begin{equation}
 d \tilde{\bm\theta}^{\alpha}=  -{\bm\Gamma^{\alpha}}_{\beta} \wedge 
\tilde{\bm\theta}^{\beta},\nonumber
\end{equation}
in pseudo-orthonormal coordinate basis that results the four below equations 
\begin{equation}
\label{equacao_de_Cartan_pseudo-ortonormal}
 \begin{cases}
  d\tilde{\bm\theta}^0 =  {\bm \Gamma}_{01}\wedge \tilde{\bm\theta}^0 + {\bm \Gamma}_{12}\wedge\tilde{\bm\theta}^2
  + {\bm \Gamma}_{13} \wedge\tilde{\bm\theta}^3 \cr
d\tilde{\bm\theta}^1  = {\bm \Gamma}_{01}\wedge \tilde{\bm\theta}^1 + {\bm \Gamma}_{02}\wedge \tilde{\bm\theta}^2
+ {\bm \Gamma}_{03}\wedge \tilde{\bm\theta}^3 \cr
d\tilde{\bm\theta}^2  = {\bm \Gamma}_{03}\wedge \tilde{\bm\theta}^0 + {\bm \Gamma}_{13}\wedge \tilde{\bm\theta}^1
+ {\bm \Gamma}_{23}\wedge \tilde{\bm\theta}^2 \cr
d\tilde{\bm\theta}^3  = {\bm \Gamma}_{02}\wedge \tilde{\bm\theta}^0 + {\bm \Gamma}_{12}\wedge \tilde{\bm\theta}^1
- {\bm \Gamma}_{23}\wedge \tilde{\bm\theta}^3
 \end{cases}
\end{equation}

The dual basis for complex null tetrad from (\ref{vielbeins_pseudo-ortonormal}), 
\begin{equation}
 \tilde{\bm\theta}^{0}  = -l_{\mu}dx^{\mu},\hspace{1cm} 
 \tilde{\bm\theta}^{1}  = -k_{\mu}dx^{\mu},\hspace{1cm}
  \tilde{\bm\theta}^{2} = \bar{m}_{\mu}dx^{\mu}\hspace{1cm}\mbox{and} \hspace{1cm}
   \tilde{\bm\theta}^{3}  = m_{\mu}dx^{\mu}\nonumber
\end{equation}
has for each component the following complex conjugate
\begin{equation}
 \overline{\tilde{\bm\theta}^{0}}  = -{l}_{\mu}dx^{\mu} = \tilde{\bm\theta}^{0},\hspace{1cm} 
 \overline{\tilde{\bm\theta}^{1}}  = -{k}_{\mu}dx^{\mu} = \tilde{\bm\theta}^{1} ,\hspace{1cm} 
 \overline{ \tilde{\bm\theta}^{2}} = \bar{\bar{m}}_{\mu}dx^{\mu}  = m_{\mu}dx^{\mu} = \tilde{\bm\theta}^{3}\hspace{1cm}\mbox{and} \hspace{1cm}
   \overline{\tilde{\bm\theta}^{3}}  = \bar{m}_{\mu}dx^{\mu} = \tilde{\bm\theta}^{2}. \nonumber
\end{equation}
The complex conjugate for first equation of system  (\ref{equacao_de_Cartan_pseudo-ortonormal}), results in
\begin{eqnarray}
  d( \overline{\tilde{\bm\theta}^0}) &=&   \overline{\bm \Gamma}_{01}\wedge  \overline{\tilde{\bm\theta}^0}
  + \overline{\bm \Gamma}_{12}\wedge \overline{\tilde{\bm\theta}^2}
  +  \overline{\bm \Gamma}_{13} \wedge \overline{\tilde{\bm\theta}^3} \nonumber  
\end{eqnarray}
where it reduces to
\begin{eqnarray}
  d \tilde{\bm\theta}^0 &=&  \overline{\bm \Gamma}_{01}\wedge  \tilde{\bm\theta}^0
  + \overline{\bm \Gamma}_{12}\wedge \tilde{\bm\theta}^3
  +  \overline{\bm \Gamma}_{13} \wedge \tilde{\bm\theta}^2 \cr
  d\tilde{\bm\theta}^0 &=&  {\bm \Gamma}_{01}\wedge \tilde{\bm\theta}^0 + {\bm \Gamma}_{12}\wedge\tilde{\bm\theta}^2
  + {\bm \Gamma}_{13} \wedge\tilde{\bm\theta}^3\nonumber  
\end{eqnarray}
and it yields
\begin{equation}
  \overline{\bm \Gamma}_{01} = {\bm \Gamma}_{01}, \hspace{1.5cm}
   \overline{\bm \Gamma}_{12}= {\bm \Gamma}_{13}. \nonumber
\end{equation}
The complex conjugate for second equation of system  (\ref{equacao_de_Cartan_pseudo-ortonormal}), results in
\begin{eqnarray}
  d( \overline{\tilde{\bm\theta}^1}) &=&   \overline{\bm \Gamma}_{01}\wedge  \overline{\tilde{\bm\theta}^1}
  + \overline{\bm \Gamma}_{02}\wedge \overline{\tilde{\bm\theta}^2}
  +  \overline{\bm \Gamma}_{03} \wedge \overline{\tilde{\bm\theta}^3} \nonumber  
\end{eqnarray}
where it reduces to
\begin{eqnarray}
  d \tilde{\bm\theta}^1 &=&   \overline{\bm \Gamma}_{01}\wedge  \tilde{\bm\theta}^1
  + \overline{\bm \Gamma}_{02}\wedge \tilde{\bm\theta}^3
  +  \overline{\bm \Gamma}_{03} \wedge \tilde{\bm\theta}^2 \cr
 d\tilde{\bm\theta}^1  &=& {\bm \Gamma}_{01}\wedge \tilde{\bm\theta}^1 + {\bm \Gamma}_{02}\wedge \tilde{\bm\theta}^2
+ {\bm \Gamma}_{03}\wedge \tilde{\bm\theta}^3 \nonumber  
\end{eqnarray}
and it yields
$  \overline{\bm \Gamma}_{02} = {\bm \Gamma}_{03}$.
The complex conjugate for third equation of system  (\ref{equacao_de_Cartan_pseudo-ortonormal}), results in 
$d\overline{\tilde{\bm\theta}^{2}}  = d\tilde{\bm\theta}^{3}$, that is
\begin{eqnarray}
  \overline{\bm \Gamma}_{03}\wedge \overline{\tilde{\bm\theta}^0} + \overline{\bm \Gamma}_{13}\wedge \overline{\tilde{\bm\theta}^1}
+ \overline{\bm \Gamma}_{23}\wedge \overline{\tilde{\bm\theta}^2} &=&  {\bm \Gamma}_{02}\wedge \tilde{\bm\theta}^0 + {\bm \Gamma}_{12}\wedge \tilde{\bm\theta}^1
- {\bm \Gamma}_{23}\wedge \tilde{\bm\theta}^3, \nonumber
\end{eqnarray}
and it results that  $  \overline{\bm \Gamma}_{23} = {\bm \Gamma}_{32}$. We verify that exchanging the indices 2 and 3 implies complex conjugation.
Thus we can write the system of equations (\ref{equacao_de_Cartan_pseudo-ortonormal}) with three equations,
\begin{equation}
\label{equacao_de_Cartan_pseudo-ortonormal_2}
 \begin{cases}
  d\tilde{\bm\theta}^0 =  {\bm \Gamma}_{01}\wedge \tilde{\bm\theta}^0 + {\bm \Gamma}_{12}\wedge\tilde{\bm\theta}^2
  + \overline{\bm \Gamma}_{12} \wedge\tilde{\bm\theta}^3 \cr
d\tilde{\bm\theta}^1  = {\bm \Gamma}_{01}\wedge \tilde{\bm\theta}^1 + {\bm \Gamma}_{02}\wedge \tilde{\bm\theta}^2
+ \overline{\bm \Gamma}_{02}\wedge \tilde{\bm\theta}^3 \cr
d\tilde{\bm\theta}^2  = \overline{\bm \Gamma}_{02}\wedge \tilde{\bm\theta}^0 + \overline{\bm \Gamma}_{12}\wedge \tilde{\bm\theta}^1
+ {\bm \Gamma}_{23}\wedge \tilde{\bm\theta}^2 
 \end{cases},
\end{equation}
where the fourth equation of system (\ref{equacao_de_Cartan_pseudo-ortonormal}) is obtained from complex conjugation of third equation of system
(\ref{equacao_de_Cartan_pseudo-ortonormal_2}). We can see that in pseudo-orthonormal basis -  complex null tetrad - the number of independent connection 1-forms $\bm\Gamma_{\alpha\beta}$ reduces from six to four.
 
\subsection{The second equation of Cartan in pseudo-orthonormal basis}

We have seen that  in pseudo-orthonormal basis there are four connection 1-forms 
${\bm \Gamma}_{01}, {\bm \Gamma}_{02}, {\bm \Gamma}_{12}$ and 
e ${\bm \Gamma}_{23}$, with complex conjugate $\overline{\bm \Gamma}_{02} = {\bm \Gamma}_{03}$, $\overline{\bm \Gamma}_{12} = {\bm \Gamma}_{13}$ and $  \overline{\bm \Gamma}_{23} = {\bm \Gamma}_{32}$. We can observe that from second equation of Cartan (\ref{curvatura_2forma4}), 
\begin{equation}
 \bm \Theta_{\alpha\beta}= d\bm\Gamma_{\alpha\beta}+
 \bm\Gamma_{\alpha\gamma} \wedge {\bm \Gamma^{\gamma}}_{\beta},\nonumber
\end{equation}
it can be solved reducing from six to four the number of independent curvature 2-forms
$\bm \Theta_{01}, \bm \Theta_{02}, \bm \Theta_{13}$
e $\bm \Theta_{32}$. 

Let us calculate the curvature 2-form $\bm \Theta_{02}$ 
\begin{eqnarray}
 \bm \Theta_{02} &=& d\bm\Gamma_{02} +  \bm\Gamma_{0\gamma} \wedge {\bm \Gamma^{\gamma}}_{2}\cr
  &=& d\bm\Gamma_{02}+ \bm\Gamma_{01} \wedge {\bm \Gamma^{1}}_{2}+ \bm\Gamma_{02} \wedge {\bm \Gamma^{2}}_{2}
  +\bm\Gamma_{03} \wedge {\bm \Gamma^{3}}_{2}, \nonumber
\end{eqnarray}
where we can see the values of  ${\bm \Gamma^{1}}_{2}, {\bm \Gamma^{2}}_{2}$ and ${\bm \Gamma^{3}}_{2}$ on matrix 
(\ref{conexoes_pseudo-ortonomais_5}), then it results in
\begin{eqnarray}
 \bm \Theta_{02} 
 &=& d\bm\Gamma_{02}+ \bm\Gamma_{01} \wedge (-{\bm \Gamma}_{02}) + \bm\Gamma_{02} \wedge (-{\bm \Gamma}_{23})
  +\bm\Gamma_{03} \wedge (0),\nonumber
\end{eqnarray}
being simplified in
\begin{equation}
\label{2-forma_pseudo-ortonomais_1}
 \bm \Theta_{02} = d\bm\Gamma_{02} + {\bm \Gamma}_{02} \wedge ( \bm\Gamma_{01}+ \bm \Gamma_{23}).
\end{equation}

The calculation for curvature 2-form $\bm \Theta_{13}$ is 
\begin{eqnarray}
 \bm \Theta_{13} &=& d\bm\Gamma_{13} +  \bm\Gamma_{1\gamma} \wedge {\bm \Gamma^{\gamma}}_{3}\cr
  &=& d\bm\Gamma_{13}+ \bm\Gamma_{10} \wedge {\bm \Gamma^{0}}_{3}+ \bm\Gamma_{12} \wedge {\bm \Gamma^{2}}_{3}
  +\bm\Gamma_{13} \wedge {\bm \Gamma^{3}}_{3} \cr
    &=& d\bm\Gamma_{13}+ \bm\Gamma_{10} \wedge(- \bm \Gamma_{13})+ \bm\Gamma_{12} \wedge (0)
  +\bm\Gamma_{13} \wedge \bm \Gamma_{23} \cr
   &=& d\bm\Gamma_{13} + \bm\Gamma_{13} \wedge (\bm\Gamma_{10} +  \bm \Gamma_{23} )\nonumber
\end{eqnarray}
where we obtain
\begin{equation}
\label{2-forma_pseudo-ortonomais_2}
 \bm \Theta_{13} = d\bm\Gamma_{13} - \bm\Gamma_{13} \wedge (\bm\Gamma_{01} +  \bm \Gamma_{32} )
\end{equation}
 
The calculation for curvature 2-form $\bm \Theta_{01}$ is
\begin{eqnarray}
 \bm \Theta_{01} 
 &=& d\bm\Gamma_{01}+ \bm\Gamma_{01} \wedge {\bm \Gamma^{1}}_{1}+ \bm\Gamma_{02} \wedge {\bm \Gamma^{2}}_{1}
  +\bm\Gamma_{03} \wedge {\bm \Gamma^{3}}_{1}\cr
  &=&d\bm\Gamma_{01}+ \bm\Gamma_{01} \wedge (-\bm \Gamma_{01})+ \bm\Gamma_{02} \wedge (-\bm \Gamma_{13})
  +\bm\Gamma_{03} \wedge (-\bm \Gamma_{12})\cr
  &=& d\bm\Gamma_{01} - \bm\Gamma_{02} \wedge \bm \Gamma_{13} - \bm\Gamma_{03} \wedge \bm \Gamma_{12}, \nonumber
\end{eqnarray}
and finally  for $\bm \Theta_{32}$, we have
\begin{eqnarray}
 \bm \Theta_{32} 
 &=& d\bm\Gamma_{32}+ \bm\Gamma_{30} \wedge {\bm \Gamma^{0}}_{2}+ \bm\Gamma_{31} \wedge {\bm \Gamma^{1}}_{2}
  +\bm\Gamma_{32} \wedge {\bm \Gamma^{2}}_{2}\cr
  &=&d\bm\Gamma_{32}+ \bm\Gamma_{30} \wedge (-\bm \Gamma_{12})+ \bm\Gamma_{31} \wedge (-\bm \Gamma_{02})
  +\bm\Gamma_{32} \wedge \bm \Gamma_{23}\cr
  &=& d\bm\Gamma_{32} + \bm\Gamma_{03} \wedge \bm \Gamma_{12} - \bm\Gamma_{02} \wedge \bm \Gamma_{13}, \nonumber
\end{eqnarray} 
here we can sum the above equations 
\begin{equation}
\label{2-forma_pseudo-ortonomais_3}
 \bm \Theta_{01} +  \bm \Theta_{32} = d( \bm\Gamma_{01}+ \bm\Gamma_{32}) + 2\bm \Gamma_{13}\wedge \bm\Gamma_{02}.
\end{equation}
Thus we can rearrange the three complex equations (\ref{2-forma_pseudo-ortonomais_1}),  (\ref{2-forma_pseudo-ortonomais_2})
and (\ref{2-forma_pseudo-ortonomais_3}) in a system of 2-forms equations of the second equation of Cartan,
\begin{equation}
\label{2-forma_pseudo-ortonomais_4}
 \begin{cases}
  \bm \Theta_{02} = d\bm\Gamma_{02} + {\bm \Gamma}_{02} \wedge ( \bm\Gamma_{01}+ \bm \Gamma_{23})\cr
   \bm \Theta_{13} = d\bm\Gamma_{13} - \bm\Gamma_{13} \wedge (\bm\Gamma_{01} +  \bm \Gamma_{32} )\cr
   \bm \Theta_{01} +  \bm \Theta_{32} = d( \bm\Gamma_{01}+ \bm\Gamma_{32}) + 2\bm \Gamma_{13}\wedge \bm\Gamma_{02}
 \end{cases}.
\end{equation}
The above system of  2-forms curvature equations has proved very useful in Newman-Penrose formalism \cite{Stephani, Griffiths}, where exact solutions of Einstein equations are obtained.

\subsection{Example of Brinkmann spacetime}

An example most important which represents gravitational waves is the Brinkmann spacetime. This spacetime can be used to explain the non-expanding waves, known as {\it plane-fronted waves with parallel rays}, or simply {\it pp}-waves. The {\it pp}-waves may represent gravitational waves, electromagnetic waves, massless radiation associated with Weyl fermions, some other forms of matter moving at the speed of light, or any combination of these. The metric for  {\it pp}-waves can be written in the form 
 \begin{equation}
  \label{metrica_Brinkmann_1}
  ds^2 =  - 2 du\,dv -2 H\, du^2 + 2d\zeta\, d\bar\zeta,
 \end{equation}
where $H=H(\zeta,\bar\zeta, u)$.

It is straightforward to notice that the shape of the Brinkmann metric is almost the same shape of the pseudo-orthonormal metric of Minkowski spacetime (\ref{metrica_pseudo-ortonormal_1})
\begin{equation}
 ds^2 = -2dudv +2 d\zeta d\bar\zeta,\nonumber
\end{equation}
where the aditional term $-2 H\, du^2$ is responsible for curvature of spacetime. When $H=0$, the metric (\ref{metrica_Brinkmann_1}) reduces to Minkowski spacetime. We can intuit that the  the easiest way to work with Brinkmann metric is with complex null tetrad. 
In non-coordinate basis pseudo-orthonormal the metric tensor is given by (\ref{metrica_pseudo-ortonormal_2}),
\begin{equation}
  \mbox{\bf g} = -2\tilde{\bm\theta}^0 \otimes \tilde{\bm\theta}^1 + 
  2\tilde{\bm\theta}^2\otimes \tilde{\bm\theta}^3, \nonumber
\end{equation}
where we can immediately identify $\tilde{\bm\theta}^2 =d\zeta$ e $\tilde{\bm\theta}^3 =d\bar\zeta$.
While the other sector of metric can be  can be identified by
\begin{equation}
 -2\tilde{\bm\theta}^0 \otimes \tilde{\bm\theta}^1 = - 2 du\,dv -2 H\, du^2, \nonumber
\end{equation}
with $\tilde{\bm\theta}^{\alpha} = {\omega^{\alpha}}_{\mu}\, dx^{\mu}$, we have that
\begin{eqnarray}
 du\otimes dv +  H\, du\otimes du &=& \tilde{\bm\theta}^0 \otimes \tilde{\bm\theta}^1 = 
 ({\omega^{0}}_{\mu}\, dx^{\mu})\otimes ({\omega^{1}}_{\nu}\, dx^{\nu}) \cr
 &=& ({\omega^{0}}_{u}\, du+{\omega^{0}}_{v}\, dv)\otimes  ({\omega^{1}}_{u}\, du+{\omega^{1}}_{v}\, dv)\cr
 &=& {\omega^{0}}_{u}{\omega^{1}}_{u}\, du\otimes du + {\omega^{0}}_{u}{\omega^{1}}_{v} du\otimes dv +
 {\omega^{0}}_{v}{\omega^{1}}_{u} dv\otimes du \cr & & +  {\omega^{0}}_{v}{\omega^{1}}_{v}dv\otimes dv , \nonumber
\end{eqnarray}
where it yields the following equations
\begin{eqnarray}
{\omega^{0}}_{u}{\omega^{1}}_{u} &=& H;\cr
 {\omega^{0}}_{u}{\omega^{1}}_{v} +  {\omega^{0}}_{v}{\omega^{1}}_{u} &=& 1; \cr
{\omega^{0}}_{v}{\omega^{1}}_{v} &=& 0.\nonumber
\end{eqnarray}
The solution for the above system of equations is displayed in a matrix,
\begin{equation}
\label{vierbeins_Brinkmann_1}
 ( {\omega^{\alpha}}_{\mu})= \begin{pmatrix}
          {\omega^{0}}_{u} & {\omega^{0}}_{v} & {\omega^{0}}_{\zeta} & {\omega^{0}}_{\bar\zeta}\cr
           {\omega^{1}}_{u} & {\omega^{1}}_{v} & {\omega^{1}}_{\zeta} & {\omega^{1}}_{\bar\zeta}\cr
            {\omega^{2}}_{u} & {\omega^{2}}_{v} & {\omega^{2}}_{\zeta} & {\omega^{2}}_{\bar\zeta}\cr
            {\omega^{3}}_{u} & {\omega^{3}}_{v} & {\omega^{3}}_{\zeta} & {\omega^{3}}_{\bar\zeta}
                             \end{pmatrix} =
\begin{pmatrix}
      H & 1 & 0 & 0\cr
      1 & 0 & 0 & 0\cr
      0 & 0 & 1 & 0\cr
      0 & 0 & 0 & 1
\end{pmatrix},
\end{equation}
thus we obtain the dual basis as
\begin{equation}
\label{formas_diferenciais_Brinkmann_1}
 \begin{cases}
 \tilde{\bm\theta}^{0}  = H du + dv\cr
 \tilde{\bm\theta}^{1}  = du\cr
  \tilde{\bm\theta}^{2} = d\zeta\cr
   \tilde{\bm\theta}^{3}  = d\bar\zeta
 \end{cases}.
\end{equation}
The matrix of vierbein is obtained from inversion of matrix $( {\omega^{\alpha}}_{\mu})$, 
\begin{equation}
\label{vierbeins_Brinkmann_2}
 ( {e_{\alpha}}^{\mu})= \begin{pmatrix}
          {e_{0}}^{u} &  {e_{0}}^{v} &  {e_{0}}^{\zeta} &  {e_{0}}^{\bar\zeta}\cr
            {e_{1}}^{u} & {e_{1}}^{v} &  {e_{1}}^{\zeta} & {e_{1}}^{\bar\zeta}\cr
            {e_{2}}^{u} &  {e_{2}}^{v} & {e_{2}}^{\zeta} & {e_{2}}^{\bar\zeta}\cr
            {e_{3}}^{u} &  {e_{3}}^{v} & {e_{3}}^{\zeta} & {e_{3}}^{\bar\zeta}
                             \end{pmatrix} =
\begin{pmatrix}
      0 & 1 & 0 & 0\cr
      1 & -H & 0 & 0\cr
      0 & 0 & 1 & 0\cr
      0 & 0 & 0 & 1
\end{pmatrix},
\end{equation}
Then the tetrad of vectors of the pseudo-orthonormal basis is 
\begin{equation}
 \begin{cases}
 \label{tetrade_Brinkmann_1}
 \hat{\bm e}_0 = {\bm k}  =  \partial_{\bm v}\cr
 \hat{\bm e}_1 =  {\bm l}  = \partial_{\bm u} - H\partial_{\bm v}\cr
  \hat{\bm e}_2 =  {\bm m} = \partial_{\bm \zeta}\cr
  \hat{\bm e}_3 =   \bar{\bm m}  =  \partial_{\bar{\bm\zeta}}
 \end{cases},
\end{equation}
the above tetrad can be expressed as
\begin{equation}
 \label{tetrade_Brinkmann_2}
 \begin{cases}
{\bm k}  =  k^{\alpha}\frac{\partial}{\partial x^{\alpha}} \cr
{\bm l}  = l^{\alpha}\frac{\partial}{\partial x^{\alpha}}\cr
{\bm m} = m^{\alpha}\frac{\partial}{\partial x^{\alpha}}\cr
  \bar{\bm m}  = \bar{m} ^{\alpha}\frac{\partial}{\partial x^{\alpha}}
 \end{cases},
\end{equation}
where 
$\frac{\partial}{\partial x^{\alpha}} =(\partial_{\bm u},\partial_{\bm v},\partial_{\bm \zeta}, \partial_{\bar{\bm\zeta}})$
 and
 \begin{equation}
  \label{tetrade_Brinkmann_3}
  k^{\alpha}=\begin{pmatrix}
           0 \cr 1 \cr 0 \cr 0
          \end{pmatrix},\hspace{0.5cm}
 l^{\alpha}=\begin{pmatrix}
           1 \cr -H \cr 0 \cr 0
          \end{pmatrix},\hspace{0.5cm}
  m^{\alpha}=\begin{pmatrix}
           0 \cr 0 \cr 1 \cr 0
          \end{pmatrix},  \hspace{0.5cm}
     \bar{m}^{\alpha}=\begin{pmatrix}
           0 \cr 0 \cr 0 \cr 1
          \end{pmatrix}.      
 \end{equation}

The exterior derivatives of 1-forms of  (\ref{formas_diferenciais_Brinkmann_1}) result that the only non-null is
$d\tilde{\bm\theta}^{0}$
\begin{equation}
 d\tilde{\bm\theta}^{0} = \frac{\partial H}{\partial \zeta}\, d\zeta\wedge du +
 \frac{\partial H}{\partial \bar\zeta}\, d\bar\zeta\wedge du +
  \frac{\partial H}{\partial u}\, du \wedge du, \nonumber
\end{equation}
here we denote $\frac{\partial H}{\partial \zeta} = H,_{\zeta}$ and the above equation reduces in
\begin{equation}
 d\tilde{\bm\theta}^{0} = H,_{\zeta}\, d\zeta\wedge du +
 H,_{\bar\zeta}\, d\bar\zeta\wedge du, \nonumber
\end{equation}
or
\begin{equation}
 d\tilde{\bm\theta}^{0} = H,_{\zeta}\, \tilde{\bm\theta}^{2}\wedge  \tilde{\bm\theta}^{1}+
 H,_{\bar\zeta}\, \tilde{\bm\theta}^{3}\wedge  \tilde{\bm\theta}^{1}.
\end{equation}
Let us put this result in the first equation of Cartan  (\ref{equacao_de_Cartan_pseudo-ortonormal_2}),
\begin{equation}
 \begin{cases}
  d\tilde{\bm\theta}^0 =  {\bm \Gamma}_{01}\wedge \tilde{\bm\theta}^0 + {\bm \Gamma}_{12}\wedge\tilde{\bm\theta}^2
  + \overline{\bm \Gamma}_{12} \wedge\tilde{\bm\theta}^3 \cr
d\tilde{\bm\theta}^1  = {\bm \Gamma}_{01}\wedge \tilde{\bm\theta}^1 + {\bm \Gamma}_{02}\wedge \tilde{\bm\theta}^2
+ \overline{\bm \Gamma}_{02}\wedge \tilde{\bm\theta}^3 \cr
d\tilde{\bm\theta}^2  = \overline{\bm \Gamma}_{02}\wedge \tilde{\bm\theta}^0 + \overline{\bm \Gamma}_{12}\wedge \tilde{\bm\theta}^1
+ {\bm \Gamma}_{23}\wedge \tilde{\bm\theta}^2 
 \end{cases},\nonumber
\end{equation}
where we obtain
\begin{equation}
 \begin{cases}
  -H,_{\zeta}\, \tilde{\bm\theta}^{1}\wedge  \tilde{\bm\theta}^{2} -
 H,_{\bar\zeta}\, \tilde{\bm\theta}^{1}\wedge  \tilde{\bm\theta}^{3} =
 -{\bm \Gamma}_{01}\wedge \tilde{\bm\theta}^0 + {\bm \Gamma}_{12}\wedge\tilde{\bm\theta}^2
  + {\bm \Gamma}_{13} \wedge\tilde{\bm\theta}^3 \cr
0  = {\bm \Gamma}_{01}\wedge \tilde{\bm\theta}^1 + {\bm \Gamma}_{02}\wedge \tilde{\bm\theta}^2
+ \overline{\bm \Gamma}_{02}\wedge \tilde{\bm\theta}^3 \cr
0  = \overline{\bm \Gamma}_{02}\wedge \tilde{\bm\theta}^0 + \overline{\bm \Gamma}_{12}\wedge \tilde{\bm\theta}^1
+ {\bm \Gamma}_{23}\wedge \tilde{\bm\theta}^2 
 \end{cases}, \nonumber
\end{equation}
the solution of this system results in
\begin{equation}
 {\bm \Gamma}_{12} = -H,_{\zeta}\, \tilde{\bm\theta}^{1} \hspace{1cm}\mbox{and} \hspace{1cm}
  {\bm \Gamma}_{13} = -H,_{\bar\zeta}\, \tilde{\bm\theta}^{1}.
\end{equation}
The only independent non-null Ricci rotation coefficient is ${\bm \Gamma}_{12} = \overline {\bm \Gamma}_{13}$.

We can solve the second equation of Cartan for Brinkmann spacetime with (\ref{2-forma_pseudo-ortonomais_4})
\begin{equation}
 \begin{cases}
  \bm \Theta_{02} = d\bm\Gamma_{02} + {\bm \Gamma}_{02} \wedge ( \bm\Gamma_{01}+ \bm \Gamma_{23})\cr
   \bm \Theta_{13} = d\bm\Gamma_{13} - \bm\Gamma_{13} \wedge (\bm\Gamma_{01} +  \bm \Gamma_{32} )\cr
   \bm \Theta_{01} +  \bm \Theta_{32} = d( \bm\Gamma_{01}+ \bm\Gamma_{32}) + 2\bm \Gamma_{13}\wedge \bm\Gamma_{02}
 \end{cases}.\nonumber
\end{equation}
Here we can see that just the second equation of the above system has the nonzero 1-form  ${\bm \Gamma}_{13}$. Then the only nonzero curvature 2-form is
\begin{eqnarray}
\label{curvatura_2forma_Brinkmann}
  \bm \Theta_{13} &=& d\bm\Gamma_{13} = d(  -H,_{\bar\zeta}\, \tilde{\bm\theta}^{1}) 
  = -\frac{\partial H,_{\bar\zeta}}{\partial u}du \wedge \tilde{\bm\theta}^{1}  -\frac{\partial H,_{\bar\zeta}}{\partial \zeta}d\zeta \wedge \tilde{\bm\theta}^{1}  -\frac{\partial H,_{\bar\zeta}}{\partial \bar\zeta}d\bar\zeta \wedge \tilde{\bm\theta}^{1}\cr
  &=&  -H,_{\zeta\bar\zeta}\,\,\tilde{\bm\theta}^2 \wedge  \tilde{\bm\theta}^1 
  -H,_{\bar\zeta\bar\zeta}\,\,\tilde{\bm\theta}^3 \wedge  \tilde{\bm\theta}^1 
\end{eqnarray}
From definition of curvature 2-form (\ref{curvatura_2forma}), we have
\begin{equation}
 {\bm \Theta^{\alpha}}_{\beta}=\frac{1}{2}  {R^{\alpha}}_{\beta\gamma\delta}\tilde{\bm\theta}^{\gamma}
\wedge \tilde{\bm\theta}^{\delta},\nonumber
\end{equation}
such that
 \begin{eqnarray}
 \bm \Theta_{13} &=& R_{1301}\, \tilde{\bm\theta}^0 \wedge  \tilde{\bm\theta}^1 +
 R_{1302}\, \tilde{\bm\theta}^0 \wedge  \tilde{\bm\theta}^2+ 
 R_{1303}\, \tilde{\bm\theta}^0 \wedge  \tilde{\bm\theta}^3+
 R_{1312}\, \tilde{\bm\theta}^1 \wedge  \tilde{\bm\theta}^2\cr
 & & + R_{1313}\, \tilde{\bm\theta}^1 \wedge  \tilde{\bm\theta}^3+
 R_{1323}\, \tilde{\bm\theta}^2 \wedge  \tilde{\bm\theta}^3 , \nonumber
 \end{eqnarray}
 by comparing the above result with  (\ref{curvatura_2forma_Brinkmann}), we have the following components of Riemann tensor
\begin{equation}
 R_{1312} = H,_{\zeta\bar\zeta}, \hspace{1cm} \mbox{and} \hspace{1cm} R_{1313} = H,_{\bar\zeta\bar\zeta}.\nonumber
\end{equation}

The components of the Ricci tensor can be computed via $R_{\alpha\beta} = \gamma^{\delta\epsilon} R_{\epsilon\alpha\delta\beta}$. 
We use the symmetry of Riemann tensor where $ R_{3121} = H,_{\zeta\bar\zeta}$. The complex conjugate of this component is
\begin{equation}
 \bar{R}_{3121} = R_{2131} = H,_{\zeta\bar\zeta}. \nonumber
\end{equation}
There is only one nonzero component of Ricci tensor in pseudo-orthonormal basis
\begin{equation}
 R_{11} = \gamma^{23} R_{3121} + \gamma^{32} R_{2131} =  2 H,_{\zeta\bar\zeta}. \nonumber
\end{equation}
The Ricci tensor is given by
\begin{equation}
 \bm{Ric} = R_{\alpha\beta}  \tilde{\bm\theta}^{\alpha} \otimes  \tilde{\bm\theta}^{\beta}, \nonumber
\end{equation}
we recall from (\ref{base_NC2}) where $\tilde{\bm\theta}^{\alpha} = {\omega^{\alpha}}_{\mu}dx^{\mu}$, to rewrite the tensor Ricci in coordinate basis as
\begin{equation}
 \bm{Ric} = R_{\alpha\beta}  ({\omega^{\alpha}}_{\mu}dx^{\mu}) \otimes  ({\omega^{\beta}}_{\nu}dx^{\nu}),\nonumber
\end{equation}
we have  only $R_{11}\neq 0$, then it results in
\begin{equation}
 \bm{Ric} = R_{11} {\omega^{1}}_{\mu} {\omega^{1}}_{\nu}\,\, dx^{\mu} \otimes  dx^{\nu},\nonumber
\end{equation}
and with ${\omega^{1}}_{\mu} = -k_{\mu}$ from (\ref{vielbeins_pseudo-ortonormal}) the  Ricci tensor in cordinate basis is given by
\begin{equation}
 \bm{Ric} = 2 H,_{\zeta\bar\zeta} \,k_{\mu}k_{\nu} \,\, dx^{\mu} \otimes  dx^{\nu},\nonumber
\end{equation}
where we can identify the components of Ricci tensor in coordinate basis as
\begin{equation}
 \label{tensor_Ricci_Brinkmann_2}
R_{\mu\nu}= 2 H,_{\zeta\bar\zeta} \, k_{\mu}k_{\nu} .
\end{equation}
If we use (\ref{tetrad_NP_2}) we have that the matrix of Ricci tensor in coordinate basis is 
\begin{equation}
(R_{\mu\nu})=  H,_{\zeta\bar\zeta} \begin{pmatrix}
                                  1 & -1 & 0 & 0\cr
                                  -1 & 1 & 0 & 0 \cr
                                  0 & 0 & 0 & 0 \cr
                                  0 & 0 & 0 & 0
                                 \end{pmatrix} .\nonumber
\end{equation}
because $  g^{\mu\nu} k_{\mu}k_{\nu} = k_{\mu}k^{\mu} = 0 $, we have that the scalar curvature $R = g^{\mu\nu} R_{\mu\nu}$ is zero.	
Then, according to Einstein' General Relativity, the curvature os spacetime is related to the distribution of matter, where specifically the components of the Ricci tensor are directly related to the local energy-momentum tensor $T_{\mu\nu}$ by Einstein's field equations. We have from (\ref{tensor_Ricci_Brinkmann_2}) that the energy-momentum tensor $T_{\mu\nu}$ of Brinkmann spacetime is given by
\begin{equation}
 \label{tensor_energia_momento_null_dust}
T_{\mu\nu}= \rho \, k_{\mu}k_{\nu} .
\end{equation}
The above tensor is the energy-momentum tensor of {\it null dust}, where $\rho$ denotes the energy density.
The energy-momentum tensor (\ref{tensor_energia_momento_null_dust}) has a similar structure as that of dust (pressure-free perfect
fluid) this explains the notation {\it null dust}. Null dust or pure radiation or incoherent radiation, represents a simple matter source in General Relativity and describes the flux of massless particles, for example photons,  with a fixed propagation direction given by the null vector field 
$k_{\mu}$.

\section{Conclusion}

The purpose of this manuscript was to provide a self-contained review of manipulate some objects such as connection coefficients  and curvature tensors of a spacetime $({\cal M},{\bm g})$ in  non-coordinate basis with concepts of vierbein field.
In a traditional manner the authors that introduce the General Relativity \cite{DInverno,Landau,Hobson,Schutz,Wald,Weinberg}, have taught  to calculate curvature tensors directly from metric tensor $g_{\mu\nu}$ in coordinate basis. However for an advanced way, some times it is necessary to calculate the connection coefficients and curvature tensors with aid of Cartan's structure equations which deal with vierbein fields ${e_{\alpha}}^{\mu}$, the 'square root' of the metric field tensor.
So, we have seen that the non-coordinate bases, orthonormal and pseudo-orthonormal (complex null tetrad),  are useful for pedagogical study of  conection coefficients and curvature in language of differential forms and also they are useful to manipulating and pratice the Cartan's structure equations. This manuscript has the  introduction of complex null tetrad, the pseudo-orthonormal basis, that is useful to the formalism of  Newman-Penrose applied to General Relativity. The Newman-Penrose formalism adopts  a tetrad in which the two axes $u$ (advanced) and $v$ (retarded) along the direction of propagation are chosen to be  lightlike, while the two complex axes $\zeta$ and $\bar\zeta$ are transverse to direction of propagation. 
It provides a particular powerful way to deal with fields that propagates at the speed of light as we have seen with the example of Brinkmann spacetime \cite{Stephani, Stewart, Newman, Griffiths}.


\bibliographystyle{elsarticle-num}
\bibliography{<your-bib-database>} 







\end{document}